\RequirePackage{lineno}
\documentclass[aps,prc,superscriptaddress,showpacs,twocolumn,floatfix]{revtex4}

\usepackage{graphicx,colordvi}
\usepackage{color}          
\usepackage{csvsimple}
\usepackage{rotating}
\usepackage{lineno} 
\usepackage{amsmath}

\def\degree{^\circ}

\newcommand{\LD}{$^{2}\mathrm{H}$}
\newcommand{\HET}{$^{3}\mathrm{He}$}
\newcommand{\HEF}{$^{4}\mathrm{He}$}
\newcommand{\Be}{$^{9}\mathrm{Be}$}
\newcommand{\C}{$^{12}\mathrm{C}$}

\newcommand{\Cu}{$^{63}\mathrm{Cu}$}
\newcommand{\Au}{$^{197}\mathrm{Au}$}

\def\gtorder{\mathrel{\raise.3ex\hbox{$>$}\mkern-14mu \lower0.6ex\hbox{$\sim$}}}
\def\ltorder{\mathrel{\raise.3ex\hbox{$<$}\mkern-14mu \lower0.6ex\hbox{$\sim$}}}

\def\etal{\emph{et al.}}

\begin{document}

\title{Measurement of the EMC effect in light and heavy nuclei}

\author{J.~Arrington}
\affiliation{Lawrence Berkeley National Laboratory, Berkeley, California 94720, USA}
\affiliation{Argonne National Laboratory, Lemont, Illinois 60439, USA}

\author{J.~Bane}
\affiliation{University of Massachusetts, Amherst, Massachusetts 01003, USA}
\affiliation{University of Tennessee, Knoxville, Tennessee 37966, USA}

\author{A.~Daniel}
\affiliation{University of Houston, Houston, Texas 77044, USA}
\affiliation{University of Virginia, Charlottesville, Virginia 22904, USA}

\author{N.~Fomin}
\affiliation{Los Alamos National Laboratory, Los Alamos, New Mexico 87545, USA}
\affiliation{University of Tennessee, Knoxville, Tennessee 37966, USA}
\affiliation{University of Virginia, Charlottesville, Virginia 22904, USA}

\author{D.~Gaskell}
\affiliation{Thomas Jefferson National Accelerator Facility, Newport News, Virginia 23606, USA}

\author{J.~Seely}
\affiliation{Laboratory for Nuclear Science, Massachusetts Institute of Technology, Cambridge, Massachusetts 02139, USA}

\author{R.~Asaturyan}
\thanks{Deceased}
\affiliation{A.I. Alikhanyan National Science Laboratory (Yerevan Physics Institute), 02 Alikhanyan Brothers Str., Yerevan 0036, Armenia}

\author{F.~Benmokhtar}
\affiliation{University of Maryland, College Park, Maryland 20742, USA}

\author{W.~Boeglin}
\affiliation{Florida International University, Miami, Florida 33199, USA}

\author{P.~Bosted}
\affiliation{Thomas Jefferson National Accelerator Facility, Newport News, Virginia 23606, USA}

\author{M.H.S.~Bukhari}
\affiliation{University of Houston, Houston, Texas 77044, USA}

\author{M.E.~Christy}
\affiliation{Hampton University, Hampton, Virginia 23669, USA}

\author{S.~Connell}
\thanks{Present address: University of Johannesburg, Johannesburg, South Africa}
\affiliation{University of Virginia, Charlottesville, Virginia 22904, USA}

\author{M.M.~Dalton}
\affiliation{University of Virginia, Charlottesville, Virginia 22904, USA}
\affiliation{Thomas Jefferson National Accelerator Facility, Newport News, Virginia 23606, USA}

\author{D.~Day}
\affiliation{University of Virginia, Charlottesville, Virginia 22904, USA}

\author{J.~Dunne}
\affiliation{Mississippi State University, Mississippi State, Mississippi 39762, USA}

\author{D.~Dutta}
\affiliation{Mississippi State University, Mississippi State, Mississippi 39762, USA}
\affiliation{Triangle Universities Nuclear Laboratory, Duke University, Durham, North Carolina 27710, USA}

\author{L.~El Fassi}
\affiliation{Argonne National Laboratory, Lemont, Illinois 60439, USA}

\author{R.~Ent}
\affiliation{Thomas Jefferson National Accelerator Facility, Newport News, Virginia 23606, USA}

\author{H.~Fenker}
\affiliation{Thomas Jefferson National Accelerator Facility, Newport News, Virginia 23606, USA}

\author{H.~Gao}
\affiliation{Laboratory for Nuclear Science, Massachusetts Institute of Technology, Cambridge, Massachusetts 02139, USA}
\affiliation{Triangle Universities Nuclear Laboratory, Duke University, Durham, North Carolina 27710, USA}

\author{R.J.~Holt}
\affiliation{Argonne National Laboratory, Lemont, Illinois 60439, USA}

\author{T.~Horn}
\affiliation{University of Maryland, College Park, Maryland 20742, USA}
\affiliation{Thomas Jefferson National Accelerator Facility, Newport News, Virginia 23606, USA}
\affiliation{Catholic University of America, Washington, DC 20064, USA}

\author{E.~Hungerford}
\affiliation{University of Houston, Houston, Texas 77044, USA}

\author{M.K.~Jones}
\affiliation{Thomas Jefferson National Accelerator Facility, Newport News, Virginia 23606, USA}

\author{J.~Jourdan}
\affiliation{Basel University, Basel, Switzerland}

\author{N.~Kalantarians}
\affiliation{University of Houston, Houston, Texas 77044, USA}

\author{C.E.~Keppel}
\affiliation{Thomas Jefferson National Accelerator Facility, Newport News, Virginia 23606, USA}
\affiliation{Hampton University, Hampton, Virginia 23669, USA}

\author{D.~Kiselev}
\thanks{Present address: Paul Scherrer Institut (PSI), 5232 Villigen, Switzerland}
\affiliation{Basel University, Basel, Switzerland}

\author{A.F.~Lung}
\affiliation{Thomas Jefferson National Accelerator Facility, Newport News, Virginia 23606, USA}

\author{S.~Malace}
\affiliation{Hampton University, Hampton, Virginia 23669, USA}

\author{D.G.~Meekins}
\affiliation{Thomas Jefferson National Accelerator Facility, Newport News, Virginia 23606, USA}

\author{T.~Mertens}
\affiliation{Basel University, Basel, Switzerland}

\author{H.~Mkrtchyan}
\affiliation{A.I. Alikhanyan National Science Laboratory (Yerevan Physics Institute), 02 Alikhanyan Brothers Str., Yerevan 0036, Armenia}

\author{G.~Niculescu}
\affiliation{James Madison University, Harrisonburg, Virginia 22807, USA}

\author{I.~Niculescu}
\affiliation{James Madison University, Harrisonburg, Virginia 22807, USA}

\author{D.H.~Potterveld}
\affiliation{Argonne National Laboratory, Lemont, Illinois 60439, USA}

\author{C.~Perdrisat}
\affiliation{College of William and Mary, Wiliamsburg, Virginia, 23185, USA}

\author{V.~Punjabi}
\affiliation{Norfolk State University, Norfolk, Virginia 23529, USA}

\author{X.~Qian}
\affiliation{Triangle Universities Nuclear Laboratory, Duke University, Durham, North Carolina 27710, USA}

\author{P.E.~Reimer}
\affiliation{Argonne National Laboratory, Lemont, Illinois 60439, USA}

\author{J.~Roche}
\affiliation{Thomas Jefferson National Accelerator Facility, Newport News, Virginia 23606, USA}

\author{V.M.~Rodriguez}
\affiliation{University of Houston, Houston, Texas 77044, USA}

\author{O.~Rondon}
\affiliation{University of Virginia, Charlottesville, Virginia 22904, USA}

\author{E.~Schulte}
\affiliation{Argonne National Laboratory, Lemont, Illinois 60439, USA}

\author{K.~Slifer}
\affiliation{University of Virginia, Charlottesville, Virginia 22904, USA}

\author{G.R.~Smith}
\affiliation{Thomas Jefferson National Accelerator Facility, Newport News, Virginia 23606, USA}

\author{P.~Solvignon}
\thanks{Deceased}
\affiliation{Argonne National Laboratory, Lemont, Illinois 60439, USA}

\author{V.~Tadevosyan}
\affiliation{A.I. Alikhanyan National Science Laboratory (Yerevan Physics Institute), 02 Alikhanyan Brothers Str., Yerevan 0036, Armenia}

\author{L.~Tang}
\affiliation{Thomas Jefferson National Accelerator Facility, Newport News, Virginia 23606, USA}
\affiliation{Hampton University, Hampton, Virginia 23669, USA}

\author{G.~Testa}
\affiliation{Basel University, Basel, Switzerland}

\author{R.~Trojer}
\affiliation{Basel University, Basel, Switzerland}

\author{V.~Tvaskis}
\affiliation{Hampton University, Hampton, Virginia 23669, USA}

\author{F.R.~Wesselmann}
\affiliation{University of Virginia, Charlottesville, Virginia 22904, USA}

\author{S.A.~Wood}
\affiliation{Thomas Jefferson National Accelerator Facility, Newport News, Virginia 23606, USA}

\author{L.~Yuan}
\affiliation{Hampton University, Hampton, Virginia 23669, USA}

\author{X.~Zheng}
\affiliation{Argonne National Laboratory, Lemont, Illinois 60439, USA}
\affiliation{University of Virginia, Charlottesville, Virginia 22904, USA}

\noaffiliation 

\date{\today}

\begin{abstract}

Inclusive electron scattering from nuclear targets has been measured to
extract the nuclear dependence of the inelastic cross section ($\sigma_A$) 
in Hall C at the Thomas Jefferson National Accelerator facility.  Results 
are presented for $^2$H, $^3$He, $^4$He, $^9$B, $^{12}$C, $^{63}$Cu and 
$^{197}$Au at an incident electron beam energy of 5.77 GeV for a range 
of momentum transfer from $Q^2$=2 to 7 (GeV/c)$^2$.  These data improve
the precision of the existing measurements of the EMC effect in the 
nuclear targets at large $x$, and allow for more detailed examinations 
of the $A$ dependence of the EMC effect.  

\end{abstract}
\pacs{13.60.Hb,25.30.Fj,24.85.+p}

\maketitle

\maketitle



\section{Introduction} \label{intro.sec}

Quantum chromodynamics (QCD) is the theory governing the strong interaction,
with quarks and gluons as elementary degrees of freedom. The interaction
between quarks is mediated by gluons as the gauge bosons. Understanding QCD in
terms of the elementary quark and gluon degrees of freedom remains the
greatest unsolved problem of strong interaction physics. The challenge
arises from the fact that quarks and gluons cannot be examined in isolation.
The degrees of freedom observed in nature (hadrons and nuclei) are different
from the ones typically used in the QCD formalism (quarks and gluons).
However, detailed studies of the structure of hadrons, mainly protons and
neutrons, provide a wealth of information on the nature of QCD. Thus, one of
the main goals of the strong interaction physics is to understand how the
fundamental quark and gluon degrees of freedom give rise to the nucleons and to
inter-nucleon forces that bind nuclei.

The investigation of deep-inelastic scattering of leptons from the nucleon is
one of the most effective ways for obtaining fundamental information on the
quark-gluon substructure of the nucleon. Nuclear structure functions are sensitive the
impact of the nucleon binding and motion in the nucleus, as well as possible
modification to the structure of a nucleon in the nuclear medium. Measurements
by the European Muon Collaboration~\cite{aubert83} showed the unexpected
result that the nuclear structure functions differed significantly from the
sum of proton and neutron distributions.  This observation was dubbed the
`EMC effect', and is still the focus of experimental and theoretical efforts
to understand the origin of these differences in detail.  We describe here
an experiment where electrons were scattered from the free proton and several
nuclear targets to better understand the possible modification of hadron
properties in the nuclear environment, with a focus on light nuclear targets.

In the remainder of this section we briefly discuss electron scattering, structure functions 
and introduce the kinematics. In section~\ref{intro.sec}, we discuss the EMC 
effect and briefly survey the findings of earlier experimental and theoretical investigations
and discuss the physics motivation behind the present
experiment. Section~\ref{apparatus.sec} gives an overview of the experimental
apparatus used to collect the presented data. Section~\ref{anal.sec} describes
the data analysis procedures and section~\ref{syst.sec} discusses the details
of the systematic uncertainties. The final results are presented in
section~\ref{results.sec} with conclusions and an overview of the results
given in section~\ref{concl.sec}.

\subsection{Kinematics and definitions} \label{kinem.ssec}

Consider electron scattering off a stationary nucleon through the
exchange of a single virtual photon,
\begin{equation}
e^{-}(k) + N(P) \longrightarrow e^{-}(k^{'}) + X ~,
\end{equation}
where $k$ and $k^{'}$ are the four momenta of the initial and scattered
electrons and $P$ is the four momentum of the target nucleon. The four
momentum of the incoming electron is $k= ( E, \overrightarrow k ) $ and of the
scattered electron is $k= ( E^{'}, \overrightarrow {k ^{'}}) $. Since the target
is at rest in the laboratory frame, its four momentum is $P = ( M,
\overrightarrow 0 ) $ where $M$ is the nucleon rest mass. Experimentally, the
produced hadrons $X$ are not detected in inclusive electron scattering. The scattering
process takes place through the electromagnetic interaction by the exchange of
a virtual photon $\gamma^{*}$, with energy, $\nu=E - E^{'}$ and momentum
$\overrightarrow q$. In the laboratory frame, ignoring the electron mass,
one can express $Q^2$, the negative of the four momentum transfer squared, as
$Q^2=4EE^{'}\sin^2\left( \theta/2\right)$, where $\theta$ is the electron scattering angle in the lab frame, and the invariant mass of the
final hadronic system as $W=\sqrt{M^2+2M\nu - Q^2}$. The Bjorken scaling
variable, $x=Q^2/2M\nu$, represents the longitudinal momentum fraction of the
hadron carried by the interacting parton in the infinite momentum frame. For
electron scattering from a free nucleon, $x$ ranges from 0 to 1. For scattering
from a nucleus of mass number $A$, $x$ ranges from 0 to
$M_A/M \approx A$.

In terms of the deep-inelastic structure functions $F_1(x,Q^2)$
and $F_2(x,Q^2)$, the differential cross section for scattering of an
unpolarized electron in the laboratory frame can be written as
\begin{eqnarray} \nonumber \label{xsec_eqn}
\frac{d^2\sigma}{d\Omega dE^{'}} &=& \frac{4\alpha^2E^{'2}}{Q^4} \cos^2(\theta/2) \big[ F_2(x,Q^2)/\nu \\
&&+ 2\tan^2(\theta/2) F_1(x,Q^2)/M \big] ~,
\end{eqnarray}
where $\alpha$ is the fine structure constant. For
brevity, this doubly differential cross section is denoted by the symbol
$\sigma$. When $Q^2$ and $\nu \rightarrow \infty$, the structure functions
will only depend on the ratio $Q^2/\nu$ or equivalently on the variable $x$
\cite{bjorken_scaling}. Thus, in this scaling region the structure functions
are simply a function $x$. In the quark parton model (QPM), this scaling
behavior is due to the elastic scattering from moving quarks inside the nucleon. In
this model, the structure function $F_2$ is given by
\begin{eqnarray} \label{qpmf2_eqn}
 F_2(x)= \sum_f e_f^2 x q_f(x) ~.
\end{eqnarray}
where the distribution function $q_f(x)$ is the expectation value of the
number of partons of flavor $f$ (up, down, strange...) in the hadron, whose
longitudinal momentum fraction lies within the interval $[x, x+dx]$ and $e_f$
is the charge of the parton.

In the region of deep-inelastic
scattering (DIS), the structure functions do not scale exactly, and instead
depend logarithmically on $Q^2$.  This is a consequence of QCD, in which the
parton distribution functions (PDFs) are not scale independent, but evolve
with $Q^2$.  The logarithmic scaling violations associated with QCD
do not break down the connection between the structure function and the
underlying PDFs, but simply reflect the scale-dependence of the PDFs.

Along with the $Q^2$ dependence associated with QCD, additional power
corrections appear at lower $Q^2$ values, mainly at large $x$.  So-called
``target mass corrections''~\cite{Schienbein_tarmass_rev} yield deviations from
scaling at finite $Q^2$ values arising from terms neglected in the high-$Q^2$
approximations used in the ideal scaling limit.  In addition, higher-twist
effects, associated with breakdown of the assumption of incoherent elastic
scattering from individual quarks at lower $Q^2$, also modify the scaling
behavior.  This is most clearly manifested in the appearance of clear
structures in the inclusive structure function associated with production of
individual resonances.

Analogous to the absorption cross section for real photons, the $ F_1$ and $
F_2$ structure functions can be expressed in terms of longitudinal
($\sigma_L$) and transverse ($\sigma_T$) virtual-photon cross sections
\begin{equation}\label{rosenrxsec_eqn}
 \frac{d^2\sigma}{d\Omega dE^{'}} =
\Gamma \left[ \sigma_T \left( x, Q^2\right) +
\epsilon \, \sigma_L \left( x, Q^2\right) \right]~,
\end{equation}
where $\epsilon =\Gamma_L/\Gamma_T= \big[1+2~(1+Q^2/4M^2x^2)
\tan^2\frac{\theta}2\big] ^{-1}$ is the virtual polarization parameter,
$\Gamma$ is the virtual photon flux, and $\Gamma_L$ and $\Gamma_T$ defines the
probability that a lepton emits a longitudinally or transversely polarized
virtual photon.

The ratio of longitudinal to transverse virtual-photon absorption cross
section is given by
\begin{eqnarray} \label{r_eqn}
R(x,Q^2) = \frac{\sigma_L}{\sigma_T} = \left[ \left(1+\frac{\nu^2}{Q^2} \right)
\frac{M}{\nu} \frac{F_2(x,Q^2) }{F_1(x,Q^2) }\right]-1 ~.
\end{eqnarray}
Using equations~\ref{xsec_eqn} and~\ref{r_eqn}, the per-nucleon cross section
(cross section divided by the total nucleon number) ratios for two different
nuclei $A_1$ and $A_2$ can be written as
\begin{eqnarray} 
\frac{\sigma_{A_1}}{\sigma_{A_2}} =
\frac{F_2^{A_1}\, (1+\epsilon \, R_ {A_1}) \, (1+ R_ {A_2}) }
     {F_2^{A_2}\, (1+\epsilon \, R_ {A_2}) \, (1+ R_ {A_1}) }~. 
\end{eqnarray}
Note that when $\epsilon=1$ or $R_{A_1} = R_{A_2}$, the ratio of the $F_2$
structure functions is identical to the cross section ratio. In this and all
previous extractions of the EMC effect, it is assumed that $R$ is target
independent, and therefore the cross section ratios correspond to the 
$F_2$ structure function ratios.

Because the structure functions depend
on $Q^2$, the ratio may also have a $Q^2$ dependence which we must account
for in comparing our data to measurements at other different $Q^2$ values.
However, the effect of QCD evolution on the ratios should be essentially
negligible as the evolution is nearly identical for all nuclei, and so cancels
in the ratio. The main effect of the target mass corrections can be applied with a simple change of variables from Bjorken-$x$ to
Nachtmann-$\xi$ when comparing measurements at different $Q^2$
(Sec.~\ref{hixcor.ssec}).  Thus, in kinematics where any remaining higher-twist
contributions are small or $A$ independent, the comparison of EMC ratios from
experiments at different $Q^2$ values is straightforward. Accounting for this
change of variables mentioned above, it has been shown that the EMC ratios are
independent of $Q^2$ down to very low values of $Q^2$ and $W^2$, well below
the typically-defined DIS regime~\cite{Arrington:2003nt, niculescu06}.

\subsection{EMC effect}\label{emc.ssec}

Nuclei consist of protons and neutrons bound together by the strong nuclear
force, with binding energies of 1--2\% of the nucleon mass, and characteristic
momenta below 200--300 MeV/c. Because DIS involves incoherent scattering from
the quarks, and the energy and momentum scales associated with nuclear binding
are small compared to the external scales in DIS, the naive assumption was
that the nuclear structure function in high-energy scattering from a nucleus
with $Z$ protons and $N$ neutrons would simply be the sum of the proton and
neutron structure functions:
\begin{equation} \label{f2asimple1_eqn}
F^A_2(x,Q^2) = Z F^{p}_2(x,Q^2) + N F^{n}_2(x,Q^2) ~.
\end{equation}

Even before the discovery of the EMC effect, Fermi motion of the nucleons
in the nucleus was known to play a role in nuclear structure functions.
While the typical scale of the Fermi momentum is small compared
to the momentum scale of the probe, the longitudinal component is directly
added to the momentum of the virtual photon and cannot be completely 
neglected. It is necessary to perform a convolution of the PDFs of the proton
and neutron with the momentum distribution of the nucleons in the
nucleus~\cite{Akulinichev:1985xq}:
\begin{equation}\label{conv_eqn}
F_2^A(x)=\int_{x}^A dz\, f_{N}^A(z,\epsilon) F_2^{N}\left(\frac{x}{z}\right) ~,
\end{equation}
where the longitudinal momentum distribution function $f_{N}^A(z,\epsilon)$
for the nucleon is given by,
\begin{equation}\label{conv2_eqn}
f_{N}^A(z,\epsilon)=
\int d^{4}p~S_N(p)~\delta\left(z-\left(\frac{p \cdot q}{M_N q_{0}}\right) \right) ~.
\end{equation}
Here, $S_N(p)$ is the spectral function of the nucleus (assumed to be
identical for protons and neutrons), $z$ is the light-cone
momentum carried by the nucleon and $\epsilon$ its removal energy. The
four-momenta of the struck nucleon and virtual photon are given by $p$
and $q$, where $q_{0}$ is the energy transferred by the virtual photon. One
can think of the convolution as ``smearing'' the nucleon PDF in $x$, yielding
little change where the PDF is relatively flat in $x$, and larger effects
where it grows or falls rapidly. Calculations showed that the effect was
minimal at low $x$ values, but that the convolution has a large impact for $x
\gtorder 0.6$, where the PDF of the nucleon falls rapidly~\cite{bodek81a,
bodek81b, frankfurt81}.

Therefore, it came as a surprise when this expectation was shown to be
incorrect by measurements which showed significant effects on the nuclear
PDF for nearly all values of $x$~\cite{aubert83}. As part of a
comprehensive study of muon scattering, the European Muon Collaboration
compared data from iron with data from deuterium by forming a per-nucleon
structure function ratio of these targets. Since the $x$ distributions of up
and down quarks differ, yielding different structure functions for the
proton and neutron, EMC ratios are usually taken as a ratio of a heavy
isoscalar target to deuterium. This cancels out the contribution due to
the difference between the proton and neutron structure function but
yields a ratio which depends on the nuclear effects in both the heavy
nucleus and the deuteron. For non-isoscalar nuclei, a correction is typically
applied to estimate the effect of the neutron excess in heavy nuclei. Note
that many calculations provide the ratio of the heavy nucleus to the sum of
free proton and neutron structure functions. This provides a more direct
measure of the nuclear effects in the nucleus, but cannot directly be compared
to the data, as the lack of a free neutron target makes direct measurements of
the free neutron structure function impossible.

When plotted as a function of $x$, the EMC ratio shows significant deviation
from unity. The deviation of this ratio from unity was unexpected, and, this
$A$ dependence in deep-inelastic scattering is known as the
EMC effect. This discovery had a significant impact on views of the
structure of nuclei, and has spurred discussion of the importance of the
concepts of quarks, gluons and QCD to nuclear physics.

Though the boundaries are somewhat arbitrary, generally the $x$ dependence of
the cross section ratios are divided into four regions in $x$. The gross
features of the data are; (1) the region $x<0.1$, where the nuclear cross sections
are suppressed (known as the shadowing region); (2) the region $0.1<x<0.3$, where
the nuclear cross sections are slightly enhanced compared to nucleon cross
sections (anti-shadowing region); (3) the region $0.3<x<0.7$, where a large
suppression of the nuclear cross section is observed (``EMC effect'' region),
and (4) the region $x>0.7$ where the $A/D$ cross section ratio increases and grows beyond unity
due to the convolution (``Fermi smearing'') effects.

\subsection{Previous measurements of the EMC effect}\label{prevexpt.ssec}

After the initial observation of an unexpected nuclear dependence in the
structure functions of heavy nuclei~\cite{aubert83}, additional measurements were
performed at both CERN~\cite{cern_na2prime1, cern_na2prime2, cern_bcdms1} and
SLAC~\cite{slace49b,slace61,slace87,slace139}. Further measurements by the
European Muon Collaboration (EMC)~\cite{aubert86} and the New Muon Collaboration
(NMC)~\cite{Arneodo:1995cs,Amaudruz:1995tq} significantly improved the
precision and kinematic range of measurements at low $x$, mapping out in
detail the shadowing region for a range of nuclei. The HERMES collaboration
also measured DIS cross sections on several nuclear targets including
$^3$He~\cite{hermes_ackerstaff,hermes_correction_airapetian}. The data
in the anti-shadowing region are consistent with unity, while data at
higher $x$ have large uncertainties.

Focusing on the high-$x$ region, SLAC experiment E139~\cite{slace139}
mapped out the EMC effect for $^4$He, Be, C, Al, Ca, Fe,
Ag, and Au in the range $0.09<x<0.9$ and $2<Q^2<15$~GeV$^2$. Examining
the target ratios, and in particular their deviations from unity,
the experiment showed no significant $Q^2$ dependence and an identical $x$
dependence for all nuclei, although the high-$x$ behavior of $^4$He appeared
to differ, but not in a significant fashion given its large uncertainties.
The $A$-dependence of the nuclear effects could be parameterized several
different ways: varying logarithmically with A, linearly with A$^{-1/3}$,
or being proportional to the average nuclear density (assuming a uniform
sphere based on the measured nuclear charge radius). Exploiting the local density approximation~\cite{antonov86}, it was found that the EMC effect scales as A$^{-1/3}$ which allowed for data from finite nuclei to be extrapolated to infinite nuclear matter~\cite{sick92}.

The universal $x$ dependence and weak A dependence for heavy nuclei makes
it difficult to evaluate models of the EMC effect~\cite{geesaman95,
Norton_emcreview, sargsian03}.  In addition, the EMC effect at very large $x$ values ($>$0.7) had not been well measured. The typical DIS requirement, $W^2 >
4$~GeV$^2$, yields extremely high $Q^2$ measurements for $x \gtorder 0.8$,
where the cross sections are extremely small. However, an extraction of
EMC ratios from JLab experiment E89008 in the resonance region
($1.2<W^2<3.0$~GeV$^2$ with $Q^2 = 3$--4 GeV$^2$) demonstrated that the
nuclear effects in the resonance region and DIS region are
identical~\cite{Arrington:2003nt}. This implies that relaxing the constraint
on $W^2$ may allow for measurements at larger $x$ values than previously
accessed.  Precise measurements at large $x$ allow for tests of the
convolution model where other effects are expected to be small, providing a
constraint on the convolution effects which must be accounted for at all $x$
values.

\subsection{Theoretical models}\label{theory.ssec}

Even though it has been almost four decades since the discovery of the EMC effect
and there are extensive data on its $x$ and $A$ dependence for $A \ge 12$,
there is no clear consensus as to its origin. The EMC effect has been under
intense theoretical and experimental study since the original observation (see
the reviews~\cite{geesaman95, Norton_emcreview, sargsian03, malace14, fomin17} and references
therein). The models used to explain the observed effect range from
traditional nuclear descriptions in terms of pion exchange or binding energy
shifts, to QCD inspired descriptions that include effects from  dynamical rescaling, multi-quark
clustering and deconfinement in nuclei, some of which involve changes to
the nucleon's internal structure when in the dense nuclear medium.

Traditional calculations begin with the convolution model, where the nucleon
motion modifies the effective $x$ and $Q^2$ values of the e--N
interaction, such that the virtual photon probes a modified quark
distribution compared to the stationary nucleons.  In general, these
convolution calculations result in a suppression of the nuclear
structure function at large $x$, but do not describe the full depletion
observed in EMC effect measurements. Another drawback of the convolution
calculations is that they often fail to describe the nuclear dependence of the Drell-Yan
reaction observed by the Fermilab E772 collaboration~\cite{Alde_DY_E772:1990}. 

Although convolution calculations can be improved with the addition of binding effects, 
Miller and Smith~\cite{Smith:2002ci, Miller:2001tg} have demonstrated that binding alone is insufficient
to reproduce the EMC effect. However, these calculations do not include off-shell effects.
The calculation by Benhar~\etal~\cite{benhar_emccalcnucmatt} uses nuclear wave functions that
include high momentum tails in the nucleon momentum distribution while adding a model to handle the off-shell nucleon cross-section effects.  The combination of these two ingredients results in a significant depletion of
the structure function at large $x$ (larger than the observed EMC effect) and the addition of contributions from ``nuclear pions'' is required to provide quantitative agreement with EMC measurements.
Kulagin and Petti~\cite{Kulagin:2004ie} also start from a convolution approach including binding effects, shadowing, and contributions from nuclear pions,  yielding roughly half of the observed EMC effect. Off-shell effects are then introduced and their contribution is tuned to give good agreement
with the experimental data.  These calculations that predict a significant role for off-shell effects are particularly interesting in light of potential explanations for the observed correlation between the size of the EMC effect and the number of short-range correlated pairs (SRCs) in nuclei~\cite{weinstein11, Hen:2012fm, arrington12c}.

Frankfurt and Strikman~\cite{Frankfurt_photon_pdf2010, Frankfurt:2012qs} account for
some of the deficit in the momentum-sum rule for the nucleons by a modification
to the Coulomb field of the nucleus.  Starting from a convolution model which
uses the separation energy, accounting for the momentum in the Coulomb
field simply accounts for loss of momentum from the nucleons; it does not
yield an additional suppression of the structure function at large $x$.
However, it would suggest that proposed modifications to the nuclear pion field,
used to explain the deficit of the momentum sum rule in some calculations, may
be overestimated in heavy nuclei where the modification of the Coulomb field is more significant.

Additional contributions that have been examined are virtual constituents of
the nucleus which are not present in a nucleon.  In the dense environment of
a nucleus, one may have color-singlet clusters of 6, 9,... valence
quarks~\cite{jaffe83, pirner11} or hidden-color
configurations~\cite{Bashkanov:2013cla, West:2020rlk}. The
PDFs of these exotic objects is expected to differ significantly from the
sum of individual nucleons, leading to a modification to the nuclear PDF. Estimates
of such clusters predict a modest contribution to the PDFs in the EMC region,
but show a larger impact that may be experimentally accessible at larger $x$, as
such configurations contribute to the structure function well beyond
$x=1$~\cite{geesaman95, sargsian03, arrington06, fomin10}.

Finally, some calculations invoke a modification to the internal structure
of individual nucleons within the dense medium of the nucleus.
Different rescaling models~\cite{close83,Nach_pirner, chanfray:1984} have been
proposed to explain the EMC effect, based on a change in the nucleon radius
due to partial deconfinement in the nuclear medium. In terms of QCD, a change
in confinement means a change in $Q^2$. Thus, QCD evolution starts at lower
$Q^2$ for a free nucleon, and, hence, the QCD radiative processes per nucleon
are larger in a bound nucleon than in a free nucleon.
In this case, scaling is referred to as ``dynamic'' because of the
evolution of the quark, anti-quark and gluon distributions.
Close~\etal~\cite{close85} shows that an increase in confinement size could
explain the data on a medium nucleus such as iron but fail to explain the data
for $x \gtorder 0.65$, since there is no inclusion of Fermi motion effects.

There are other models involving medium-modified nucleons that do not use
a rescaling of $Q^2$.  In such models the quark wave function of a nucleon is
modified by external fields generated by the surrounding nucleons. Quark-meson
coupling models~\cite{Saito:1994ki} include the effect of the nuclear medium
by allowing quarks in nucleons to interact via meson exchange and additional
vector and scalar fields. These models have been applied to the study the EMC
effect in unpolarized and polarized \cite{Cloet:2005rt, Cloet:2006bq}
structure functions, as well as other observables for nuclei and nuclear
matter~\cite{Saito:2005rv}. In addition, calculations for finite
nuclei~\cite{Cloet:2006bq} show a significant difference between the polarized
and unpolarized EMC effect and also predict flavor dependent effects~\cite{Cloet:2009qs}. 
Recent work by Miller and Smith use a chiral
soliton model to relate nucleon form factor modification~\cite{Smith:2004dn},
the EMC effect in polarized~\cite{Smith:2005ra} and
unpolarized~\cite{Smith:2002ci} structure functions.

\subsection{Physics motivation behind E03103}\label{motivation.ssec}

The experiment reported here, JLab E03103, was designed to precisely map out
the $x$, $Q^2$ and $A$-dependence of inclusive electron scattering from
light to medium heavy nuclei, with emphasis on light nuclei and the large $x$
region~\cite{E03103proposal}. Results for the EMC ratios for the light nuclei
have been reported in reference~\cite{seely09}.  The analysis presented in this
work uses an updated isoscalar correction prescription (described in Sec.~\ref{iso.ssec})
as well as a slightly modified radiative correction scheme (see Sec.~\ref{rc.ssec}) as compared
to reference~\cite{seely09}.  The impact of these modifications on the light target results
is not large (at most ~1\% for the isoscalar correction and 0.6\% for the radiative 
corrections), but does result in slightly different cross section ratios.

While the EMC effect has been well measured in heavy nuclei, the SLAC E139
ratios for $^4$He have large uncertainties and there were no previous measurements on
$^3$He in the valence region. Data on light nuclei are important in
understanding the microscopic origin of the EMC effect as they allow direct
comparison to detailed few-body calculations with minimal nuclear structure
uncertainties. Data on light nuclei can also help constrain nuclear effects in
the deuteron which are critical to the extraction of the neutron structure
function from measurements on the deuteron~\cite{whitlow92, Arrington:2008zh,
accardi10, accardi11, arrington12b}.  Light nuclei allow for better tests
of the $A$ dependence of the EMC effect, while also providing measurements of
nuclei more similar to the deuteron in mass and density.

In addition, studies of short-range correlations~\cite{frankfurt93,
arrington99, egiyan03, egiyan06, shneor07, subedi08, fomin12, arrington12a}
suggest that high-density configurations play an important role in nuclei,
which could potentially yield a modification of the nucleon structure function
in overlapping nucleons~\cite{geesaman95, arrington01, sargsian03,
arrington04, arrington06, fomin10}.  If two-body effects have a significant
contribution to the EMC effect, then the EMC effect could be different in
few-body nuclei than it does in heavy nuclei, where the effects may be
saturated. There were also models which predicted a very different $x$
dependence for the EMC effect for $A$=3,4~\cite{smirnov99, burov99,
afnan03}, so the inclusion of light nuclei was considered
important as a way to look for two-body effects as a possible source of medium
modification in nucleon structure.

Beyond the focus on light nuclei, E03103 emphasized large $x$,
where Fermi motion and binding effect dominate. Because of the lack of data
in this region and the limited data for few-body nuclei, many calculations of
the EMC effect are performed for nuclear matter and extrapolated to lower
density when comparing to nuclear parton distributions.  In such cases,
the important contributions of binding and Fermi motion are not modeled
in detail, making it difficult to isolate contributions beyond these more
conventional effects.

While many models mentioned in the previous section have had some
success, most are incomplete.  They may work only in a limited $x$ range,
conflict with limitations set by other measurements, or explain the data while
neglecting Fermi motion and binding. However, it is clear that the effects of
binding and Fermi motion are important and contribute over the entire $x$
region, not just at the largest $x$ values.  The large $x$ data are
particularly sensitive to these effects and to the details of nuclear
structure.  As such, precise high-$x$ data for both light and heavy nuclei can
help to constrain these effects.

\section{Experimental Apparatus}\label{apparatus.sec}

Experiment E03103 was carried out in Hall C in 2004 at the Thomas Jefferson National
Accelerator Facility (JLab)~\cite{Leemann:2001dg}. The unpolarized
electron beam from the Continuous Electron Beam Accelerator Facility was
incident on solid, liquid and high pressure gas targets. The High Momentum Spectrometer 
(HMS) (a magnetic focusing spectrometer) was used to detect the scattered electrons. The nominal electron
beam energy ($E$) was measured with the Hall C arc energy
measurement~\cite{johna_thesis}, the scattered momentum ($E^{'}$) and angle ($\theta$)
are reconstructed from the particle trajectory in the HMS.

\subsection{Experiment kinematics}\label{exptkinem.ssec}

\begin{figure}[htb]
\begin{center}
\includegraphics[width=85mm, height=60mm, trim={8mm 2mm 15mm 15mm}, clip]{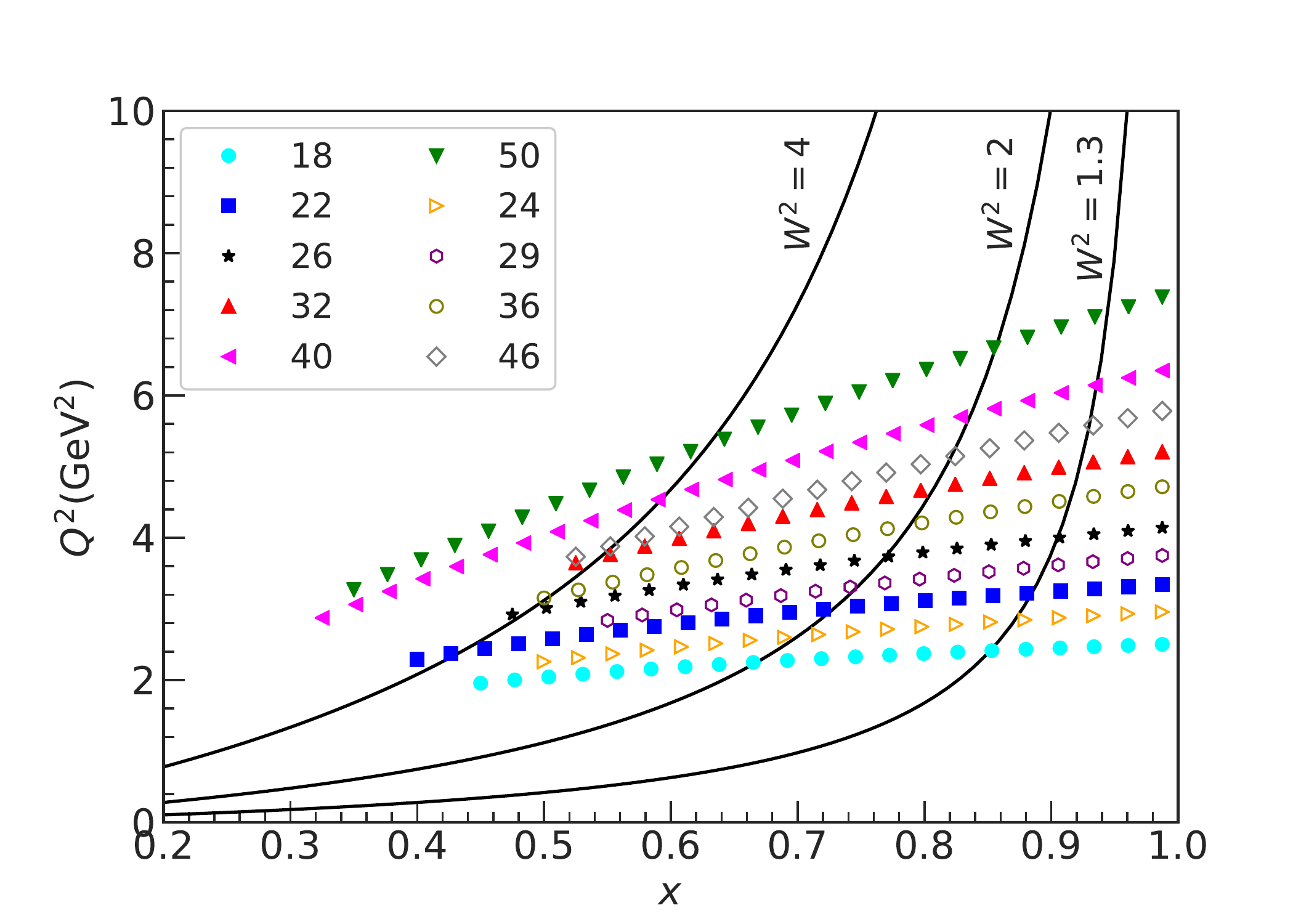}
\caption{(Color online) Kinematic coverage for the experiment. Contours of
constant invariant mass squared are shown with black lines. Different colors
represent different angles, given in the legend. Closed symbols were taken at
$E$=5.776~GeV beam energy and open symbols at 5.01~GeV.}
\label{xemkinem_fig}
\end{center}
\end{figure}

Most of the data for the experiment were taken at 5.776 GeV beam energy with
beam currents of 30--80~$\mu A$. The cryogenic targets $^2$H, $^3$He,$^4$He
and solid targets $^9$B, $^{12}$C, $^{63}$Cu and $^{197}$Au were used for EMC ratio measurements while $^1$H was used primarily for calibration. Data
on all targets were taken at 40$^\circ$ and
50$^\circ$, and the cross section ratios with respect to
deuterium were extracted. At high $x$, the kinematics were not in the
conventional DIS region ($W^2>4\,\mathrm{GeV^2}$), so additional data were
taken for $^{12}$C and $^2$H at 8 additional kinematic settings, half at
$E$=5.776~GeV and half at 5.01~GeV, as shown in Fig.~\ref{xemkinem_fig}.

\subsection{Targets}\label{target.ssec}

\begin{figure}[htb]
\begin{center}
\includegraphics[width=85mm, angle=0, trim={5mm 5mm 5mm 10mm}, clip]{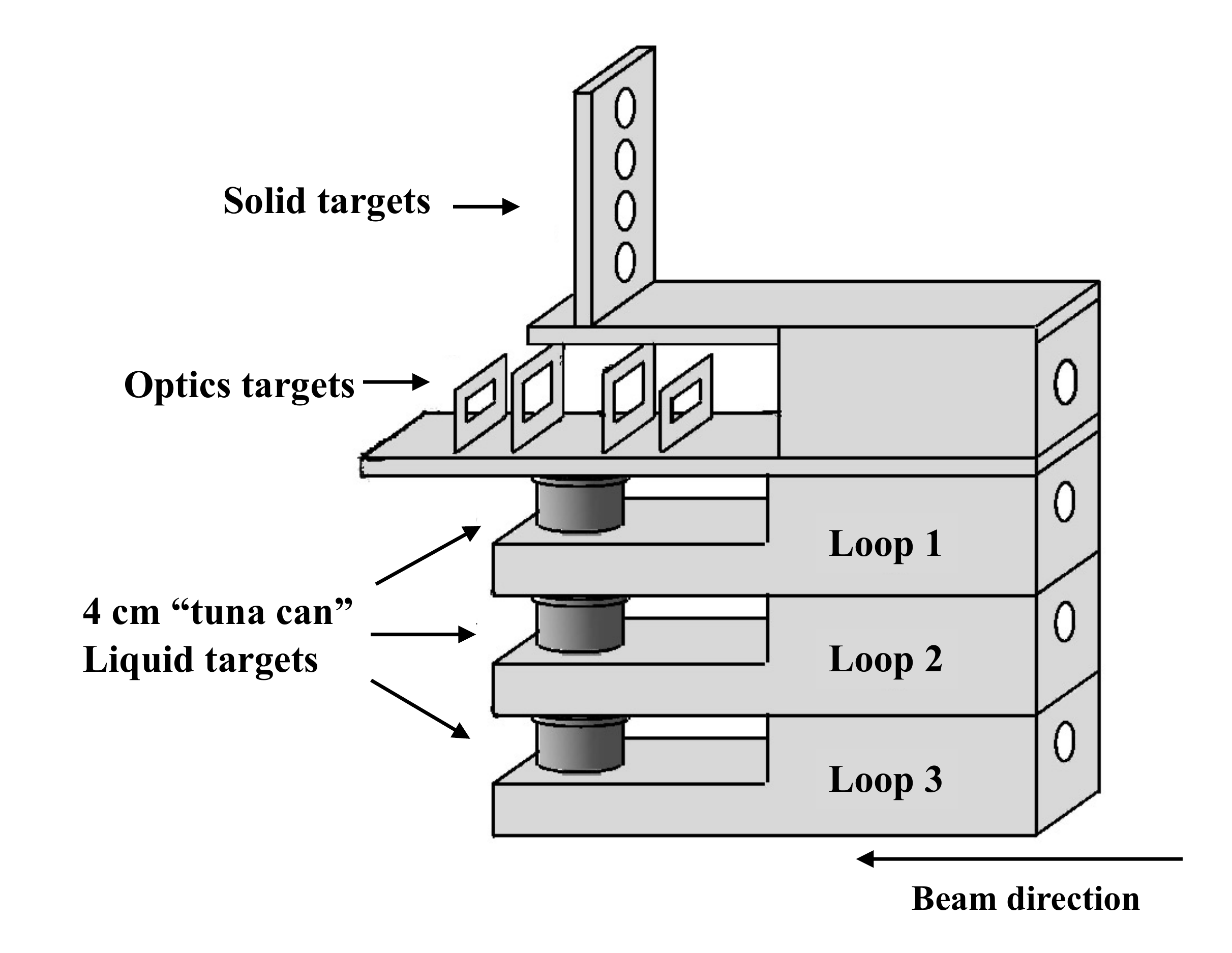}
\caption{A schematic side view of Hall C target ladder.}
\label{tarladder_fig}
\end{center}
\end{figure}

E03103 measured inclusive electron scattering from a wide range of nuclei
using both cryogenic and solid targets. This experiment used the
standard Hall C target ladder (see Fig.~\ref{tarladder_fig}) which was placed
inside a vertical cylindrical vacuum scattering chamber. The scattering
chamber had entrance and exit openings for the beam as well as a vacuum
pumping port and several view ports. The beamline was connected directly to the
scattering chamber, so the beam did not pass through any solid entrance
window. There were two cutouts on the chamber for the two spectrometers to
detect the scattered particles, which are covered with thin (0.41 mm) aluminum windows.

\begin{table}[htb]
\begin{center}
\caption{Nominal cryotarget dimensions. Here, $\left<t\right>$ represents the
average offset-corrected cryogen in the path of the beam and R.R.L is the
relative radiation length (material thickness as a fraction of its radiation length \label{cryotarthick_tab})}
\begin{tabular}{|l|c|c|c|c|c|}
\hline
Target & $\left<t\right>$ & Density & Areal thickness &R.R.L & Purity\\ 
& (cm) & (g/cm$^3$) & (g/cm$^2$) &($\%$) &($\%$)\\
\hline
$^1$H  & 3.865 & 0.0723 & 0.2794(36) & 0.456 &99.99\\
$^2$H  & 3.860 & 0.167  & 0.6446(83) & 0.526 &99.95\\
$^3$He & 3.865 & 0.0708 & 0.2736(51) & 0.419 &99.9\\
$^4$He & 3.873 & 0.135  & 0.5229(85) & 0.554 &99.99\\
\hline
\end{tabular}
\end{center}
\end{table}

The target assembly contained several loops for cryogenic targets and the solid
target ladder was attached above the optics sled. The target stack could be
raised or lowered by an actuator in order to put the desired target in the
beam path. The cryogenic targets were contained in vertical cylindrical Al cans
with a diameter of $\approx 4$ cm. Each loop consisted of a circulation fan, a
target cell, heat exchangers and high powered heaters. The target liquid in
each loop was cooled with helium gas using a heat exchanger. The liquid moved
continuously through the heat exchanger, to the target cell and back. A high
power heater regulated the temperature of the cryogenic targets, compensating
for the power deposition by the beam during low current or beam off periods.
Solid targets were attached above the optics sled and all the foils in the
solid target ladder were separated vertically.

The optics sled contained a dummy target, which consisted of two aluminum
foils (aluminum alloy Al-6061-T6 - identical to the cryotarget endcaps)
placed $\sim4$ cm apart. These dummy targets
mimicked the cell walls of the cryogenic target and facilitated the
measurement of the background originating from the cell walls. The dummy
targets were flat aluminum foils and were approximately 8 times thicker than
the walls of liquid targets to reduce the time needed for background
measurement.

\begin{table}[htb]
\begin{center}
\caption{Solid target dimensions, relative radiation length, and purity. Here, Al(1)
and Al(2) represents the aluminum foils which mimicked the cell walls of
cryogenic target.}
\label{solidltarthick_tab}
\begin{tabular}{|l|c|c|c|c|}
\hline
Target & Density &Areal thickness & R. R. L. & Purity\\ 
& (g/cm$^3$) & (g/cm$^2$) &($\%$) & ($\%$)\\
\hline
Be    & 1.848 & 1.8703(94) & 2.87 & 99.0 \\
C     & 2.265 & 0.6667(40) & 1.56 & 99.95\\
Cu    & 8.96  & 0.7986(40) & 6.21 & 99.995 \\
Au    & 19.32 & 0.3795(38) & 5.88 & 99.999\\
Al(1) & 2.699 & 0.2626(13) & 1.09 & 98.0\\
Al(2) & 2.699 & 0.2633(13) & 1.10 & 98.0\\
\hline
\end{tabular}
\end{center}
\end{table}

Areal thicknesses of the cryotargets were computed (see
Table~\ref{cryotarthick_tab}) from the target density and the length of the
cryogen in the path of the beam. Since the target cans were cylindrical, the
effective target length seen by the beam differed from the diameter of the can
if the beam did not intersect the geometrical center of the targets, and a
correction accounting for beam offset was applied run-by-run. The target
density was calculated using the knowledge of temperature and pressure.

Thicknesses of the solid targets were calculated using measurements of the
mass and area of the targets. For solid targets, there is an uncertainty in
the effective thickness due to uncertainty in angle of the target relative to
beam direction, but this is estimated to be $<0.01\%$. Solid targets used in
the experiment and their dimensions are given in
Table~\ref{solidltarthick_tab}.  No correction is applied for the $\sim$1\%
impurities in the $^9$Be target, as the cross section per nucleon for $^9$Be
and heavier nuclei differs at the few percent level, so the correction is
typically $\ll$0.1\%.

\subsection{High-Momentum Spectrometer}\label{hms.ssec}

\begin{figure}[htb]
\begin{center}
\includegraphics[width=0.48\textwidth]{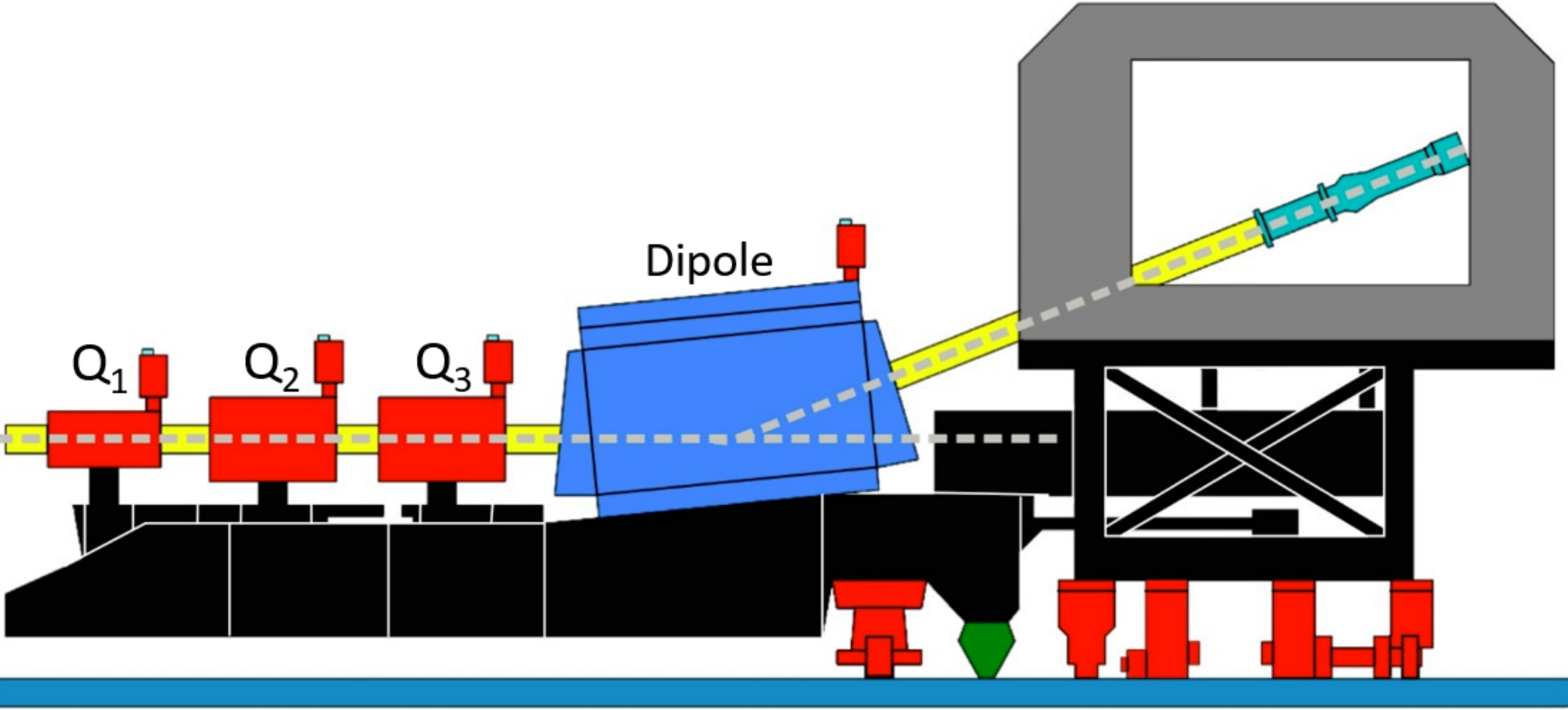}
\caption{(Color online) Schematic side view of the HMS. The first
three magnets (red) are Q1, Q2, and Q3; the blue magnet is the Dipole. Adapted from ref.~\cite{johna_thesis}.}
\label{hmsside_fig}
\end{center}
\end{figure}

E03103 used the HMS to detect the scattered electrons. 
The HMS is a $25\degree$ vertical-bend spectrometer that consists of
three quadrupole magnets, one dipole magnet and a detector package. The
detectors are housed inside a concrete enclosure and this shield hut, along 
with the HMS magnets, are
mounted on a steel carriage which can be rotated on a pair of concentric rails
to the desired scattering angle.  An octagonal collimator is placed before the
entrance to the first magnet which is used to define the acceptance for a
short target for particles within approximately 10\% of the central momentum
setting. A schematic side view of the HMS is shown in
Figure~\ref{hmsside_fig}. All magnets in the HMS are superconducting and are
cooled with 4K liquid helium. The focusing properties and acceptance of the
HMS are determined by the quadrupole magnets, and the central momentum is
determined by the dipole.  The spectrometer volume is under vacuum with thin
(0.5 mm) mylar-kevlar windows at the entrance (before the collimator) and exit (after
the dipole, in the detector hut). See Ref.~\cite{johna_thesis, Dutta:2003yt} for more
details on the spectrometer and detector package.

There are two drift chambers in the HMS located at the front of the
detector stack~\cite{baker95}. The drift chambers are used to find the
position and trajectory of the particle at the focal plane, which are used to
reconstruct the position and momentum of the scattered particle at the
interaction vertex. Two sets of $x-y$ scintillators hodoscopes were used for
triggering and time-of-flight measurements~\cite{johna_thesis}. The detector stack also
contains a threshold gas $\mathrm{\check{C}erenkov}$ counter used for electron
identification~\cite{johna_thesis}. The HMS $\mathrm{\check{C}erenkov}$ detector is a large
cylindrical tank (inner diameter $\approx 150$ cm and length $\approx 165$
cm). It has two front reflecting mirrors which focus the light onto two PMTs.
The circular ends of the tank are covered with 0.1 cm aluminum windows. For
E03103, the detector was filled with 5.15 psi ($\sim 0.35$ atmospheres) of
Perfluorobutane ($\mathrm{C_4F_{10}}$) at room temperature. At this pressure
and temperature, the index of refraction of the gas is 1.00050, yielding
a threshold momentum of 16 MeV for electrons and 4.4 GeV for pions. The pion
threshold was above the momentum range of E03103 except for the lowest
angles, where the $\pi$/e ratio is small and the separation between electrons
and pions in the calorimeter is sufficient to yield a negligible pion
background.

A lead glass calorimeter detector~\cite{mkrtchyan13} was used in conjunction
with the $\mathrm{\check{C}erenkov}$ detector for electron identification. The
HMS calorimeter consists of 10 cm$\times$10 cm$\times$70 cm blocks of TF-1 lead
glass, positioned at the rear of the detector hut. The blocks are arranged in
four layers with 13 blocks per layer for a total thickness of 14.6 radiation
lengths along the particle direction. The calorimeter blocks are calibrated by using the gas $\mathrm{\check{C}erenkov}$ detector to identify a clean sample of electrons, and the scale factor applied ADC signals from the individual blocks are adjusted to provide a spectrum peaked at the electron momentum as determined from the tracking. Thus, Electrons (or positrons) entering the calorimeter deposit their entire energy and the normalized energy spectrum, $E_\text{cal}/E^{'}$, is peaked around 1. Pions typically deposit $\sim 300 $ MeV in the calorimeter and the $E_\text{cal}/E^{'}$ distribution peaks around 0.3
GeV/$E^{'}$.

\section{Data Analysis}\label{anal.sec}

The data acquisition system used for E03103 was the CODA (CEBAF Online Data
Acquisition) software package. CODA events from the individual run files were
decoded by the standard Hall C replay software (ENGINE). It reads the raw data 
written by the data acquisition system, decodes the detector hits, locates possible
tracks and particle identification information for each event, and calculates
different physics variables. Input and output of the ENGINE are handled using
the CEBAF Test Package (CTP). ENGINE makes use of CERN HBOOK libraries and
provides output as ASCII report files (scalers, integrated charge $\ldots$),
histogram files (ADC/TDC spectra for different detectors) and the
reconstructed event-by-event data as ntuples. Detailed cuts, corrections and
other analysis details will be discussed in the following sections.

\subsection{Methodology of Cross Section Extraction}\label{proc.ssec}

The measured inclusive electron scattering cross section at scattered electron
energy $E^{\prime}$ and a central angle $\theta_{c}$ was extracted using a simulation
of the electron scattering process via the ratio method,
\begin{equation}\label{csmaster1.eq}
 \sigma^\text{Born}_\text{data}(E^{\prime},\theta_{c}) = \frac{Y_\text{data}}{Y_\text{sim}}\, \sigma^\text{Born}_\text{model}(E^{\prime},\theta_{c})
\end{equation}
where $\sigma^\text{Born}_\text{data}(E^{\prime},\theta_{c}) $ denotes the differential
cross section $\frac{d^2\sigma(E^{\prime},\theta_{c})}{dE^{\prime}d\Omega}$,
$Y_\text{sim}$ represents the simulated yield which includes the features of the
detector acceptance and the model radiated cross section, $Y_\text{data}$ is the
charge normalized yield integrated over the acceptance of the experiment and
$\sigma^\text{Born}_\text{model}(E^{\prime},\theta_{c})$ represents the Born model cross
section.  To the extent that the simulation properly includes the corrections,
efficiencies, and acceptance, the ratio of experimental to simulated yield
will simply reflect the error in the initial cross section model.

$Y_\text{data}$ is the number of detected electrons, averaged over the
kinematics, divided by the efficiency- and deadtime-corrected luminosity of the
measurement, so that $Y_\text{data}$ represents the normalized yield for an ideal
detector averaged over the acceptance of the experiment. The calculation of
$Y_\text{sim}$ must result in the same acceptance-averaged normalized yield,
and so must include a detailed model of the acceptance as well as all of the
physics effects required to go from the starting Born cross section model to
the final observed counts, i.e. radiative effects, multiple scattering,
energy loss, etc....  In addition, because this is the integrated yield over
the acceptance, the cross section model must do a reasonable job of accounting
for the cross section variation across the acceptance. Note that the
\textit{position-dependent} inefficiencies are applied to the simulation,
rather than the data, as discussed in Sec.~\ref{cercut.sssec}. Energy loss
is included event-by-event in the simulation, to yield a realistic distribution.
A single correction for the median energy loss was applied to both data and
simulation to remove the average kinematic offsets.

\subsubsection{Extraction of experimental yield}\label{datayield.sssec}

Each kinematic setting contains data taken over one or more runs. Each run is
analyzed separately, with detector and acceptance cuts applied and the
efficiency and other experimental correction factors calculated run-by-run.
The efficiency-corrected and charge-normalized yield for all the runs in a
given setting, with
\begin{equation}\label{ydata.eq}
Y^\text{tot}_\text{data}=\frac{\sum_i N(i)}{N_\text{sc}~\sum_i C_\text{data}(i) \,Q_\text{tot}(i) },
\end{equation}
where $N_i$ is the total number of events that passed all cuts for the $i^{th}$
run in the given setting, $Q_\text{tot}(i)$ is the total accumulated charge and
$N_\text{sc}$ is the number of scattering centers in the target; $N_\text{sc}=\rho t
N_A/M$ where $\rho$ is the density, $t$ is the thickness, $M$ is the atomic
mass of the target and $N_A$ is Avogadro's number. The factor $C_\text{data}(i) $
in Eq.~\ref{ydata.eq} is the correction factor which includes experimental
efficiencies and live times (fraction of time that the DAQ and computer readout systems are active); $C_\text{data} = PS/(\varepsilon_\text{trig} \times
\varepsilon_\text{track} \times \varepsilon_\text{det} \times t_\text{comp} \times t_\text{elec})$
where PS is the prescale factor used to control the trigger rate when the data
is taken, $\varepsilon_\text{trig}$ corrects for the events lost due to
inefficiency at the trigger level, $\varepsilon_\text{track}$ is the tracking
efficiency, $\varepsilon_\text{det}$ denotes the global detector efficiencies, and
$t_\text{comp}$ and $ t_\text{elec}$ are the computer and electronic live time,
respectively.

Because we are only interested in primary beam electrons which scatter
in the target, we have to subtract the contribution of electrons which
scatter in the target entrance and exit windows (for the cryogenic targets)
and secondary electrons which come from other processes. The subtraction of
the cryotarget endcap contribution is discussed in Sec.~\ref{bg.sssec}, and the
secondary electrons in Sec.~\ref{csbg.sssec}.

\subsubsection{Extraction of simulated yield}\label{simyield.sssec}

In order to evaluate $Y_\text{sim}$ one needs to account for the finite acceptance
of the HMS using a detailed model of the spectrometer acceptance. Cuts
are applied to the measured and simulated distributions to limit the data to
events where the momentum acceptance is well understood.  These cuts, given in
Table~\ref{acccuts.tab} are large enough in angle so that the collimator
defines the angular acceptance, but are effective in removing in-scattering events.
These are electrons that are outside of the nominal acceptance but which reach the
detectors because of scattering from an aperture in the spectrometer. Because of
the scattering inside the spectrometer, these events tend to reconstruct to trajectories outside of the acceptance and are thus removed by the acceptance cuts.

\begin{table}[htb]
\caption{Acceptance cuts used in the analysis for data and simulation. Here,
$\delta$ is the relative deviation from the central momentum and $x'_\text{tar}$
and $y'_\text{tar}$ are the out-of-plane and in-plane angles of the reconstructed
tracks at the target.}
 \begin{center}
 \begin{tabular}{|l|c|}
	 \hline
	 Variable & cut value \\
	 \hline
 abs($\delta$) & $<$~9\% \\
 abs($x'_\text{tar}$) & $<$~120 mr\\
 abs($y'_\text{tar}$) & $<$~40 mr \\
	 \hline
 \end{tabular}
 \end{center}
\label{acccuts.tab}
\end{table}

The Hall C single arm Monte Carlo is used to extract the simulated yield.
Each event is randomly generated in the target coordinates ($x$, $y$, $z$), while the
quantities $\delta, y_\text{tar}', x_\text{tar}'$ are randomly chosen within their allowed
limits. Then the particles are projected forward and transported to the
detector hut using transport matrix elements calculated by the COSY INFINITY
program~\cite{berz_cosy}, which models magnetic transport properties of the
spectrometer. Events that fail to pass through the different apertures defined
in the Monte Carlo are rejected.  Multiple scattering is simulated as the electrons
pass through material in the spectrometer, and so the simulation is run 
for each spectrometer momentum setting to account for the energy-dependence of the scattering. If the particle successfully traverses the spectrometer and passes 
all the criteria in the detector then it is accepted.

After applying cuts and binning the Monte Carlo counts in the same manner as
data, the simulated yield is given by, 
\begin{multline}
Y_\text{sim} = \mathcal{L} \sum_{events} \varepsilon^{\prime}_\text{det}\left( \frac{d\sigma}{d\Omega dE'}\right)^\text{rad}_\text{model}\\ 
 \times J(\Omega \rightarrow x_\text{tar}' y_\text{tar}') \Delta E' \Delta x_\text{tar}' \Delta y_\text{tar}'
\end{multline}
where $\mathcal{L}$ is the Monte Carlo luminosity, $\varepsilon^{\prime}_\text{det}$
accounts for any position-dependent efficiencies in the detectors, and $\left(\frac{d\sigma}{d\Omega dE'}\right)^\text{rad}_\text{model}$ is the cross section model (including radiative effects).
$J(\Omega \rightarrow x_\text{tar}' y_\text{tar}')$ is the Jacobian that transforms between the spherical
solid angle ($d\Omega$) and the spectrometer angles, $x'$, and $y'$, which is required
since the Monte Carlo event generation is performed in spectrometer coordinates.
In this analysis $5\times 10^{6}$ events were generated for each kinematic setting
with generation limits $\delta=\pm 15\%$, $x^{'}_\text{tar}=\pm 100$~mr and 
$y^{'}_\text{tar}=\pm 50$~mr.  Once the measured and simulated yields have been obtained, their ratio is applied as a correction factor to the initial Born cross section used in the simulation to extract the final cross sections (Eq.~\ref{csmaster1.eq}).

\subsection{Efficiencies}

In the cross section analysis, we apply particle identification (PID) cuts on
the signals from the gas $\mathrm{\check{C}erenkov}$ counter and lead-glass
calorimeter to distinguish electrons from other negatively charged particles.
Because of this, we must also correct for losses of real electron events when these
cuts are applied arising from detector-related inefficiencies. There are
additional losses due to trigger and tracking related inefficiencies.  

\subsubsection{Trigger efficiency}\label{trigeff.sssec}

The trigger was designed to be efficient for electrons while suppressing other
particle types. The electron trigger is described in detail
elsewhere~\cite{johna_thesis, blok08, aji_thesis}, and the key points are
summarized here. There are two main electron triggers. The first (ELHI) requires
signals from 3/4 hodoscope layers, and both preshower and total calorimeter energy exceeding fixed thresholds. The second (ELLO) requires a $\mathrm{\check{C}erenkov}$ signal and two out of three of the following: 3/4 hodoscope planes, 2/4 planes (one from the front and one from the back), or a calorimeter signal exceeding a threshold that is lower than used for ELHI. The final electron trigger is the combination of ELLO and ELHI signals.  This trigger provides modest pion rejection while being relatively insensitive
to possible lower efficiency in a particular component of the trigger, i.e., the $\mathrm{\check{C}erenkov}$, calorimeter, or hodoscopes.

Because there were no problems with the operation of the detectors, the
final trigger level efficiency was extremely high. The efficiency for
a good event to give a signal for ELHI was determined
run-by-run, and found to be 99.2\% on average, while the efficiency for ELLO was 99.7\%. Although ELLO required both a signal from the calorimeter
and $\mathrm{\check{C}erenkov}$ detectors, ELHI required only one PID signal, making the
trigger efficiency high even if one of the detectors had a low efficiency.
Accounting for all of these effects, the trigger efficiency is
99.7\%~\cite{aji_thesis}, and was largely rate and kinematic independent,
yielding a negligible uncertainty in the cross section ratios.

\subsubsection{Tracking efficiency}\label{track.sssec}

The normalized yields are also corrected for tracking inefficiency. In some
cases, real events do not yield a good track because of noise, hardware
inefficiency, or imperfections in the tracking algorithm. In other cases, events are
recorded for which there is not a good electron track going through the drift
chambers, in which case the lack of a track does not represent an inefficiency. A
series of cuts are applied to identify events for which an electron passed
through the drift chambers and should have yielded a good track. The fraction of those 
events which fail to give a track is taken to be the tracking inefficiency.

First, we select electrons by requiring that the event yielded a large signal in the 
$\mathrm{\check{C}erenkov}$ and calorimeter detectors. We exclude events which hit 
scintillator paddles near the edges of each plane, to suppress events which may have
missed the chamber but still hit the hodoscope and generated a trigger. Finally, we 
exclude events with more than 25 hits per chamber, as previous studies indicate that
these come from electrons hitting apertures near the entrance of the detector, yielding 
a shower of particles. Because they hit an aperture near the entrance, they are not within
the nominal acceptance of the detector and should not be treated as good tracks that were lost. 

This tracking efficiency correction was applied on a run-by-run basis. At low rates,
the inefficiency was approximately 2\%, with a small reduction at high rates
(up to 4\% total inefficiency) which is consistent with the expected loss due to 
rejection of events with real multiple tracks. 
When there are two real tracks in the event, only one trigger is registered and read
out, so one track is corrected for in the deadtime corrections, and the other is
treated as a tracking inefficiency.

\subsubsection{Calorimeter cut efficiency}\label{caloricut.sssec}

To reject pions, we require that the energy deposited in the calorimeter be at
least 70\% of the reconstructed momentum ($E_{cal}/E^{'}>$0.7).  It is
important to know how many otherwise valid events are lost when we place a cut
on the calorimeter distribution. To determine the fraction of electrons lost
due to the calorimeter cut, we need to identify a clean and unbiased sample of
electrons.  For this analysis, we used elastic scattering data, where the
initial fraction of pions is small, and then apply a cut on the
$\mathrm{\check{C}erenkov}$ detector to yield a pure electron sample. While
elastically scattered electrons tend to populate a limited region in the
acceptance of the spectrometer, this region can be moved across the acceptance
by changing either the angle or central momentum of the spectrometer, allowing us to
map out the response of the spectrometer throughout the acceptance. We use
these scans to verify that the cut efficiency is uniform across the
acceptance. The efficiency is found to be constant for $E^\prime$ above 1.7
GeV ($99.89\%$), but below this momentum, the efficiency starts to decrease
mainly due to decreasing resolution of the calorimeter. This falloff is
approximately linear, dropping the efficiency by 0.3\% for $E^\prime \approx
0.7$~GeV/c~\cite{aji_thesis} and is parameterized as a function of the
scattered electron momentum and is used to correct data in the analysis. The
efficiency measured with elastics is consistent with the efficiency extracted
using inelastic kinematics which populate the full acceptance, where the
kinematics have few enough pions for the $\mathrm{\check{C}erenkov}$ to yield
a pure electron sample.

\subsubsection{$\mathrm{\check{C}erenkov}$ cut efficiency}\label{cercut.sssec}

Another cut was applied on the number of photo-electrons collected by the
$\mathrm{\check{C}erenkov}$ detector in order to distinguish electrons from
pions. In addition to the pion-rejection cut in the calorimeter, we also
require the $\mathrm{\check{C}erenkov}$ detector sees at least 1.5
photo-electrons.  To measure the electron efficiency of this cut, we
identify a pure sample of electrons using elastic scattering kinematics
along with a cut on the calorimeter.

During the analysis it was found that the signal from the
$\mathrm{\check{C}erenkov}$ detector was lower near the vertical center of the
detectors, corresponding to $\delta =0$.  This is due to the gap between the
upper and lower mirrors. In addition to this $\delta$-dependent inefficiency,
the $\mathrm{\check{C}erenkov}$ has a momentum-dependent inefficiency that
was parameterized in terms of both $\delta$ and the HMS momentum
setting. The efficiency is close to 100\% for momenta above the spectrometer central
momentum ($\delta>0.5$\%), 1--2\% lower on the low-momentum side of the acceptance,
with loss of up to 2--4\% efficiency in the central $\pm$0.5\% of the momentum
acceptance (the inefficiencies are larger at low momentum settings).  For
details, see Ref.~\cite{aji_thesis}.

\subsection{Backgrounds}\label{bg.ssec}

In addition to the scattered electrons, there are secondary electrons that are
in the acceptance of the detector due to other physical processes which
constitute a background for the measurement. This background mainly consists
of scattered electrons from the cryotarget cell wall, pions that survive the
nominal PID cuts and are treated as scattered electrons, and secondary
electrons from pair production after bremsstrahlung in the target or $\pi^0$
which decay to photons. The following subsections discuss each of these
processes, and how we estimate and correct for them in the analysis.
 
\subsubsection{Background from target cell wall}\label{bg.sssec}
Since the cryogenic targets were contained in aluminum cells, electrons
scattered from the cell walls also contribute to the total number of detected
events. This contribution is measured and subtracted from the total
detected events. The cryocells were made of Al 7075 which has a density of
2.7952~g/cm$^3$ and the thickness of the cell walls was $\sim$0.12~mm. The
electrons traverse two cell walls, and since the cryotarget thickness varies
between 0.2 to 0.6~g/cm$^2$, the typical size of the background contribution
is between $10\%$ to $20\%$. We used a dummy aluminum target to directly
measure the cell wall contribution to the total yield. The dummy target
consists of two Al foils (Al 6061- T6) separated by $\sim$4~cm which are
$\sim$8 times thicker than the cryocell walls, thus allowing a higher
luminosity and a smaller data acquisition time. During the experiment dummy
data were taken at the same kinematics as the cryotarget data. Dummy data are
treated in the same way as cryotarget data and the normalized dummy yield is
subtracted from the cryotarget yield. Thus the total yield is
\begin{equation}
Y = Y_\text{cryo}- \left[ \frac{R^\text{ext}_\text{dummy}}{R^\text{ext}_\text{walls}}\,
\frac{T_\text{walls}}{T_\text{dummy}} \right] \times Y_\text{dummy},
\end{equation}
where $T_\text{walls}$ and $T_\text{dummy}$ are the thicknesses of the cell walls and the
dummy respectively, $Y_\text{cryo}$ and $Y_\text{dummy}$ are the measured cryotarget yield and
dummy yield respectively, and the ratio of $R^\text{ext}_\text{dummy}$ and
$R^\text{ext}_\text{walls}$ represents a correction factor which is applied to 
account for differences in radiative effects between the dummy target and the
cryotarget cell walls.  The correction was found to be about 5\% for
larger scattering angles at low $x$ values and smaller for other angles.

\subsubsection{Charge symmetric  background (CSB)}\label{csbg.sssec}

Most of the electrons observed in the spectrometer are beam electrons that scattered in the target. However, the incident electron can also interact with the target nuclei and produce neutral pions in the target. These pions can decay into high energy photons which produce an equal number of positrons and electrons and these electrons can be detected in the HMS and treated as scattered electrons. This contribution is generally small, but it can be a significant at kinematics corresponding to scattering at low $x$ and high $Q^2$.

\begin{figure}[htb]
\begin{center}
\includegraphics[width=85mm, angle=0, trim=10mm 20mm 15mm 30mm, clip]{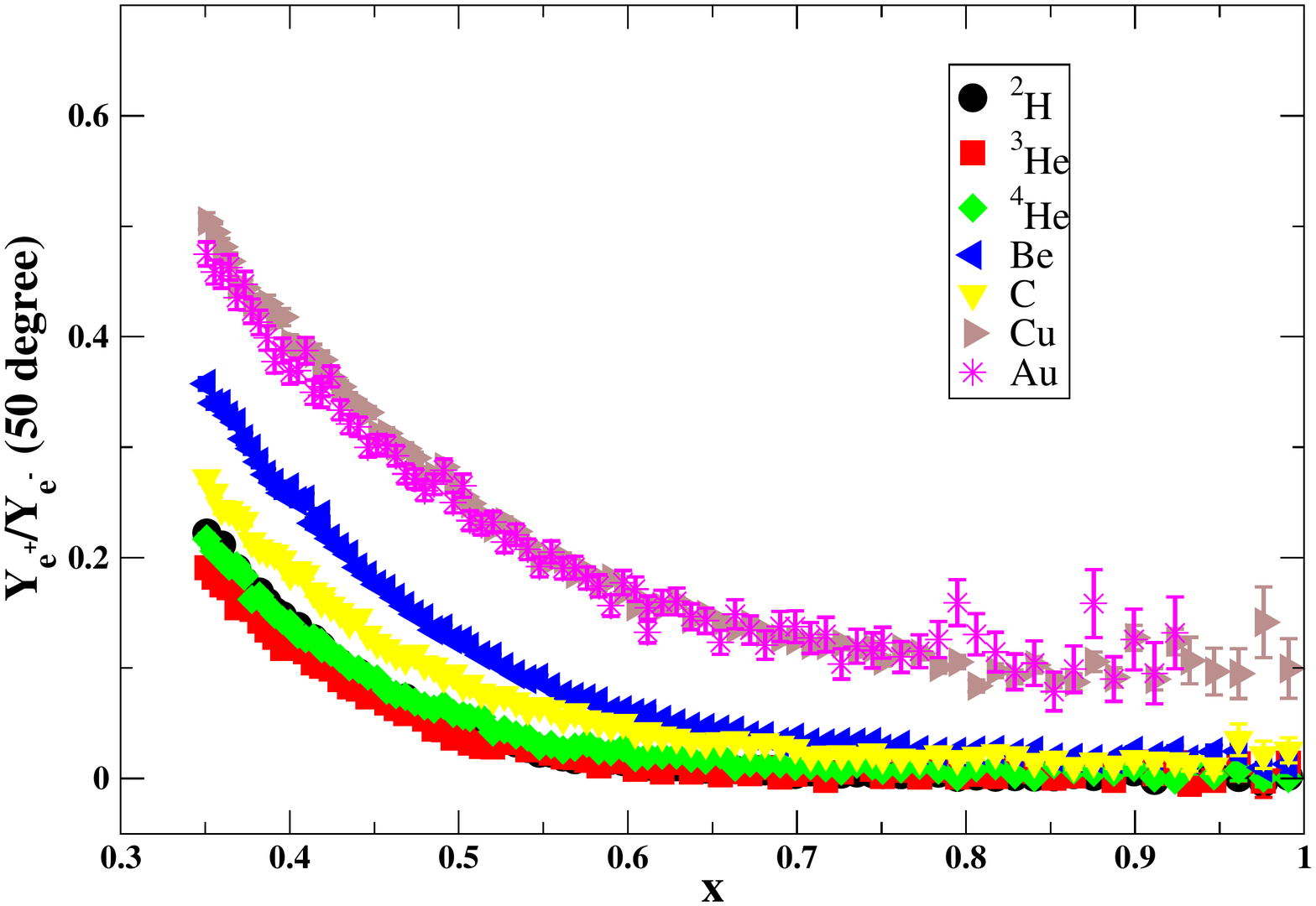}\\
\includegraphics[width=85mm, angle=0, trim=10mm 20mm 15mm 30mm, clip]{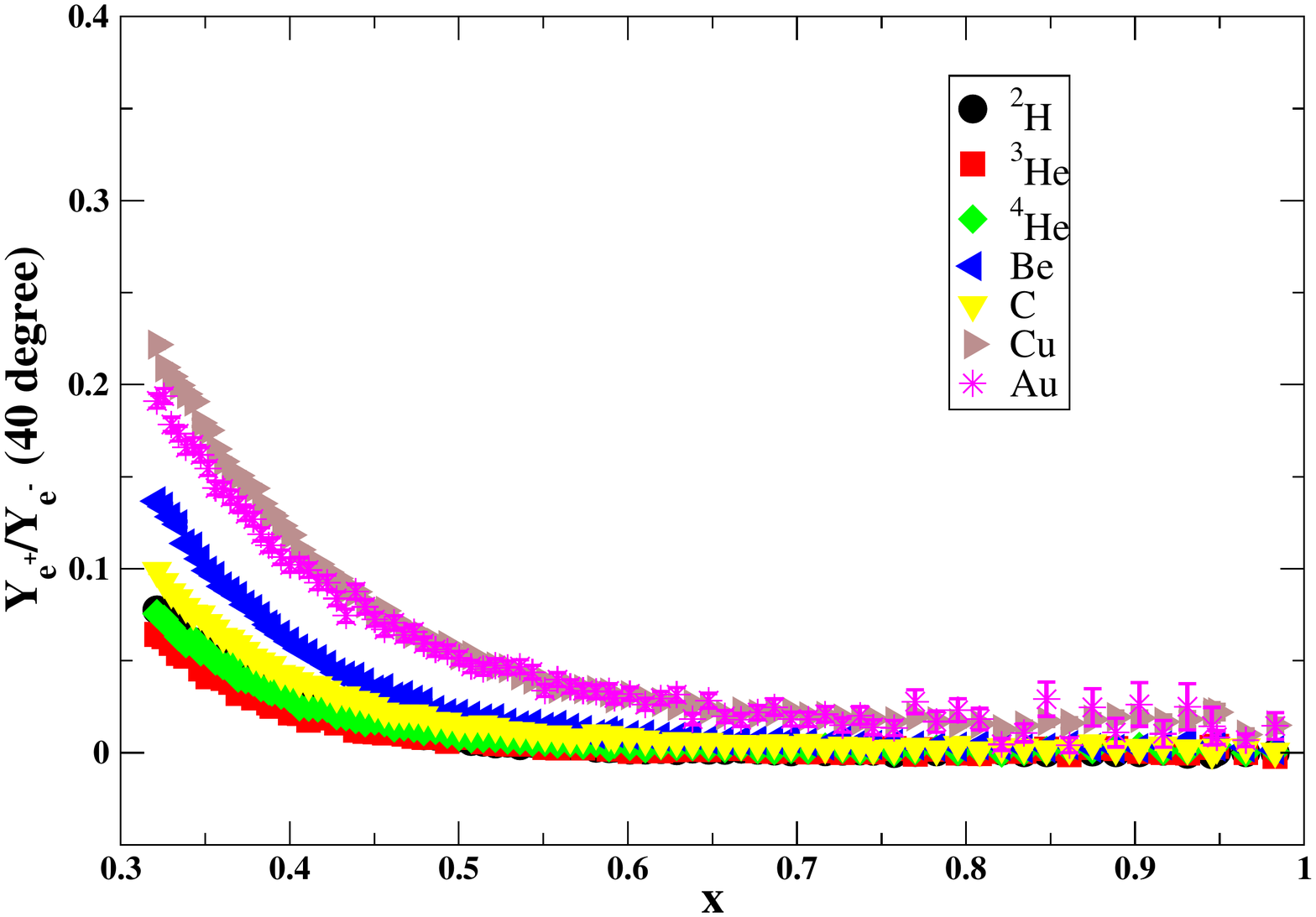}
\caption{The charge symmetric background as a function of $x$ for data taken
at 50 degrees (top) and 40 degrees (bottom).
\label{csbg_fig}}
\end{center}
\end{figure}

The total number of
electrons detected in the spectrometer is $e^-_\text{detected} = e^-_\text{primary} +
e^-_\text{background}$. Since an equal number of positrons and electrons are
produced, the yield is charge symmetric. This allows us to estimate the number
of secondary background electrons by running the spectrometer with positive
polarity and detecting the positrons. During E03103, we used the HMS to take
positron data for each target at all kinematic settings where the CSB was
significant (the 40 and 50 degree settings at 5.78~GeV and the 46 degree setting
at 5.01~GeV) allowing for a direct subtraction of the background by assuming
$e^-_\text{background} =e^+_\text{detected}$.  Luminosity normalized yields are used
to subtract the CSB, with identical cuts applied to the positron and electron
data. $R_\text{CSB}=\frac{ Y_{e^+}}{Y_{e^-}}$ is the fraction of the detected
electrons associated with CSB, and 
is shown in Fig.~\ref{csbg_fig} as a function of $x$ for the 50 and 40 degree
data. Note that our final EMC ratios are formed from the 40 degree data, and
so the correction is below 10\% except for the smallest values of $x$ and the
high-$Z$ targets.

\subsubsection{Pion backgrounds}\label{picont.sssec}

Pion rejection factors for the $\mathrm{\check{C}erenkov}$ and calorimeter
detectors are always greater than 500:1 and 100:1, respectively.  Nonetheless,
for runs with a high $\pi/e$ ratio, there could still be a small contamination
of pions after the PID cuts.

\begin{figure}[htbp]
\begin{center}
\includegraphics[width=85mm,trim={4mm 4mm 15mm 15mm}, clip]{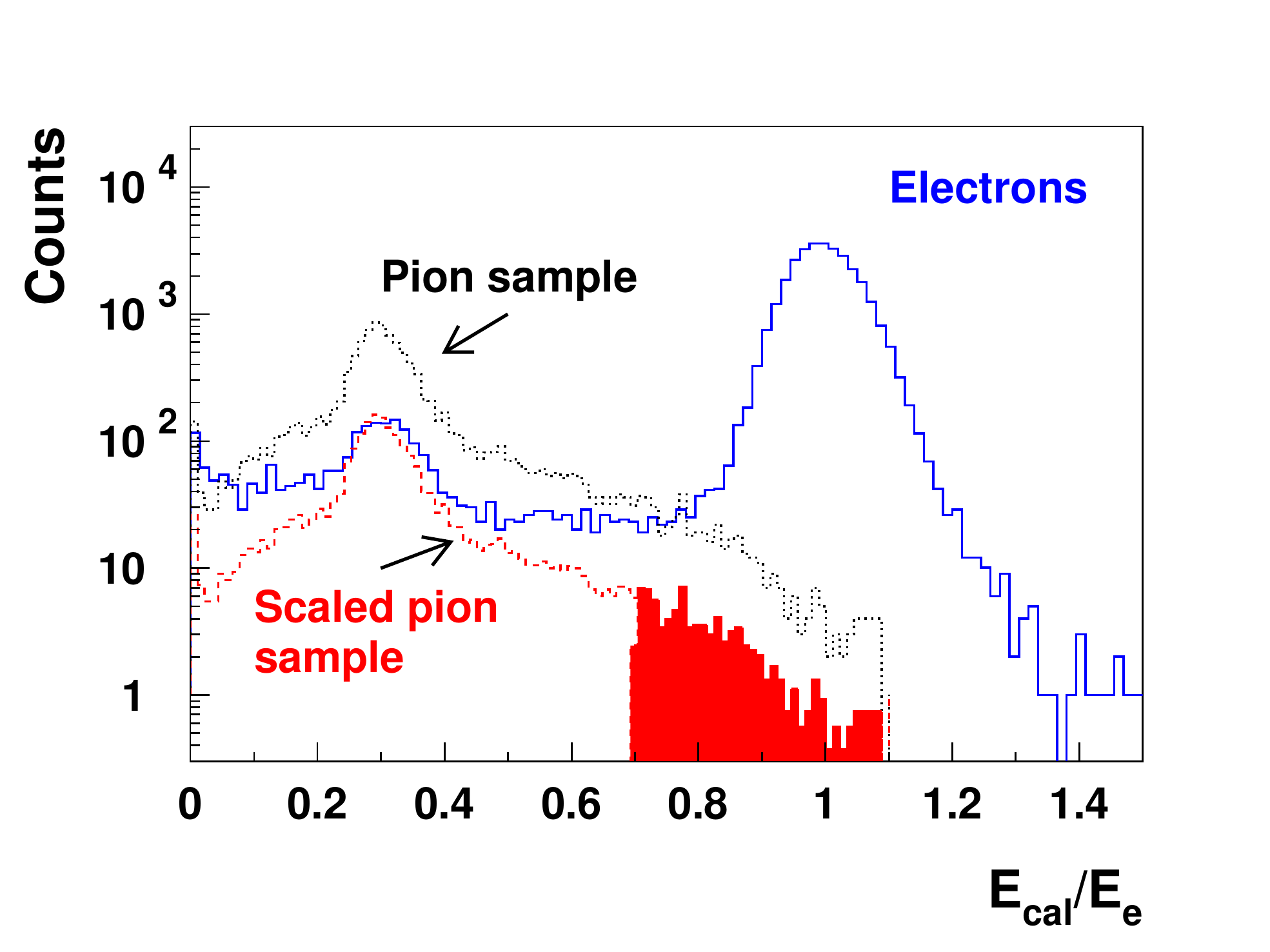}
\caption{(Color online) Illustration of the extraction of pion contamination using calorimeter spectra. A pion data sample (black dotted curve) is renormalized to the number of counts in an electron data sample (blue solid curve) at low $E_{cal}/E^{\prime}$.  The resulting spectrum (red dashed curve) is then used to estimate the pion contamination in the region $E_{cal}/E^{\prime}$ $>$0.7 (solid red).\label{pion_contam_fig}}
\end{center}
\end{figure}

To estimate the pion background, we generate calorimeter spectra first for a data sample
using electron PID cuts and then for a sample that is almost entirely pions.
The pion spectrum at low $E_{cal}/E^{\prime}$ is renormalized to match the electron spectrum in
that region.  The pion contamination is then determined from the number of (renormalized) counts
in the pion spectrum in the region $E_{cal}/E^{\prime}$ $>$0.7 (our nominal calorimeter electron cut). This technique is illustrated in Fig.~\ref{pion_contam_fig}.

It was found that
the final pion contamination is always below $0.5\%$.  This is further
suppressed as the subtraction of the positive-polarity data intended to remove
charge-symmetric backgrounds (see section ~\ref{csbg.sssec}) will have a
nearly identical contribution from positive pions.  We estimate that any
residual pion contamination is extremely small, and so we do not apply any
correction, but assign a 0.2\% point-to-point uncertainty to
allow for a small net contribution of pions.

\subsection{Target boiling corrections}\label{tarboil.ssec}
When the electron beam passes through the target material of cryogenic
targets, it deposits energy in the form of heat. This causes local density
fluctuations, ``target boiling'', along the path of the beam. The boiling
effects depend on the beam current, beam raster size and the thermal
properties of targets. We perform luminosity scans, measurements of the yield
at fixed kinematics with varying beam currents, to estimate the boiling
effects. In addition to measuring the effect on the cryogenic targets, we also
take data on carbon as a reference measurement, to ensure that corrections for
rate-dependent effects do not introduce variations which are misinterpreted as
density fluctuations.

A small current dependence was observed for the carbon target, even after
correcting for all known rate-dependent effects. Because the beam-current
monitors have an uncertainty in their DC offset, an error in that offset will
produce an error in the charge that goes like the inverse of the beam current.
The effect in carbon was small enough to be consistent with the uncertainty
in the BCM offset uncertainty, and so a correction to the BCM offset was inferred from 
the current dependence of the carbon yield. The hydrogen and deuterium targets did not show any residual slope after correcting for the BCM offset, but the helium targets show a linear reduction in the yield. For $^{3}$He, the measured density loss was $(-3.10 \pm 0.64)\% $ at
$100\,\mu A$ and for $^{4}$He, $(-1.27 \pm 0.50)\% $ at $100\,\mu A$. The
yield for each run is divided by a correction factor which depends linearly on
the average current (excluding periods with no beam).

\subsection{Computer and electronics deadtime}

Events are also lost due to the finite time it takes to either form a trigger
for an event or read out the data. During the time the trigger or
DAQ systems are busy, no new events can be taken. The dead time induced by the trigger electronics is monitored on a run-by-run basis by looking at the number of events generated with final trigger module gate widths of 50, 100, 150 and 200~ns. The electronic deadtime scales with the trigger rate and nominal gate width except for the 50~ns measurement,
which has an effective latency time of 60~ns.
While the typical gate widths are 40~ns, the coincidences formed between
different hodoscope planes have variable widths, typically 50-60~ns, so our
final trigger module is set to 60~ns to minimize the event-to-event variation
of the effective latency time. We calculate and apply a deadtime correction of
60~ns time the raw pretrigger rate, giving a maximum correction of 1.5\% with typical values well below 0.5\%.

Computer deadtime occurs when the DAQ computers are busy processing events
(either digitizing fastbus information or sending the data to the DAQ
computers), and are not available for processing new events. Because the
events are buffered in the fastbus and VME modules, there is not a fixed
latency period for each event, so we make a direct measurement of the
computer deadtime and apply the correction on a run-by-run basis. We take
the number of events recorded to disk divided by the number of generated
triggers which should have been read out and take the ratio to be the live
time. The deadtime was kept below 20\% by adjusting the prescale factors,
although previous tests have shown reliable operation and corrections for
deadtimes well over 90\%~\cite{johna_thesis}.

\subsection{Cross Section Model}\label{modelxsec_sec}

A cross section model is required for the bin centering corrections as well as modeling radiative effects and Coulomb distortion. The Born cross section
model (known as the XEM model) is broken down into contributions from inelastic
and quasielastic scattering:
\begin{equation}
\label{Bornmodel_eqn}
\sigma_\text{Born} =\sigma_\text{inel} + \sigma_\text{qe} ~.
\end{equation}

For the quasi-elastic contribution $\sigma_\text{qe}$, we use a $y$-scaling model~\cite{day_fymodel}.
The scaling variable $y$ can be interpreted as the minimum momentum of the
struck nucleon in the direction of the virtual photon. The scaling function,
$F(y)$, is an energy and momentum integral of the spectral function and is
defined as the ratio of the measured nuclear cross section to the off-shell
cross section for a nucleon, multiplied by a kinematic
factor~\cite{day_fymodel, arrington99}:
\begin{equation}
\label{Fy_eqn}
F(y) = \frac{d\sigma}{d\Omega d\nu}\frac{1}{\mathrm{Z}\sigma_p +
\mathrm{N}\sigma_N} \frac{q}{\sqrt{M^2 +(y+q)^2}},
\end{equation}
where $Z$ is the number of protons in the nucleus, $N$ is the number of neutrons,
$q$ is the three-momentum transfer, and $M$ is the proton mass. $F(y)$ is
expected to scale in $y$ on the low energy loss side of the quasielastic peak
where inelastic contributions and final state interactions are minimal. The
scaling function used for \LD\ is:
\begin{equation}
\label{Fy_eqn2}
F(y) =(f_0-B) \frac{\alpha^2\, e^{-(ay)^2}}{\alpha^2+y^2} + B \, e^{-b|y|}.
\end{equation}
For heavier targets the high-momentum components is modified and we take:
\begin{equation}
\label{Fy_eqn3}
 F(y) =(f_0-B) \frac{\alpha^2\, e^{-(ay)^2}}{\alpha^2+y^2} + B \, e^{-(by)^2},
\end{equation}
where the parameters $a, b, f_0, B $ and $\alpha$ are fit to the $F(y)$,
extracted from the data for each target.  The model parameters were varied
to reproduce the data from this measurement, along with the measurements
covering $x \gtorder 1$ on the same targets from Refs.~\cite{fomin10,fomin12}.
The model was also compared to low $Q^2$ quasielastic data, taken from
Ref.~\cite{Benhar:qe_archive}. This is important because a reliable model in this
region is needed when applying radiative corrections, as events from low $Q^2$
quasielastic scattering, which has a large cross section, can radiate
photons and contribute to higher $Q^2$, lower $x$ distributions.

$F(y)$ was extracted from the data in the QE region, taken as part of E03103 and
E02019~\cite{fomin10, fomin12} after subtracting the model
inelastic contribution (everything except the QE contribution)~\cite{nadia_thesis}.
After fitting $F(y)$, the updated model was used as
the input for the cross section extraction, and the process was repeated until
good agreement between data and the model was achieved for all settings. A small
additional correction was added to improve the agreement to the QE data at large
$x$ values~\cite{nadia_thesis}.

The inelastic contribution to the cross section is evaluated separately and
added to the quasielastic contribution.
For the deuteron, parameterizations of the proton and neutron structure
functions (developed by P. Bosted and E. Christy~\cite{bosted_model}) are used
for the full $x$ range. They are smeared using the momentum distribution based
on the fit to our QE peak~\cite{nadia_thesis}.

For heavier nuclei, the inelastic cross section is computed somewhat differently. 
As for the deuteron, the model cross section is the sum of the proton and neutron 
structure functions smeared by the momentum distribution based on the fit to the
QE peak for the nucleus~\cite{nadia_thesis}. 
In addition, for $x<0.8$, this inelastic model is then multiplied by a target-dependent 
polynomial function to improve the agreement between data and model (this is required 
since a pure-smearing calculation will not reproduce the size or shape of the nuclear EMC
effect correctly).  This is smoothly joined to the full smearing prescription (with
no correction) for $x>0.9$, using an $x$-weighted average for $0.8<x<0.9$.
An additional polynomial correction is applied to both the deuteron and the heavier targets to
slightly suppress the inelastic cross section at large $x$ above the QE peak~\cite{nadia_thesis}.

\begin{figure}[htbp]
\begin{center}
\includegraphics[width=90mm, trim=0mm 0mm 0mm 16mm, clip]{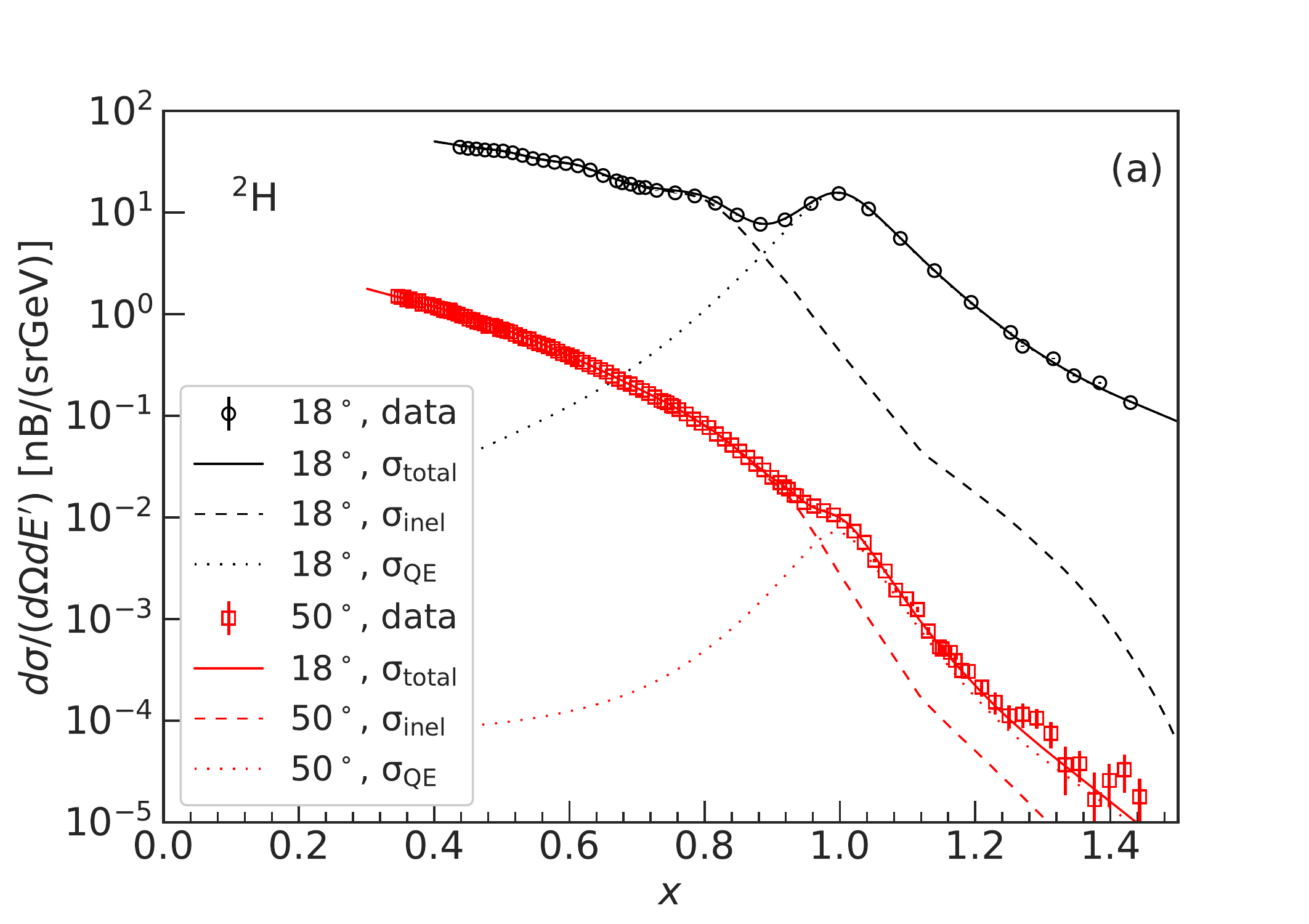}\\
\includegraphics[width=90mm, trim=0mm 0mm 0mm 16mm, clip]{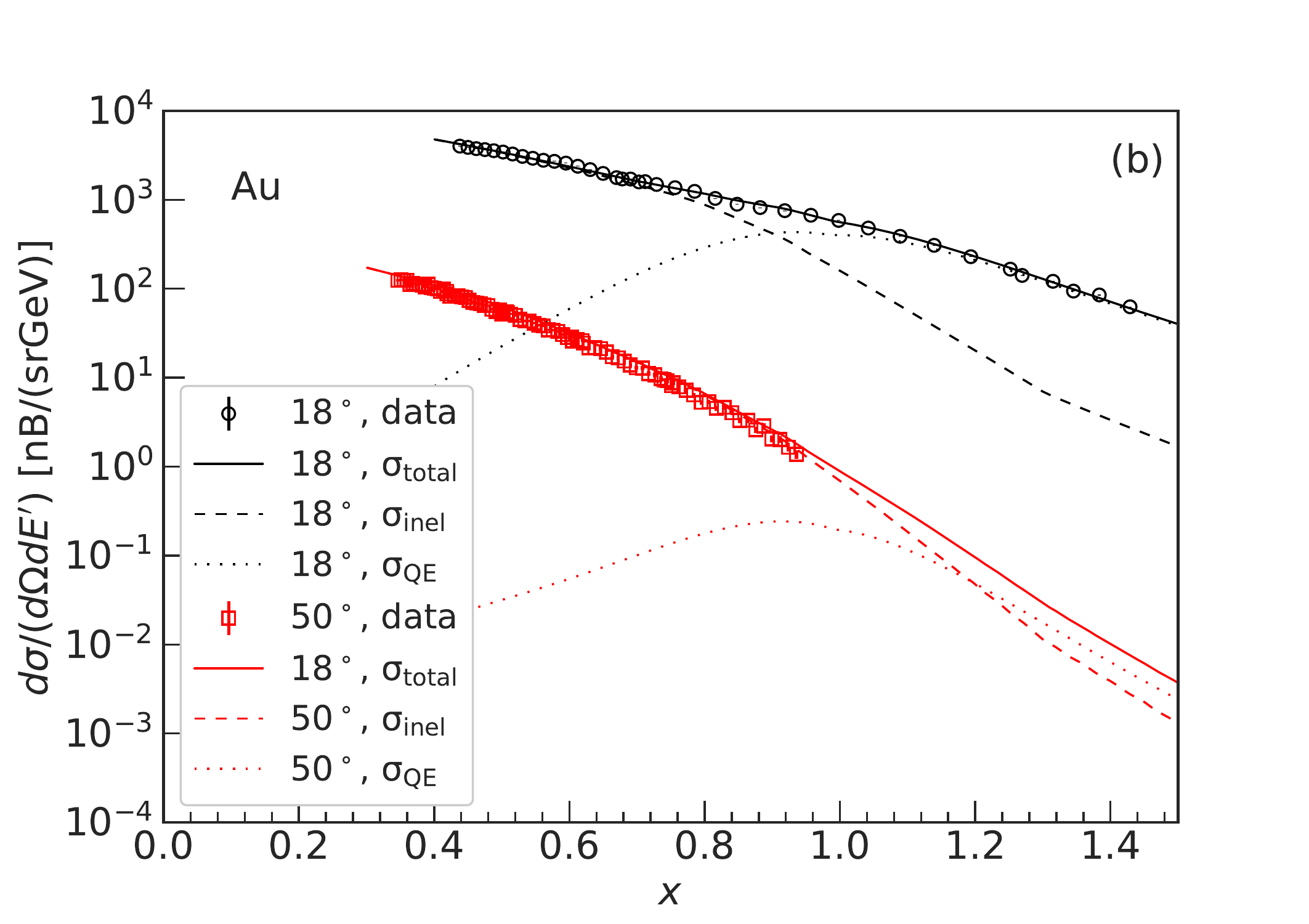}
\caption{(Color online) Data and model cross section for $^2$H and $^{197}$Au at selected
kinematics. Here, the circles show 18 degree data and the squares show 50
degree data. Relative contribution from inelastic (dashed line) and
quasielastic (dotted line) to the total cross section (solid line) are also
shown in the figure.}\label{all_model_compare_fig}
\end{center}
\end{figure}

The smearing  calculation described above, when performed in combination with our full radiative 
corrections procedure is quite time consuming. Therefore, the full radiative correction
was only calculated at the central spectrometer angle for a given setting.  Since a radiated 
model is also required to describe the variation of the cross section across the spectrometer 
acceptance, a simplified, approximate form of the radiative correction was used in combination with the 
smearing calculation when calculating the two-dimensional grid in $E^{'}$ and $\theta$ used for
Monte-Carlo weighting.  An additional ad-hoc correction (a polynomial in $x$) was applied to this latter
calculation, to compensate for the approximate form of the radiative corrections used.
The data as well as the model
cross sections, including the relative contributions from the inelastic piece
and the quasielastic piece for $^2$H and $^{197}$Au are shown in
Fig.~\ref{all_model_compare_fig}.

\begin{figure}[htb]
\begin{center}
\includegraphics[width=85mm, angle=0, trim=5mm 15mm 5mm 30mm, clip]{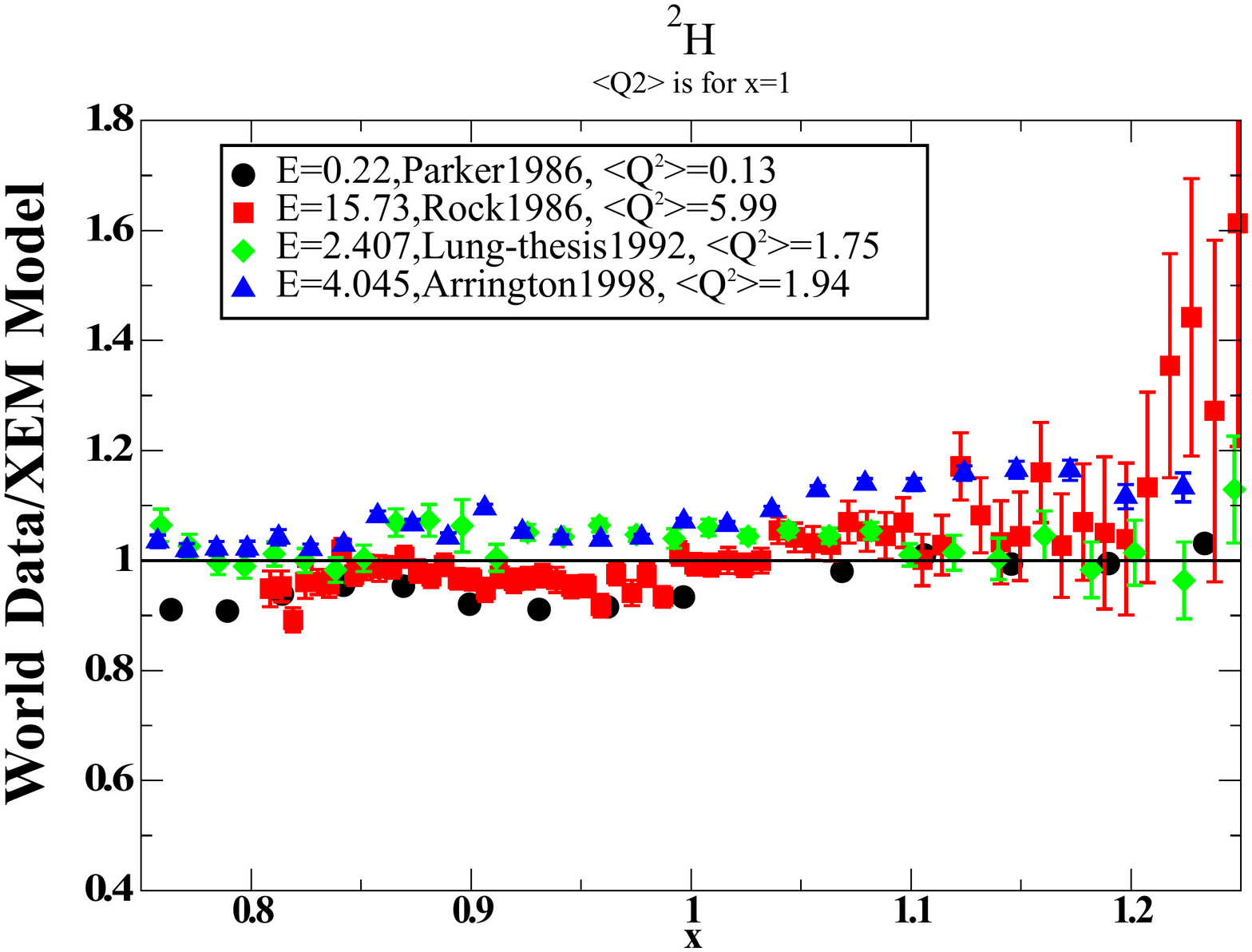}\\
\includegraphics[width=85mm, angle=0, trim=5mm 15mm 5mm 30mm, clip]{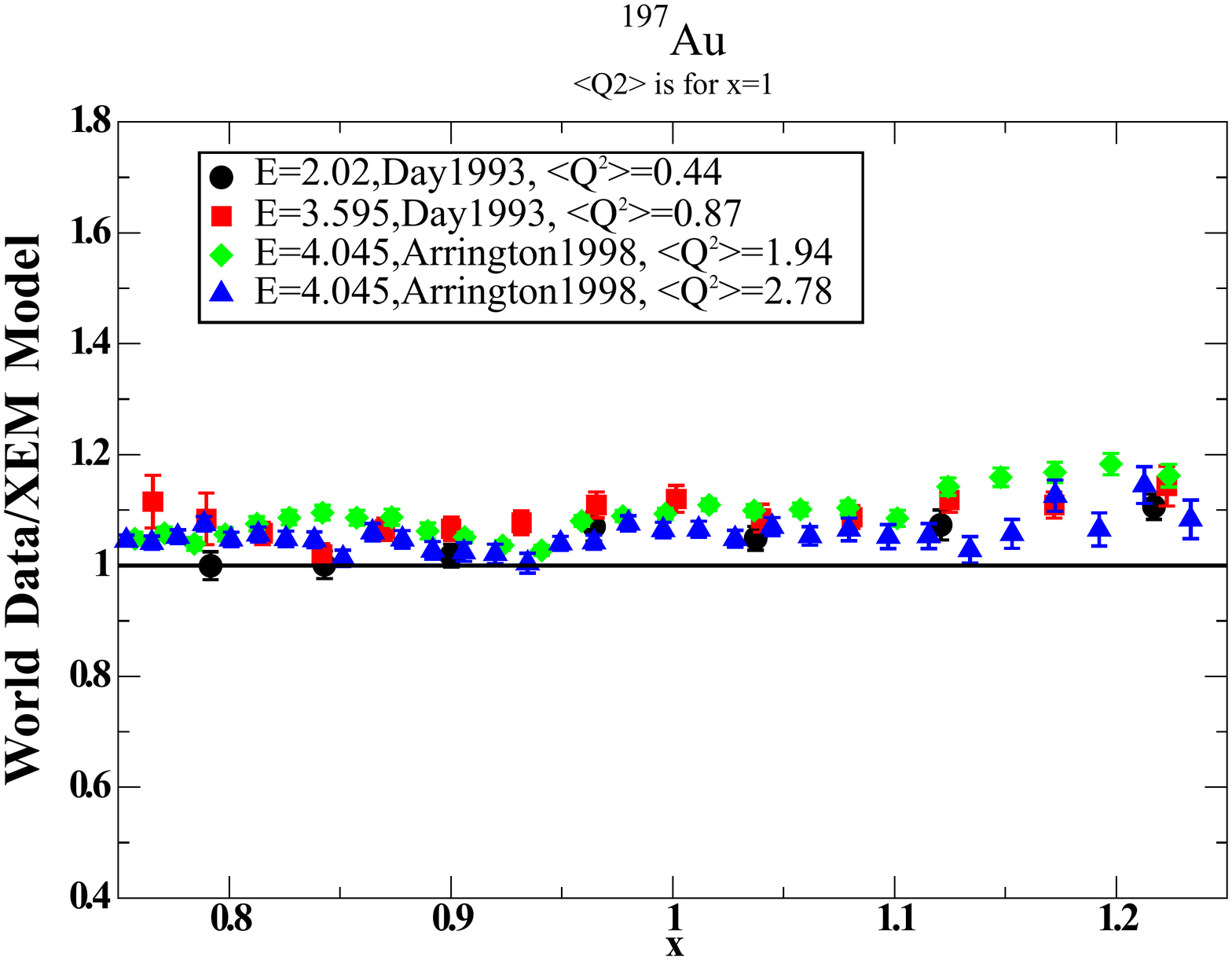}
\caption{(Color online) Comparison of world data from the
quasielastic electron nucleus scattering archive~\cite{Benhar:qe_archive} and
our cross section model for a variety $Q^2$ settings (quoted $Q^2$ value 
corresponds to $x=1$) for the $^2$H (top) and $^{197}$Au (bottom)
targets. The shape of the QE peak is well reproduced for both targets
at both low and high $Q^2$, yielding a nearly flat ratio of data/model over
the entire $x$ range.
\label{all_worlddata_compare_fig}}
\end{center}
\end{figure}

At low $Q^2$ values, the quasi-elastic peak accounts for a significant portion
of the total cross section at large $x$.  The low-$Q^2$ QE cross section also
has a large impact on the radiated model at low $x$ and high $Q^2$. We have
done extensive studies and compared our model with the data available from the
quasielastic electron nucleus scattering archive~\cite{Benhar:qe_archive}. For
heavy nuclei, our model cross section was compared with world QE
data down to $Q^2=0.5$~GeV$^2$, and the agreement between data and model was
found to be at the 10\% level near the quasi-elastic peak, as shown in
Fig~\ref{all_worlddata_compare_fig}.

\subsection{Other corrections}\label{othercor.ssec}

The XEM cross section model is in the Born or one-photon exchange
approximation. However, higher order processes in $\alpha$ also contribute to
the measured cross sections~\cite{motsai_rc,tsai_rc} and must be applied to
the starting model. To compare to the measured cross sections, all significant
contributions from higher order processes must be estimated and corrected for
in the measured cross section. These include traditional radiative effects,
as well as the Coulomb distortion associated with the long-range
interaction of the electron with the charge of the nucleus.

\subsection{Radiative corrections}\label{rc.ssec}

Radiative corrections need to be applied to account for higher order QED
processes, the most significant of which are the emission of one or more
real photons by the incoming or outgoing electron or the struck quark (in the
DIS regime), exchange of a virtual photon between the incoming and outgoing
electron, and the fluctuation of the exchange photon into a lepton-antilepton
pair.  Because the elastic and quasielastic cross sections are very large
at low $Q^2$, one must also account for low-$Q^2$ interactions which, due to
radiation of a hard photon, end up at low $x$ and high $Q^2$ values.
Thus, we express the total measured radiated cross section as:
\begin{equation}
\sigma_\text{meas}= \sigma^\text{rad}_\text{inelastic}+
\sigma^\text{rad}_\text{quasielastic}+ \sigma^\text{rad}_\text{elastic}~.
\end{equation}
Since the radiative tails from the QE and elastic processes are small, ($<$20\%), as
are the contributions from large $x$, low $Q^2$ inelastic processes,
we used the multiplicative radiative correction method. 
For the kinematics of this analysis, our studies indicate that the nuclear 
elastic tail contributes less than 0.1\% to the total cross section for $^2$H, 
and significantly smaller contributions for heavy nuclei, and so are neglected in the analysis.

The program used to compute the radiative effects for this analysis was
developed at SLAC and is described in detail in~\cite{dasu_thesis}. For
E03103, the external corrections are computed using a complete calculation of
Mo-Tsai~\cite{motsai_rc} with a few approximations. Note that, in particular,
the energy-peaking approximation is not used for the computation of external
contributions. This approach, ``MTEQUI'', uses the equivalent radiator
approximation~\cite{dasu_thesis}. In the equivalent radiator method, the
effect of ``internal'' Bremsstrahlung is calculated using two hypothetical
radiators of equal radiation length, one placed before and one after the
scattering. The internal contribution in ``MTEQUI'' method is evaluated by
setting the radiation length of the material before and after the scattering
point to zero, and ignoring the target length integral. Then the radiated
model cross section is given by the sum of the internal and external
contributions.

\begin{figure}[htbp]
\begin{center}
\includegraphics[width=90mm, angle=0, trim={0mm 0mm 0mm 0mm}, clip]{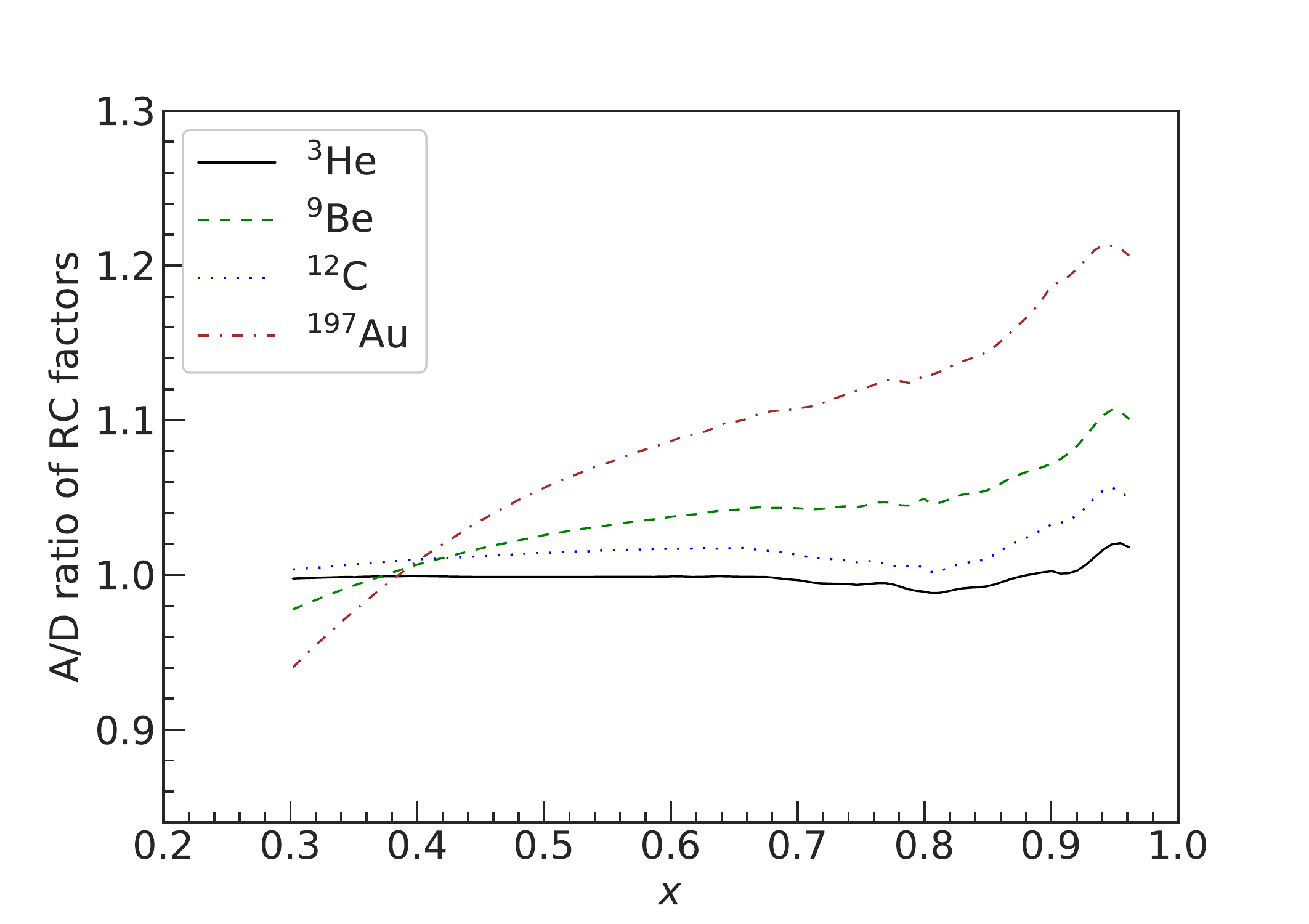}
\caption{(Color online) Radiative correction factor to the A/D cross section ratios for a
range of targets at 40 degrees; the correction at 50 degrees is nearly
identical.
\label{rc_superrat_fig}}
\end{center}
\end{figure}

Our simulations are performed using the radiated model,
\begin{equation}
\label{rc1_eqn}
\sigma^\text{model}_\text{rad}=\text{external}\otimes \text{internal} \otimes \sigma^\text{model}_\text{Born},
\end{equation}
The convolution involves integrating over the ``internal'' and ``external''
bremsstrahlung photon momenta and angles, and the target dimensions.
To obtain $\sigma^\text{model}_\text{rad}$, one needs to know the cross sections
over the entire kinematic range (from elastic threshold up to the kinematic
point being calculated, see Figure C.1 in reference~\cite{dasu_thesis}). The
effect of radiative correction on measured cross sections varied from a few
percent to about $40\%$, depending on the kinematics and targets. Because the
structure functions of nuclei are very similar, the internal radiative
corrections and some of the external corrections cancel, yielding smaller
corrections in the target ratios which depend mainly on the difference in the
targets' radiation length, as shown in Fig.~\ref{rc_superrat_fig}.

\subsection{Coulomb corrections}\label{cc.ssec}

The incoming electron will interact with the Coulomb field of the nucleus
prior to interacting with the nucleus. Classically, once the electron enters
the electron cloud of the atom, the screening of the nuclear potential is
no longer perfect, and the electron will be accelerated towards the nucleus,
increasing its momentum at the interaction vertex. After the scattering,
there will be a similar interaction as the electron leaves the nucleus.
This change in the kinematics can have a significant effect on the measured
cross sections if either the Coulomb potential is large compared to the energy
of the initial or final electron, or when the cross section varies rapidly
with the kinematics. In addition to the modification of the scattering
kinematics there is also a `focusing' of the incoming electron plane wave which
also impacts the scattering cross section. For the present analysis, we account
for these effects using the improved version of the Effective Momentum
Approximation (EMA)~\cite{aste2_ccor}, following the approach given in Ref.~\cite{nadia_thesis}.

The charge of the nucleus has two effects on the electron wave function. The
initial and final state electron momenta ($\vec{k}_{i,f}$) are modified in the
vicinity of the nucleus due to the attractive electrostatic potential.
Secondly, the attractive potential leads to focusing of the electron wave
function in the interaction region. The distorted electron wave can be
approximated by~\cite{rosenfelder_ccor, aste2_ccor},
\begin{equation}
\label{ccor1_eqn}
\psi_{\vec{k}_{i,f}}= \frac{|(\vec{k}_{i,f})_\text{eff}|}{|\vec{k}_{i,f}|} \,
\psi_{ (0)}\, \exp\left(i\,\vec{k}_{i,f}\cdot \vec{ r}\right),
\end{equation}
where $\psi_{(0)}$ is the Dirac-spinor with
$|(\vec{k}_{i,f})_\text{eff}|=|(\vec{k}_{i,f})| - \overline {V}$, and $\overline
{V}$ is the average electrostatic potential of the nucleus.

Treating the nucleus as a spherical charge distribution, radius $R_0$, 
central potential is given by:
\begin{equation}
\label{ccor2_eqn}
V_{(0)} = -\frac{3\alpha (Z-1)}{2 R_0}~.
\end{equation}
Because the standard convention is to neglect Coulomb corrections in
$Z=1$ targets, we use a factor $Z-1$ rather than $Z$ to account only
for the additional charge in the nucleus compared to scattering from
the proton or deuteron.

The central potential is an upper limit, as the potential is smaller everywhere
else in the nuclear volume, so it is necessary to determine an appropriate average
potential for scattering from the nucleus. This effect is incorporated in
the EMA approach by an average potential $0.75$-$0.80$ times the central
potential, $ V_{(0)}$~\cite{aste2_ccor}. For E03103, we take $\Delta E=
\overline {V} = 0.775 V_{(0)}$ and estimate this potential to be known
at the 10\% level. Note that Ref.~\cite{aste2_ccor} uses $Z$ rather than 
$Z-1$ in determining the average potential, but this has minimal impact
on their extraction of the optimal potential, as this is obtained from
calculations for heavy nuclei.

\begin{table}[htb]
\caption{The average effective potential $\Delta E$ and the
values of the charge radii for the different targets used in the analysis. The
radii for $^{3,4}$He are measured values while the rest are calculated from the
approximation $R_0(A) =1.1\,A^{1/3} +0.86\,A^{-1/3}$~\cite{aste2_ccor}.}
 \begin{center}
 \begin{tabular}{|c|c|c|}
 \hline
 Target & $R_0 $ (fm) & $\Delta E$ (MeV) \\
 \hline 
 \HET & 2.32 & 0.66 \\
 \HEF & 2.17 & 0.77  \\
 \Be  & 2.70 & 1.88 \\
 \C   & 2.89 & 2.92 \\
 \Cu  & 4.59 & 10.2 \\ 
 \Au  & 6.55 & 19.9 \\ 
 \hline
 \end{tabular}
 \end{center}
\label{ccor_table}
\end{table}

In the EMA approach, the focusing factor of the incoming wave,
$F_{i}=|(\vec{k}_{i})_\text{eff}|/|\vec{k}_{i}|$, enters quadratically in the cross
section calculation and produces an enhancement in cross section strength.
However, the focusing factor of the outgoing wave cancels with the enhanced
phase space factor in the effective cross section. The Coulomb correction
factor in the EMA approach is given by the ratio of the model cross sections
with nominal and shifted kinematics, scaled by the square of the focusing
factor:
\begin{equation}
\label{ccor4_eqn}
F_\text{ccor} =\frac{\sigma_{(E,E^\prime)} }{\sigma_{(E+\Delta E,\, E^\prime+\Delta
E)}}{\left[\frac{E}{ E+\Delta E}\right]}^2,
\end{equation} 
where the $\sigma$s are the Born model cross sections. The measured cross sections
are then multiplied by $F_{ccor}$, to get the Coulomb-corrected cross sections.

\begin{figure}[htbp]
\begin{center}
\includegraphics[width=85mm, angle=0, trim=10mm 20mm 10mm 29mm, clip]{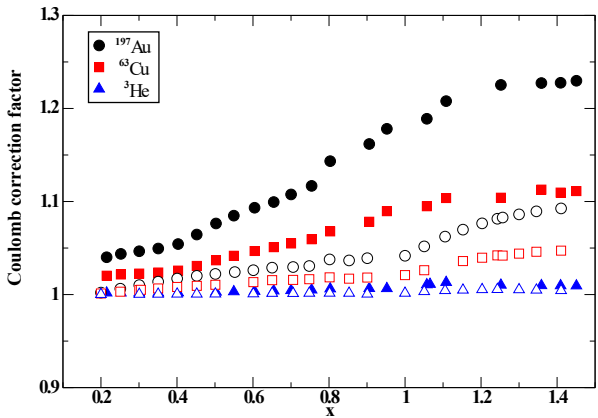}
\caption{Coulomb correction factors as a function of $x$ for several targets as noted in the legend for a few selected kinematics for 5.776 GeV beam energy. The filled symbols show the correction for the 50 degree data while the open symbols represent 18 degree data.}
\label{ccor_fac_fig}
\end{center}
\end{figure}

Table~\ref{ccor_table} shows the values for the RMS charge radii, and the
magnitude of the average energy boost for the targets used in E03103. The
Coulomb correction factors as applied to the data are shown in
Figure~\ref{ccor_fac_fig}. This figure shows the importance of the Coulomb
distortion effects for the cross section and cross section ratio extractions
in the medium energy range. These are relatively small for light nuclei, but
for the heavy nuclei and near the quasi-elastic peak, these corrections are
significant. The largest corrections are for the Au data at 40 and 50 degrees.
With no Coulomb corrections applied, the EMC ratios are systematically 3-5\%
lower for the 50 degree data than the 40 degree data.  After applying the EMA corrections described
above, they are in excellent agreement, suggesting that the correction
yields agreement at the 2\% level or better, given the uncertainties in the
comparison.  This supports the idea that the EMA does a good job estimating
this correction, though it assumes that no other effect modifies the cross
section ratios in going from 40 to 50 degrees.  This will be discussed further
in Sec.~\ref{xdepresult.ssec}.

Since this is a target- and $x$-dependent correction, neglecting the effect
will modify both the extracted size of the EMC effect and the overall A
dependence. In addition, for a given $x$ value the angular dependence of the
Coulomb correction factor implies a $Q^2$ dependence in the correction. Thus
one should be careful about $Q^2$ averaging of the cross section or cross
section ratios and the correction factor needs to be properly accounted for
before applying such an averaging procedure.  While Coulomb corrections
were not applied to previous EMC measurements, the effect was estimated to be
$\ltorder$3\%~\cite{Arrington:2003nt} for SLAC E139~\cite{slace139}, owing to
the higher beam energy and smaller scattering angles. Nonetheless, neglecting
this correction would imply some overestimate of the EMC effect in medium-heavy
nuclei. We will discuss this further in the results section.

\subsection{Isoscalar corrections}\label{iso.ssec}

EMC ratios are expressed as the cross section ratio (per nucleon) of a target
nucleus with an equal number of protons and neutrons (isoscalar nucleus) to
that of deuterium. Thus, the EMC ratio for an isoscalar nuclei is just
$\sigma^A/\sigma^D$. Since the protons and neutrons have different cross
sections, the cross sections for  nuclei with $Z \neq N$ will significantly
differ from that of nuclei with $Z=N$. Thus, one typically applies a correction function
to convert the measured $F_2^A$ to a hypothetical isoscalar nucleus with the same mass number:
\begin{equation} \label{iso1_eqn}
(F_2^p+F_2^n)/2 = f_\text{iso}^A (Z F_2^p+N F_2^n) / A.
\end{equation} 
This correction function reduces to a function of $F_2^n/F_2^p$, the neutron to
proton structure function ratios of the nucleus under investigation:
\begin{equation} \label{iso2_eqn}
f_\text{iso}^A = 
\frac{(F_2^p+F_2^n)/2}{(Z F_2^p + N F_2^n )/A} =
\frac{A (1+ F_2^n/F_2^p)}{2 (Z + N F_2^n/F_2^p)}.
\end{equation} 
The measured cross section ratios are multiplied by $f_\text{iso}^A$, which
depends only on $N$, $Z$, and the neutron-to-proton structure function ratio, to get
the isoscalar-corrected cross section ratios.  Note that the structure
functions in Eq.~\ref{iso1_eqn} correspond to the proton and neutron
contributions to the heavy nucleus, as one is trying to convert from a
non-isoscalar heavy nucleus to the isoscalar equivalent.  In the past, these
were simply replaced with the free neutron and proton structure function
ratio.

\begin{figure}[htb]
\begin{center}
\includegraphics[width=85mm,angle=0, trim=10mm 20mm 20mm 20mm, clip]{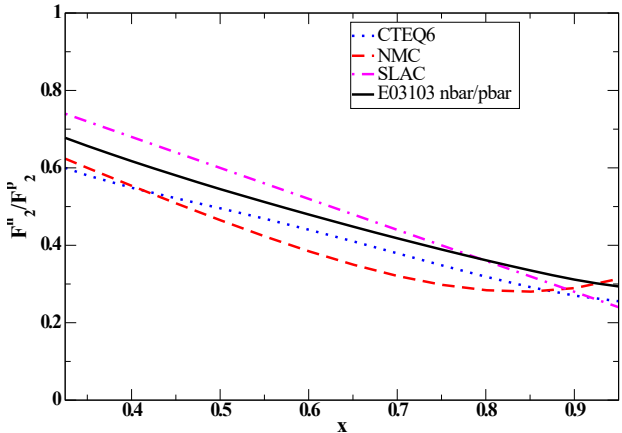}
\caption{(Color online) The ratio $F_2^n/F_2^p$ vs. $x$ for various parameterizations of the
free nucleon structure functions along with the ratio of the smeared structure functions in deuterium~\cite{arrington09, arrington12b} extracted for the 40 deg kinematics of the E03013 experiment.
\label{f2nf2p_fig}}
\end{center}
\end{figure}

There is significant uncertainty in the free neutron cross section in the large
$x$ region and so the extracted EMC ratios are sensitive to the choice of 
isoscalar correction factor. The $F_2^n/F_2^p$ ratio has been extracted from
proton and deuteron DIS measurements by SLAC~\cite{slacf2nf2p} and NMC
\cite{nmcf2nf2p1,nmcf2nf2p2}. Since there is no free neutron target, the
extraction of $F_2^n$ is always model-dependent.  The SLAC extraction included
Fermi motion while the NMC $F_2^n/F_2^p$ ratios were extracted neglecting
all nuclear effects (including binding) in the deuteron. The EMC effect
results from SLAC E139~\cite{slace139} took $\sigma_n=\sigma_p(1-0.8~x)$ when
calculating the isoscalar correction. Figure~\ref{f2nf2p_fig} shows different
representative parameterizations for $F_2^n/F_2^p$ along with $F_2^n/F_2^p$
constructed from parton distributions from CTEQ~\cite{cteq6_pdf} computed at
$Q^2=10$~GeV$^2$. The CTEQ fit also neglects the Fermi motion of nucleons. 
NMC mostly had data in the low $x$ region, however, the $x$ range covered by
SLAC data is mainly in the large $x$ region and overlaps with $x$ range
covered by E03103.  All of these extractions are based on measurements of the
deuteron-to-proton ratios in different $Q^2$ regions, and so any $Q^2$
dependence in the ratio would be expected to generate scatter in these
results, beyond that associated with differences in the assumptions made in
the extraction.

\begin{figure}[htb]
\begin{center}
\includegraphics[width=85mm, height=65mm, angle=0, trim=2mm 5mm 20mm 15mm, clip]{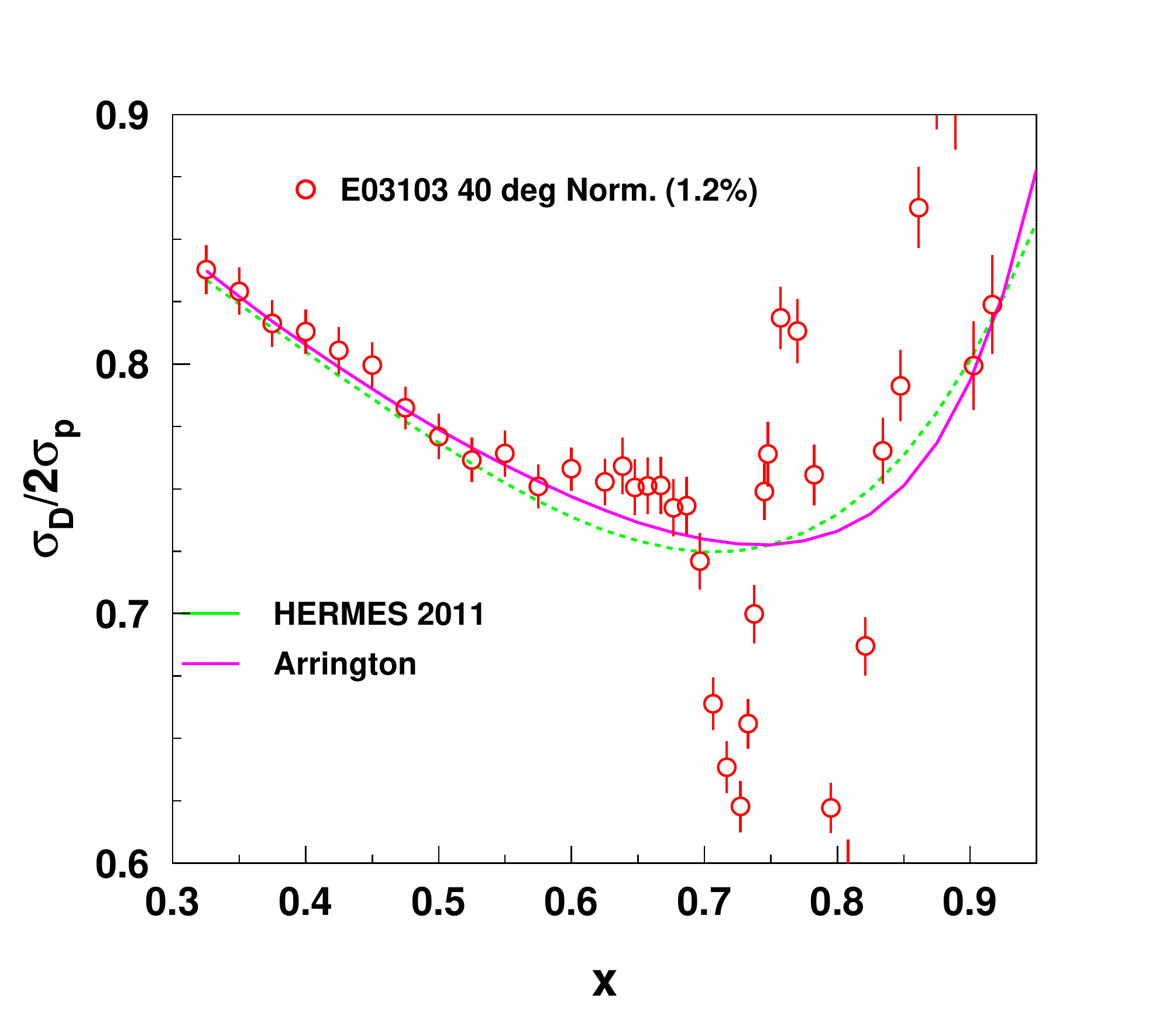}
\caption{(Color online) Our extracted $\sigma^D/2\sigma^p$ ratio along with 
calculations based on different $F_2^n/F_2^p$ extractions (dashed line
from~\cite{hermes_dbyp} and solid line using~\cite{arrington09, arrington12b}).
The structure above $x \approx 0.65$, is mainly due to the resonance in the
proton structure function.}
\label{dprat_fig}
\end{center}
\end{figure}

In our analysis we make a modified isoscalar correction. Instead of using
free proton and neutron structure functions, we have used the contributions of
$F_2^p$ and $F_2^n$ in $^2$H, $\overline{F_2^n}$ and $\overline{F_2^p}$,
in the above equation to correct the nuclear cross
sections.  As such, we are converting the deuteron structure in the 
denominator to a non-isoscalar deuteron, with the same $Z/N$ ratio as the
nucleus.  The alternative would be to evaluate the neutron-to-proton ratio
for all nuclei, which would involve significantly larger model dependence
in heavier nuclei.  In addition, we use the $F_2^n/F_2^p$ ratio at the kinematics of
our experiment, rather than taking the result from a high-$Q^2$ analysis.
We determine the in-deuteron $F_2^n/F_2^p$ ratio following the approach of
Refs.~\cite{arrington09, arrington12b}.  The extraction was performed 
taking the average of the values obtained using the different NN potentials
and off-shell effects evaluated in Ref.~\cite{arrington12b}, using the
calculated value of $\overline{F_2^p}$ in the deuteron, and taking
$\overline{F_2^n}/\overline{F_2^p} = (F_2^d -
\overline{F_2^p})/\overline{F_2^p}$. This does not involve removing the
nuclear effects to extract the free neutron structure function, as is 
usually the case, and so this procedure is somewhat less model dependent
than the extraction of the free $F_2^n/F_2^p$ ratio.  We note that these
analyses also demonstrated that the model-dependence is smaller than assumed
in some previous comparisons where the nuclear effects evaluated at a
fixed $Q^2$ were applied to extract $F_2^n/F_2^p$ spanning a range in $Q^2$.  A similar result was seen in the analysis of the impact of
nuclear effects on the extraction of the proton PDFs~\cite{accardi11}.

\begin{figure}[htbp]  
\includegraphics[width=85mm, height=50mm, trim=0mm 10mm 20mm 30mm, clip]{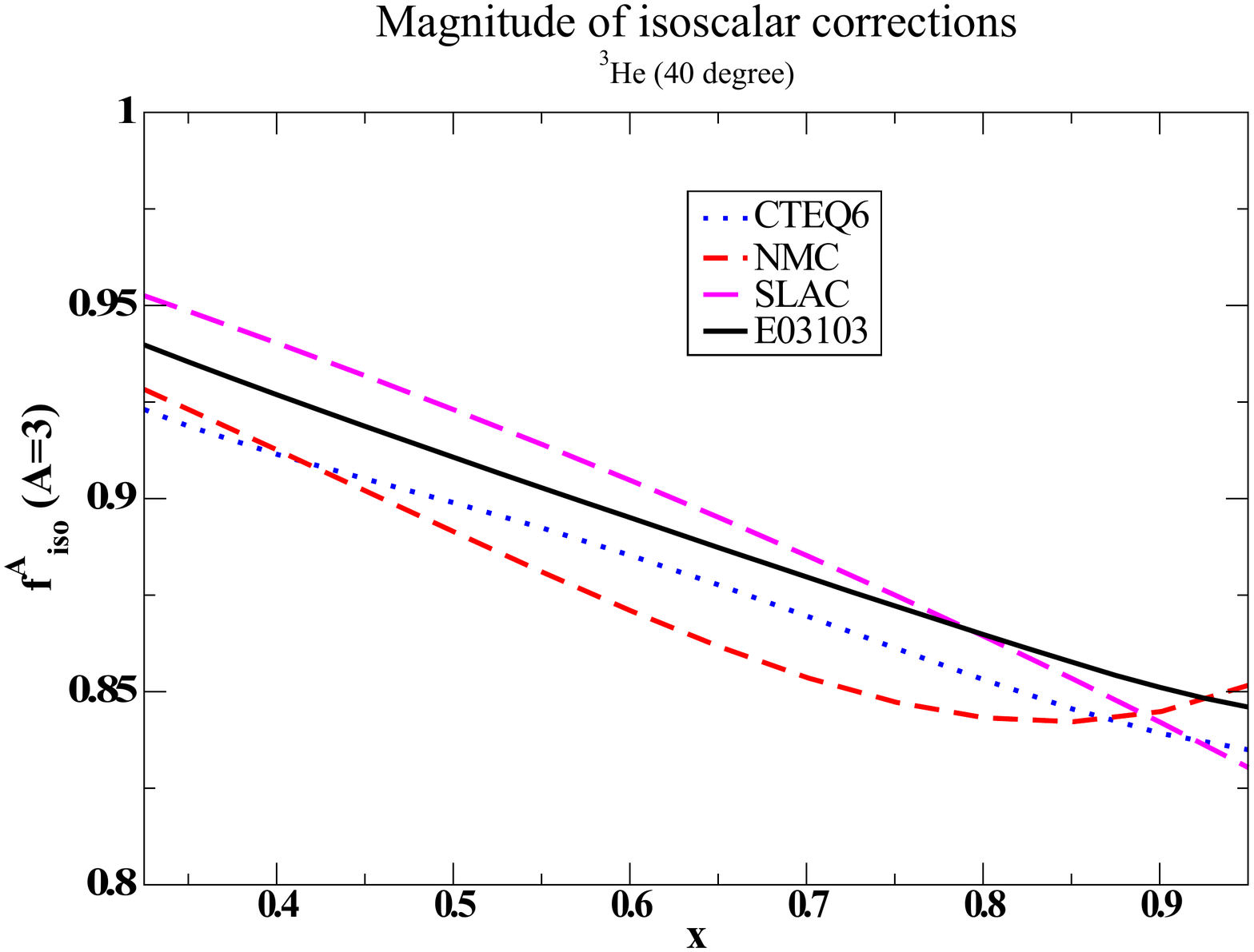} \\
\includegraphics[width=85mm, height=50mm, trim=0mm 10mm 20mm 30mm, clip]{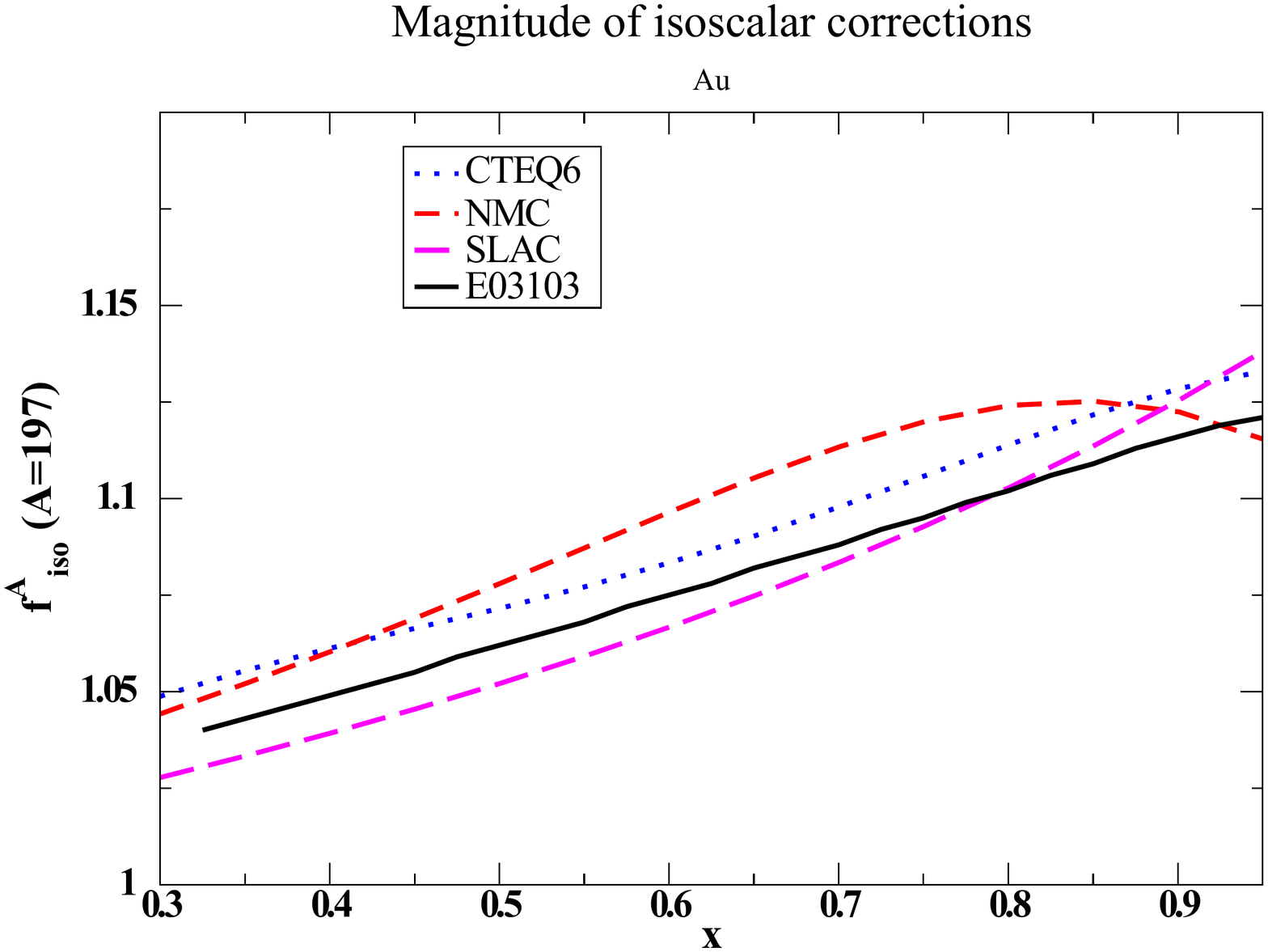}
\caption{Magnitude of isoscalar corrections for $^3$He (top) and Au
(bottom) targets for the 40 degree data for the different
parameterizations of $F_2^n/F_2^p$ as discussed in the text. The solid black
line represents the multiplicative correction factors obtained using the
smearing method discussed in the text and was used in the E03103 analysis
for the EMC ratio extraction.
\label{f2nf2psize_fig}}
\end{figure}

Figure~\ref{dprat_fig} shows the  $\sigma^D/2\sigma^p$ cross section ratios
extracted from the E03103 data for the 40 degree kinematics. Representative
extractions~\cite{hermes_dbyp, arrington12b} of the same ratio are also shown
in the figure. It should be noted that the isoscalar correction depends on
$Q^2$~\cite{arrington09, arrington12b}, and this effect is not negligible at
large $x$. The correction factors derived using various parameterizations for
$^3$He and Au are shown in Figure~\ref{f2nf2psize_fig}.

In the case of $^3$He, one can avoid the uncertainty associated with the
isoscalar corrections by extracting the ratio of $^3$He to ($^2$H+$^1$H).
This ratio and the comparison to the isoscalar-corrected $^3$He/$^2$H ratio
are presented in section~\ref{xdepresult.ssec}.

\subsection{Scaling violation effects at high $x$}\label{hixcor.ssec}

As discussed in Sec.~\ref{kinem.ssec}, deviations from the scaling of the
simple quark parton model arise due to QCD evolution of the PDFs, target-mass
corrections which involve finite-$Q^2$ corrections to the approximations
made in the infinite $\nu$, $Q^2$ limit, and higher twist contributions which
go beyond incoherent scattering from individual partons.

The kinematic effects due to target mass corrections were first calculated in
the framework of the operator product expansion OPE in Ref.~\cite{georgi_tmc}.  In
the nucleon case, the measured structure function $F_2^\text{meas}$ can be related
to the massless limit structure function $F_2^{(0)}$~\cite
{Schienbein_tarmass_rev} via
\begin{eqnarray} \nonumber  \label{tarmass_eqn}
F_2^\text{meas}(x,Q^2)&=&\frac{x^2}{\xi^2 r^{3}}F_2^{(0)}(\xi,Q^2) + \frac{6M^2x^3}{Q^2 r^{4}}h_2(\xi,Q^2)\\
&& +\frac{12M^{4}x^4}{Q^{4} r^{5}}g_2(\xi,Q^2)
\end{eqnarray}
where $h_2(\xi,Q^2) =\int_{\xi}^{1}du~u^{-2} F_2^{(0)}(u,Q^2)$, $g_2(\xi,Q^2)
=\int_{\xi}^{1}dv~(v-\xi)v^{-2} F_2^{(0)}(v,Q^2)$, $r=\sqrt{1+\frac{Q^2}{
4x^2M^2}}$ and $\xi = \frac{2x}{1+r}$. $F_2^{(0)}$ does not contain target mass
effects and this is the function which obeys the QCD evolution effects in the
absence of higher twist effects. It should be noted that there are different
prescriptions~\cite {Schienbein_tarmass_rev, Accardi_tarmass2008plb,
Accardi_tarmass2008jhep} available for these kinematical corrections with
slightly different results, however, the appropriate prescription for
target mass corrections in nuclei is not well defined.

In the extraction of EMC effect, $A$-independent scaling violations will cancel
in the cross-section ratios. If the $h_2$ and $g_2$ corrections are negligible or
target independent, then $F_2^\text{meas}(x,Q^2)$ is directly connected to
$F_2^{(0)}(\xi,Q^2)$ (see Eqn.~\ref{tarmass_eqn}) through a simple relation. In
that case, the target mass effects on cross section ratios can be well
approximated by the substitution $x\rightarrow \xi$. Our investigations show that
the $h_2$ and $g_2$ terms yield significant corrections to the structure function
for lower $x$ and $Q^2$ data, but that these are nearly target independent. 
Up to $x=0.7$, the impact of neglecting these additional model-dependent corrections
is below 1\%.

Higher-twist effects can also lead to scaling violations, although it has been
argued based on quark-hadron duality~\cite{niculescu00a, melnitchouk:2005zr}
that for nuclei, the Fermi motion of the nucleons samples a sufficient
kinematic region that the observed structure function reproduces the DIS limit
even down to extremely low $Q^2$ and $W^2$ values~\cite{Arrington:2003nt}. 
This will be examined in Sec.~\ref{results.sec} using the extensive measurements taken
to examine the $Q^2$ dependence of the EMC ratio.

\begin{figure}[htb]
\begin{center}
\includegraphics[width=85mm, angle=0, trim=10mm 18mm 20mm 30mm, clip]{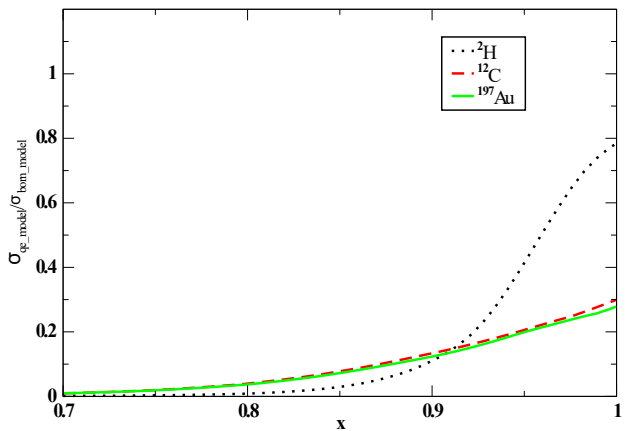}
\caption{Fractional quasielastic contribution to the cross section based on
our model at 40 degrees for $^2$H, $^{12}$C and $^{197}$Au. Here,
$\sigma_\text{qe}$ is the contribution from the quasielastic piece of the model
(in the Born approximation) and $\sigma_\text{Born}$ is the total Born cross section.
\label{qe_contri_fig}}
\end{center}
\end{figure}

It is unclear if the extended scaling of the EMC ratio will hold true in the
presence of significant contributions from quasielastic
scattering~\cite{melnitchouk:2005zr, niculescu06, Osipenko:2010sb}. 
Figure~\ref{qe_contri_fig} shows the quasielastic contribution,
$\sigma_\text{qe}/\sigma_\text{Born}$, based on our cross section model for the 40
degree kinematics.  In our model, the quasielastic contribution is negligible
for $x \ltorder 0.7$, and $\ltorder$10\% for all nuclei up to $x=0.9$, with
further suppression when examining target ratios.  In the next section, we will
compare our results at large $x$ to those from SLAC~\cite{slace139} and the CLAS collaboration at Jefferson Lab~\cite{Schmookler19}.  At SLAC kinematics, the QE contribution is highly suppressed due to the large values of $Q^2$ at large $x$.
For the CLAS data~\cite{Schmookler19}, the QE contribution is larger but because of the limited $x$ range, its contribution is $\ltorder 0.5$\% for all targets, small enough that we do not apply a correction.

\section{Systematic uncertainties}\label{syst.sec}

Statistical uncertainties for the cross section ratios presented here are 
$\approx$0.5\% per bin (size 0.025) up to $x\approx0.75$, with gradually increasing
uncertainty as $x$ increases. The total systematic uncertainty in
the cross section extraction is taken as the sum in quadrature of all
systematic uncertainties of the quantities that contribute to the cross
section. The components of the systematic uncertainty can be broadly divided
into two groups: point-to-point uncertainties and normalization uncertainties.
Point-to-point uncertainties are due to effects which may vary with time,
kinematic conditions, or detector location, and so their effect is 
(or at least can be) uncorrelated between different data points. Normalization
(scale) uncertainties affect the measurement globally (e.g., target
thickness).  Most corrections involve a mixture of point-to-point and
normalization uncertainties. The resulting overall uncertainty in the cross
section ratios is less than the total uncertainty in the cross section itself
because many of the scale uncertainties and some point-to-point type
errors cancel in the ratios. Table~{\ref{syst_short_table}} summarizes the
systematic uncertainties in extracting the cross section ratios. The dominant
remaining contributions from the scale uncertainties are those associated with
the absolute target thicknesses, radiative and background corrections.  These
range from 1.5--2.0\% on the EMC ratios, and are provided for each target
ratio in the supplementary data tables~\cite{online_material}. Individual
contributions are discussed below.

\begin{table}[htb]
\caption[]{Typical sources and magnitude of the systematic
uncertainties in extracting cross section ratios. These are added in
quadrature with the statistical uncertainties to get the total random
uncertainties.}
    \begin{tabular}{lcc}
	 \hline
	 \hline
	 Item & Absolute & $\delta R/R$ ($\pm$\%) \\
       & uncertainty($\pm$) &        \\
	 \hline
Beam Energy (offset)    & 5$\times$10$^{-4}$ 	& --   \\
Beam Energy (tgt-dep)	& 2$\times$10$^{-4}$ 	& 0.08    \\
HMS Momentum (offset)  	& 5$\times$10$^{-4}$ 	& --  \\
HMS Momentum (tgt-dep) 	& 2$\times$10$^{-4}$ 	& 0.0-0.12  \\
HMS angle (offset)   	& 0.5 mr      	& -- \\
HMS angle (tgt-dep)	    & 0.2 mr		& 0.29-0.60 \\
Beam Charge       	    & 0.5\%       	& 0.31 \\
Target Boiling     	    & 0.45\%       	& 0.0-0.1 \\
End-cap Subtraction   	& 2--3\%       	& 0.28-0.45 \\
Acceptance          	& 1\%        	& 0.3 \\
Tracking Efficiency   	& 0.7\%       	& 0.3 \\
Trigger Efficiency   	& 0.3\%       	& 0.0 \\
Electronic Dead Time  	& 0.06\%       	& 0.0 \\
Computer Dead Time   	& 0.3\%       	& 0.3 \\
Charge Symmetric BG   	&          	& 0.0-1.0 \\ 
Coulomb corrections   	& 0.2\%       	& 0.1 \\
Pion Contamination   	& 0.2\%       	& 0.1 \\
Detector Efficiency   	& 0.2\%       	& 0.0 \\
Radiative Corrections  	& 1\%        	& 0.5 \\
Bin-centering          	& 0.2\%       	& 0.1 \\
\hline
Quadrature sum         &             & 0.90-1.11 \\
    
\hline
\hline
\end{tabular}
\label{syst_short_table}
\end{table}

Kinematic offsets in the beam energy, spectrometer momentum, and spectrometer
angle can yield errors in our extracted cross sections. We use our model
cross section to assess the uncertainty in the cross sections due to these effects.
The cross section ratios, however, are largely insensitive to such offsets.

The point-to-point uncertainty in the beam charge measurement was estimated to
be $0.5\%$ via studies of the residuals to calibration fits taken throughout
the experiment. A scale uncertainty of $0.2\%$ was assumed for the charge measured,
due to the uncertainty in the calibration against the UNSER parametric beam current
calibration~\cite{aji_thesis}.

As mentioned in section~\ref{target.ssec}, thicknesses of the solid targets
were calculated using measurements of the mass and area of the targets.
Thicknesses of the cryotargets were computed from the target density and the
length of the cryogen in the path of the beam. The absolute uncertainty in the
$^2$H thickness is estimated to be 1.29\%. When comparing to other cryogenic targets,
part of this uncertainty cancels and the overall uncertainty in the cross section
ratio ($A/D$) is 1.59\% and 1.29\% for $^3$He/D and $^4$He/D respectively. For heavy 
nuclei, the scale uncertainty in the cross section ratio due to target thickness 
is found to be between 1.4\% to 2.4\%.  In addition to the nominal target densities, 
there are corrections associated with beam heating effects and fluctuations in the 
pressure and temperature. The uncertainty associated with this correction comes from the
uncertainties in the fits to target luminosity scans. Though no boiling
correction is made in the case of the deuterium target, the uncertainty from the 
luminosity scan data is still included in the $A/D$ ratios. We assign a scale uncertainty of 0.24\% (solid targets) to 0.38\% (helium targets) for the target ratios.

The scale uncertainty of the acceptance in the HMS was estimated to
be $1\%$ from the elastic cross section studies, while the 
point-to-point uncertainty comes from the comparison of the model in the inelastic
region (where the cross section is smoothly varying) to data, and is
estimated to be $0.5\%$. In the cross section ratios, these uncertainties partially cancel. The scale uncertainty in the solid target ratios is estimated to be 0.5\% and 0.2\% for helium target ratios. The point-to-point uncertainty is estimated to be 0.3\% for both.

The normalization uncertainty of the tracking efficiency is determined to be
$0.7\%$, mainly due to the limitations of the algorithm used for tracking and the 
efficiency calculation algorithm. A point-to-point uncertainty of $0.3\%$ is assigned to the tracking efficiency in the target ratios, primarily due to differences in rates 
between the targets. 

At very low $x$ values, the structure functions are expected to scale, and any
deviation is possibly due to the charge symmetric background (since this is
the dominant uncertainty for heavy nuclei at small $x$ and large scattering
angles). A comparison of 40 and 50 degree data suggests that scaling is
satisfied if the CSB varies by no more than $5\%$. A polynomial fit was made
to the charge symmetric background as a function of $x$, and $5\%$ of the
magnitude of the charge symmetric background is applied as the point-to-point
uncertainty in the charge symmetric background subtraction.

The model dependence in the radiative correction was
studied by varying the strength of the DIS and QE contributions to our model 
independently, and by comparing to a completely independent fit by Bosted and 
Mamyan~\cite{Bosted:2012qc}. The change in extracted cross section was rather
pronounced  in the low $x$ region when comparing to the Bosted-Mamyan fit (several
percent for heavier nuclei). 
This was primarily due to contributions to the radiative tail from the QE process.  
Investigations comparing the QE cross section used in the model described here and the 
Bosted-Mamyan fit showed similar levels of agreement with existing data at low $Q^2$, 
although both models displayed deviations at the 10\% level.  In the end, the final 
results were generated by taking the average of the target ratios generated with both 
models with an additional (correlated) $x$-dependent uncertainty added due to the 
difference in the models. In addition to this $x$-dependent uncertainty (coming from
differences in the QE model), an additional $1\%$ uncertainty in the cross section is 
assigned due to the inelastic model, and additional point-to-point uncertainties
are assigned to account for kinematic dependent differences.
The point-to-point uncertainty for the target ratios is estimated to be $0.5\%$.
An additional scale uncertainty, associated with the difference in radiation 
lengths between the targets, is taken to be 0.1\% except for the high-radiation 
length targets (Cu and Au) for which it is 1\%.

The efficacy of the model used to describe the Monte Carlo yield across the
acceptance of the spectrometer was studied by
varying the shape of the model. This is done by supplying artificial $x$ and
$Q^2$ dependencies as input to the individual DIS and QE pieces in the model
cross section. The variation was found to be most pronounced for the $x>$0.8
region, and we estimate a point-to-point uncertainty of $0.2\%$ for the
cross sections, and $0.1\%$ for the cross section ratios.
Uncertainties in the Coulomb corrections are mainly due to the knowledge of
the energy shift, $\Delta E$, used in the EMA calculation. We estimate this to
be known at the $10\%$ level. For the Au target at 40 degrees, this
uncertainty ranged from $0.5\%$ at low $x$ to $1.5\%$ at high $x$.

\section{Results and discussion}\label{results.sec}

\begin{figure}[htbp]
\includegraphics[width=83mm, height=70mm, trim={8mm 5mm 18mm 20mm}, clip]{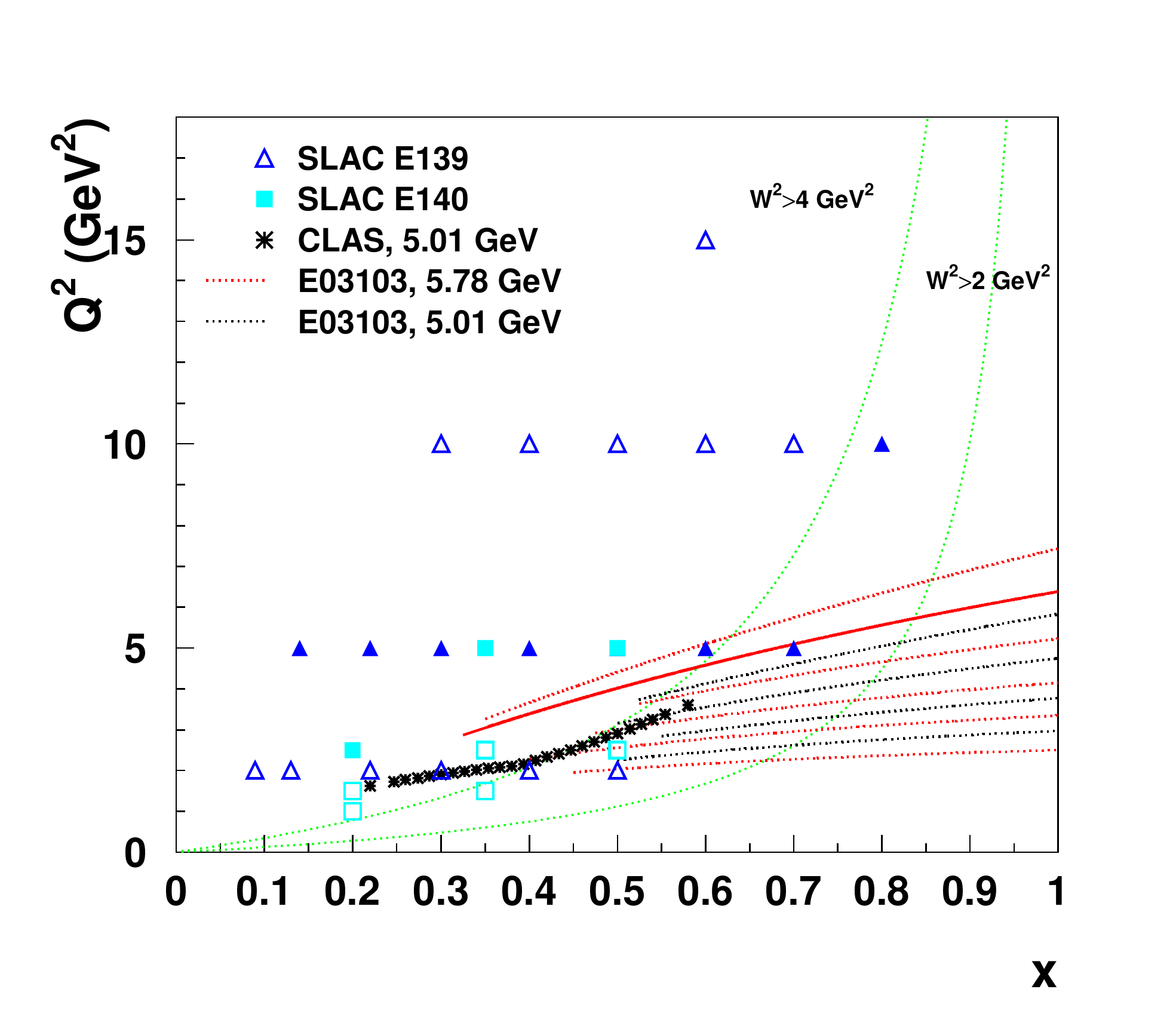}
\caption{(Color online) The E03103 kinematics, indicated with dashed and
dotted lines, along with the SLAC experiments E139~\cite{slace139} (triangles)
and SLAC E140~\cite{slace140} (squares).  Kinematics are shown for the target
with maximum coverage (Fe for the SLAC measurements, C for E03103). The solid
line and filled symbols represent the kinematics used in the main comparison
of the results. Contours of constant invariant mass squared are also shown in
the figure.
\label{slac_e0103_kinem_fig}}
\end{figure}

Before presenting the results, it is instructive to compare our kinematics to
the earlier SLAC experiments. This will help identify potential issues in the
comparison of the EMC ratios and elucidate the possible role of the $Q^2$ dependent
effects when comparing data from different experiments. 
Figure~\ref{slac_e0103_kinem_fig} shows kinematics for our measurement and SLAC E139 and
E140, as well as the recent results from CLAS.

E03103 took data on all targets at 40$^\circ$ and 50$^\circ$, and the cross
section ratios with respect to deuterium were extracted.  The EMC ratios are
extracted from the 40 degree angle (solid line in
Fig.~\ref{slac_e0103_kinem_fig}) where the data have better statistics and
more complete kinematic coverage. Data were also collected for a detailed $Q^2$
dependence study at 8 additional kinematic settings on C and $^2$H.

In the cross section ratio plots, representative world data is displayed
with the corresponding nuclei where available. In the kinematics
comparison plot we chose to display kinematics of SLAC experiments because of
the overlap in kinematics with our experiment at high $x$. For comparison of
the EMC ratios we use the SLAC data averaged over all $Q^2$ values at each $x$; note that at the highest $x$ measured by SLAC ($x=0.8$), only $Q^2=10$~GeV$^2$ is available. For each $x$, $Q^2$ value, the published SLAC E140 results are averaged over several $\epsilon$ points - this point is addressed later in this section.

To be consistent, the SLAC data are presented with updated Coulomb and isoscalar
corrections using the same prescriptions used for the analysis of E03103
data. The updated data points and corrections factors are available in the
supplementary online material~\cite{online_material}.

\subsection{$Q^2$ dependence of the ratios}\label{q2depresult.ssec}

\begin{figure}[htbp]
\includegraphics[width=85mm ,height=62mm, trim=4mm 2mm 15mm 5mm, clip]{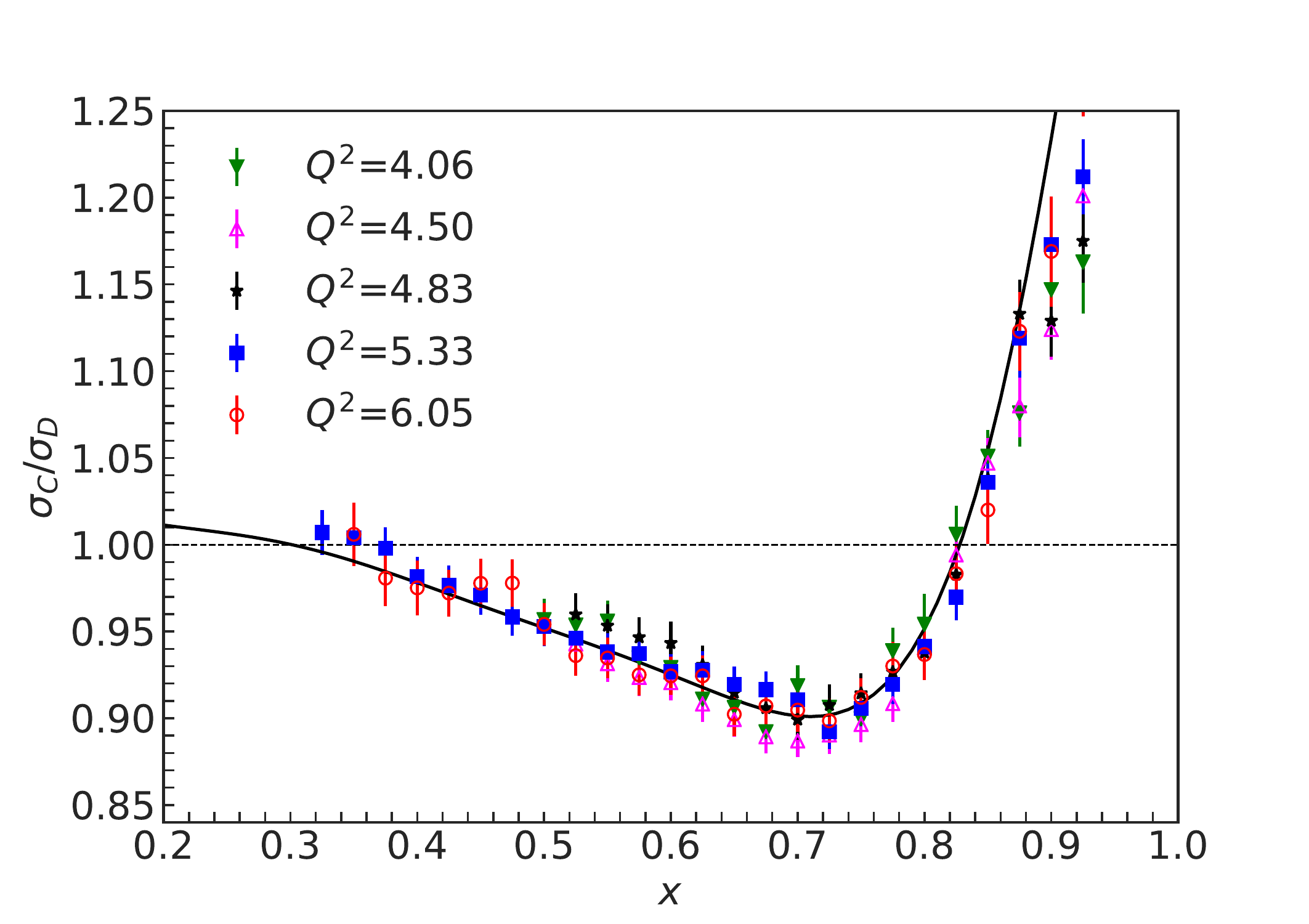}
\includegraphics[width=85mm, height=62mm, trim=4mm 2mm 15mm 5mm, clip]{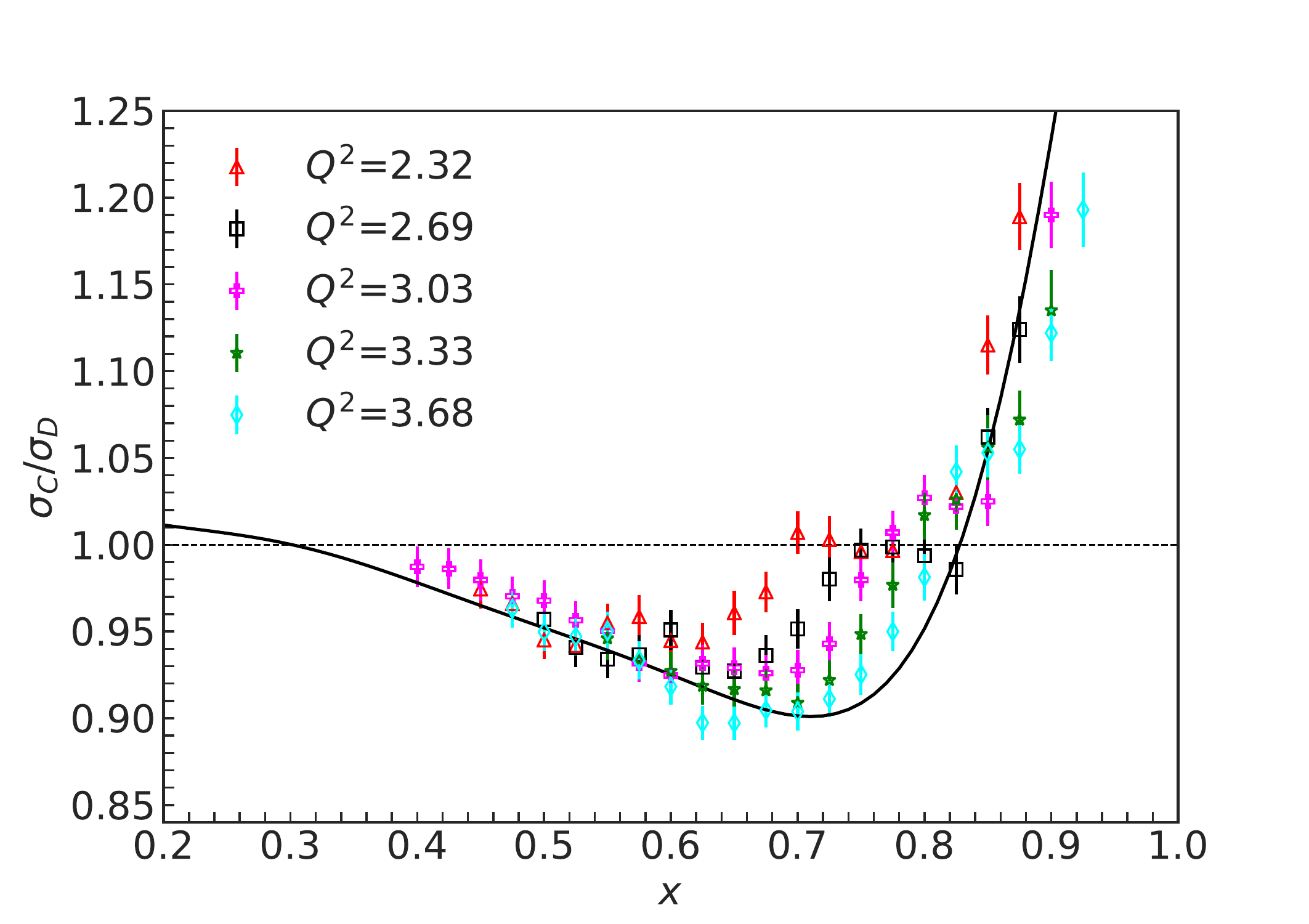}
\caption{(Color online) Ratio of C and $^2$H cross sections for the five largest $Q^2$ (top
panel) and five lowest $Q^2$ (bottom panel) settings as a function of $x$.
Uncertainties are the combined statistical and point-to-point systematic. The
$Q^2$ values quoted are for $x=0.75$, and the data labeled $Q^2$=5.33
correspond to our primary results, taken at 40$^\circ$. The solid black line is the
SLAC parameterization of the EMC effect for carbon~\cite{slace139}.
\label{emc_x_C_hiq2dep_fig}}
\end{figure} 

The scaling of the structure functions for nucleons is expected to hold 
in the conventional DIS region ($W^2>4$ and $Q^2>1$), where the
non-perturbative, resonance structure is no longer apparent and QCD
evolution is the only source of $Q^2$ dependence. At smaller values
of $W^2$, corresponding to large $x$, additional scaling violations can
originate from resonance contributions. For E03103, the data are in the conventional DIS region up to $x \approx 0.6$.  There are indications
\cite{Arrington:2003nt} that the nuclear structure functions in the resonance
region, down to very low $W^2$ values ($W^2 > 1.5$~GeV$^2$ for
$Q^2>3$~GeV$^2$), shows the same global behavior as in the DIS region. 
Therefore, we took data at large $x$ extending below $W^2=4$~GeV$^2$, and made
detailed measurements of the $Q^2$ dependence of the ratios to ensure that
there was no indication of any systematic deviation from the DIS limit.

The EMC ratios for carbon at several $Q^2$ values are compared in
Fig.~\ref{emc_x_C_hiq2dep_fig}.  The top panel shows the EMC ratios for the
five highest $Q^2$ settings from our experiment, along with the fit to the EMC effect
from~\cite{slace139}.  The data do not show any systematic $Q^2$ dependence, and the scatter at the largest $x$ values is consistent with the uncertainties in the individual
measurements.  This suggests that any $Q^2$ dependence in the structure
function is either small or cancels in the target ratios.  The bottom figure
shows the low $Q^2$ measurements, where there is a clear difference in the
$Q^2$ dependence of carbon and deuterium below $Q^2 \approx 3$~GeV$^2$ and
$x>0.6$, corresponding to $W^2$ values below 2--3~GeV$^2$, where one expects
large resonance contributions.

\begin{figure}[htbp]
\includegraphics[width=85mm, height=60mm, trim=8mm 2mm 15mm 15mm, clip]{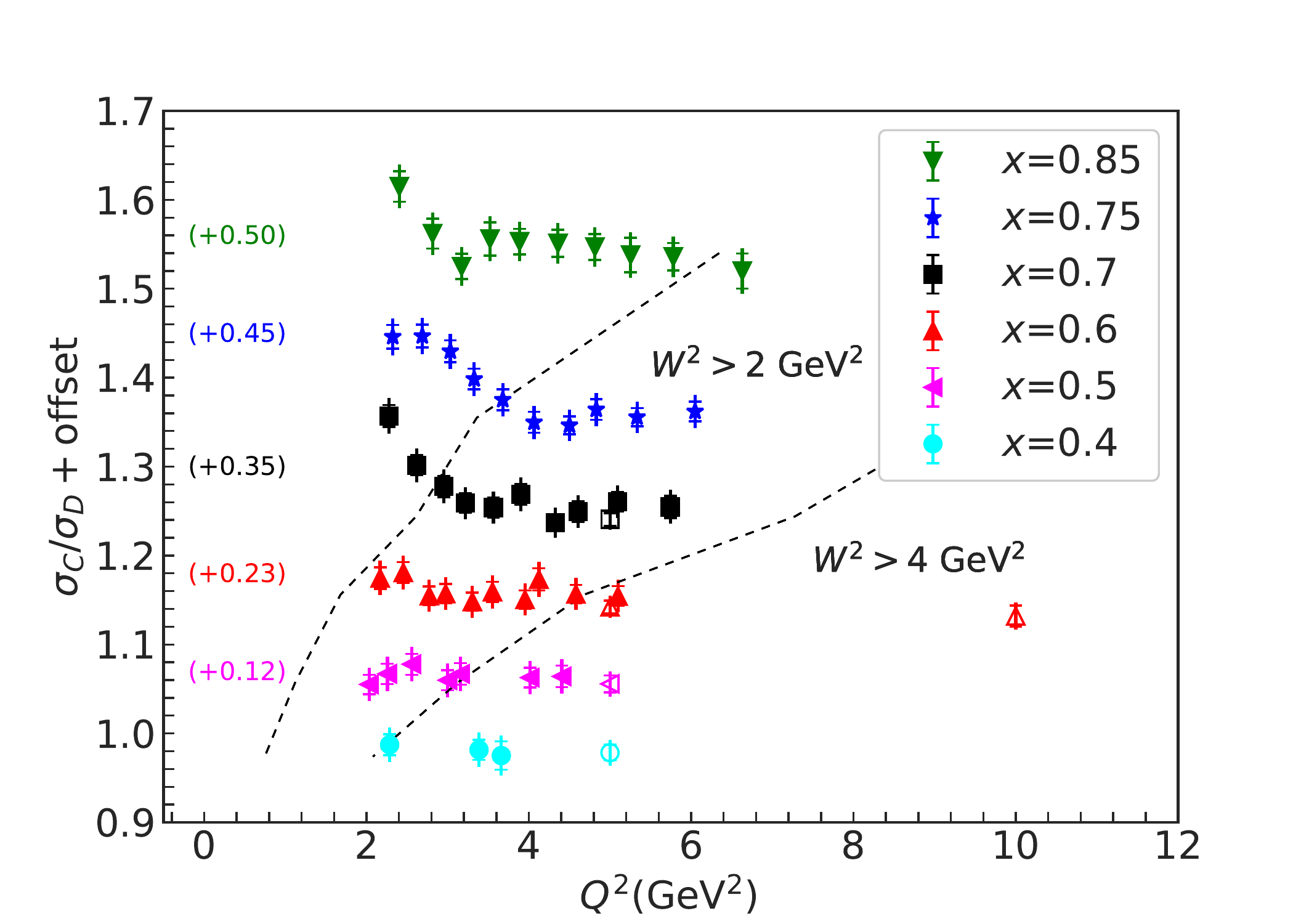} \\
\includegraphics[width=85mm, height=60mm, trim=8mm 2mm 15mm 15mm, clip]{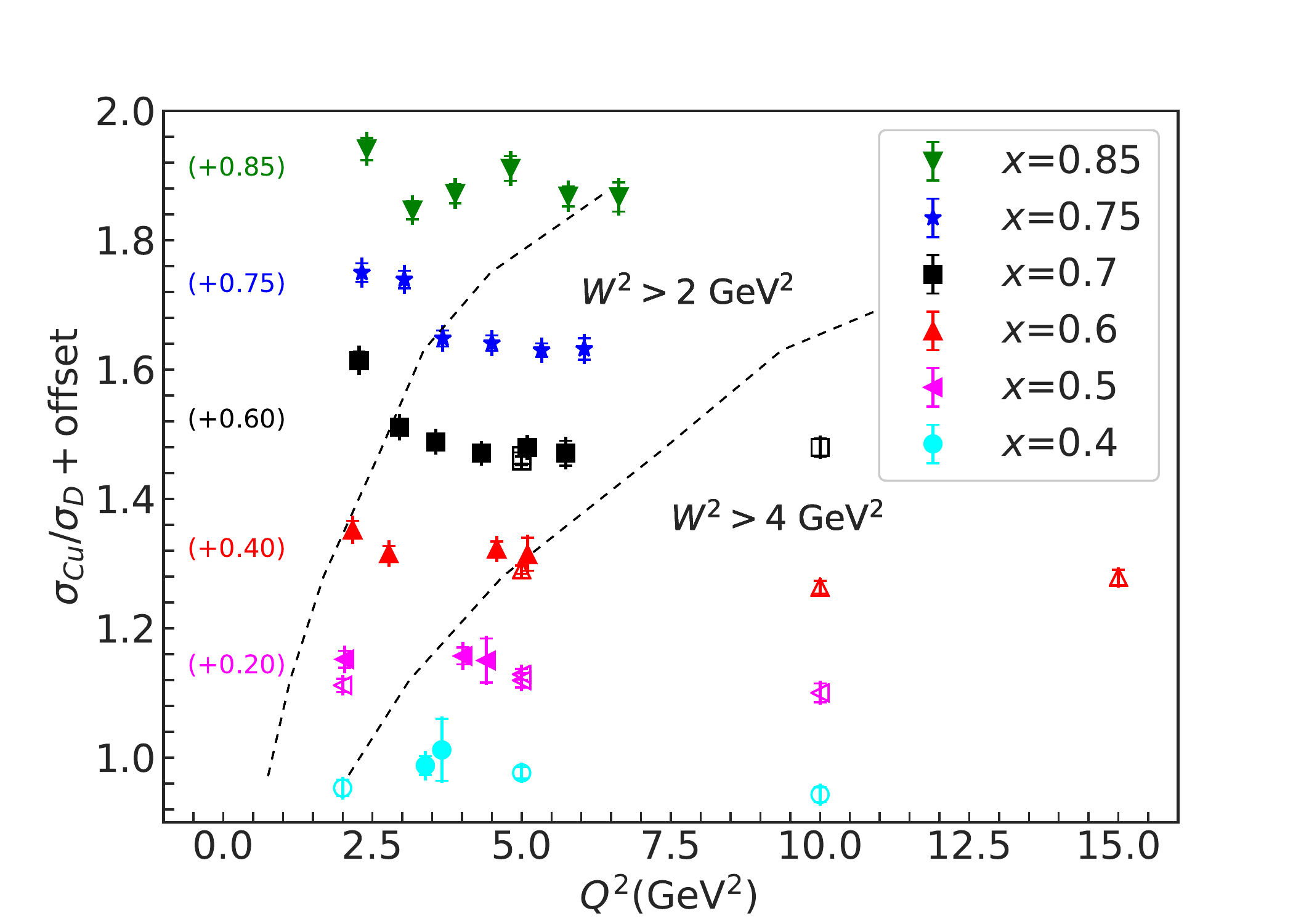}
\caption{(Color online) EMC ratios for C (top) and Cu and Fe (bottom) as a
function of $Q^2$ at fixed $x$ values as indicated in legend. For clarity, an
additive offset is applied along the y axis. Open symbols are from updated
SLAC E139~\cite{slace139} results while the closed symbols are E03103 values.
Inner error bars show the combined statistical and point-to-point systematic
while the outer error bars represent the total uncertainty including the
normalization uncertainties. The dashed lines indicate the values of $W^2=2$, 4~GeC$^2$ for each x value.
\label{emc_q2_c_fixed_x_fig}}
\end{figure}

Figure~\ref{emc_q2_c_fixed_x_fig} shows the $Q^2$ dependence of the structure
functions for C (top) and Cu or Fe (bottom) at several $x$ values, to allow
for a more careful examination of the $Q^2$ dependence as a function of $x$. 
The carbon data have additional $Q^2$ values for E03103, due to the data
taken using a lower beam energy, while the Cu data have more high-$Q^2$
data from the SLAC measurements. There is a fair agreement with the SLAC data
over the kinematic regions where data are available, and clear deviations from
a constant ratio are visible below $Q^2=4$~GeV$^2$ and at large $x$ values.

\subsection{$x$ dependence of the ratios}\label{xdepresult.ssec}

\begin{figure}[htbp]
\begin{center}
\includegraphics[width=85mm, trim=0mm 0mm 0mm 0mm, clip]{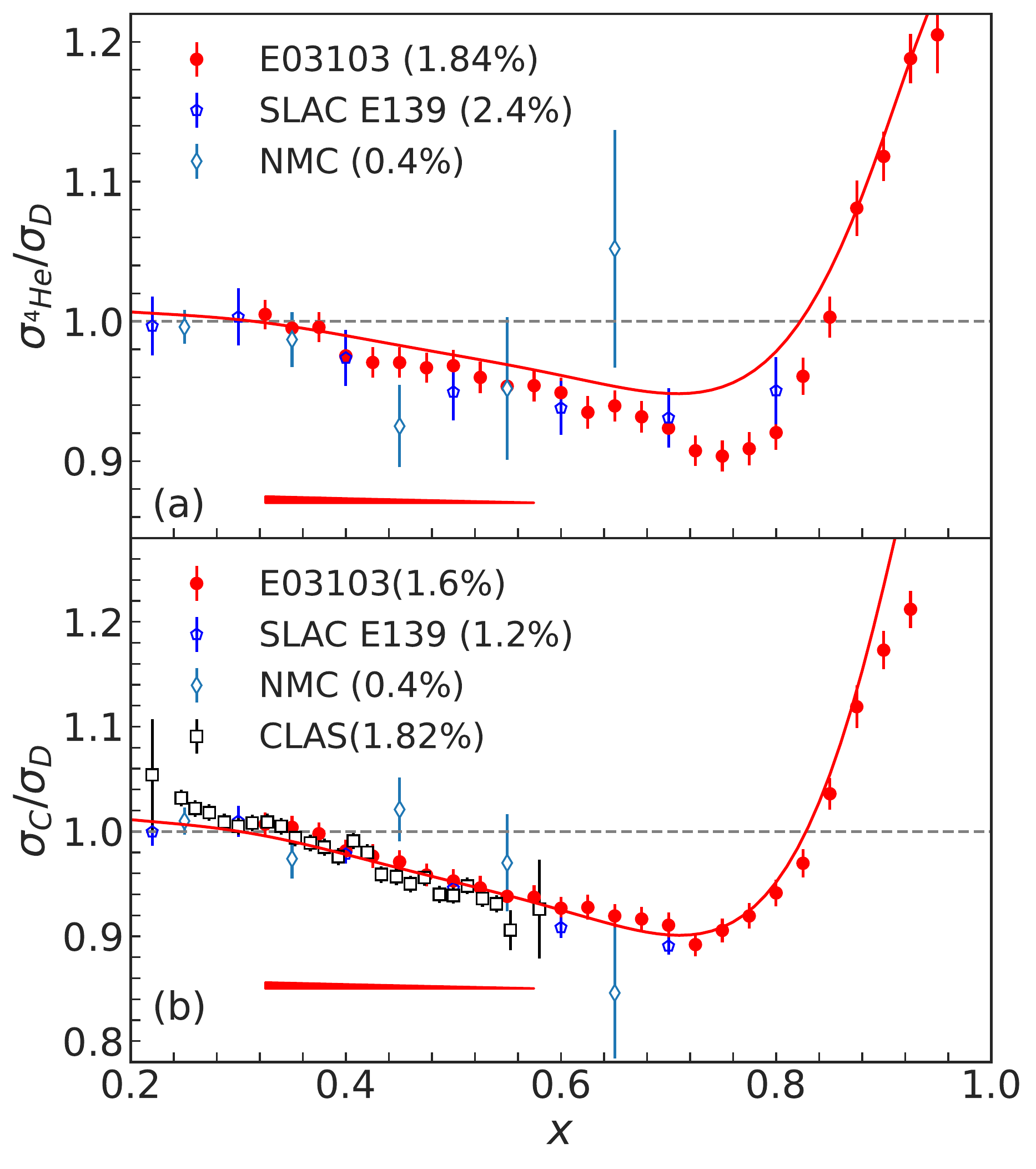}
\caption{(Color online) EMC ratios for $^4$He (a) and $^{12}$C (b) as a function of $x$ for the 40 degree results. Error bars show the combined statistical and point-to-point
systematic uncertainties. The solid error band denotes the correlated uncertainty due to the size of the quasielastic tail in the radiative corrections; overall normalization uncertainties are shown in the parenthesis. Also shown are the updated SLAC E139~\cite{slace139} and NMC data
\cite{Arneodo:1995cs, Amaudruz:1995tq}. The solid curves show the $A$ dependent  fit to the EMC effect from~\cite{slace139}.
\label{emc_x_40deg_he4_fig}}
\end{center}
\end{figure}

We now examine the $x$ dependence of the EMC ratios for all of the targets
from E03103, SLAC, and CLAS, including Coulomb corrections and our updated 
isoscalar corrections. We first discuss the cross section ratios for C and 
$^4$He, as these ratios have no isoscalar correction, and the Coulomb distortion 
effects are small ($<$1\%) for these nuclei. Figure~\ref{emc_x_40deg_he4_fig} 
shows the cross section ratios for $^4$He and $^{12}$C, along with the updated 
SLAC E139 data and the NMC data~\cite{Amaudruz:1995tq,Arneodo:1995cs}. Note that the red curve is a global fit to the A dependence from SLAC~\cite{slace139}, which yields a smaller EMC effect for $^4$He than seen in their data or
our updated measurement. CLAS results~\cite{Schmookler19} are also
shown for carbon. There is overall good agreement between the data
sets.  Both the CLAS and E03103 results are of high precision, with
E03103 extending to larger $x$, although at a lower $W^2$ 
than previous measurements.

\begin{figure}[htbp]
\begin{center}
\includegraphics[width=0.46\textwidth,  trim=0mm 0mm 0mm 0mm, clip]{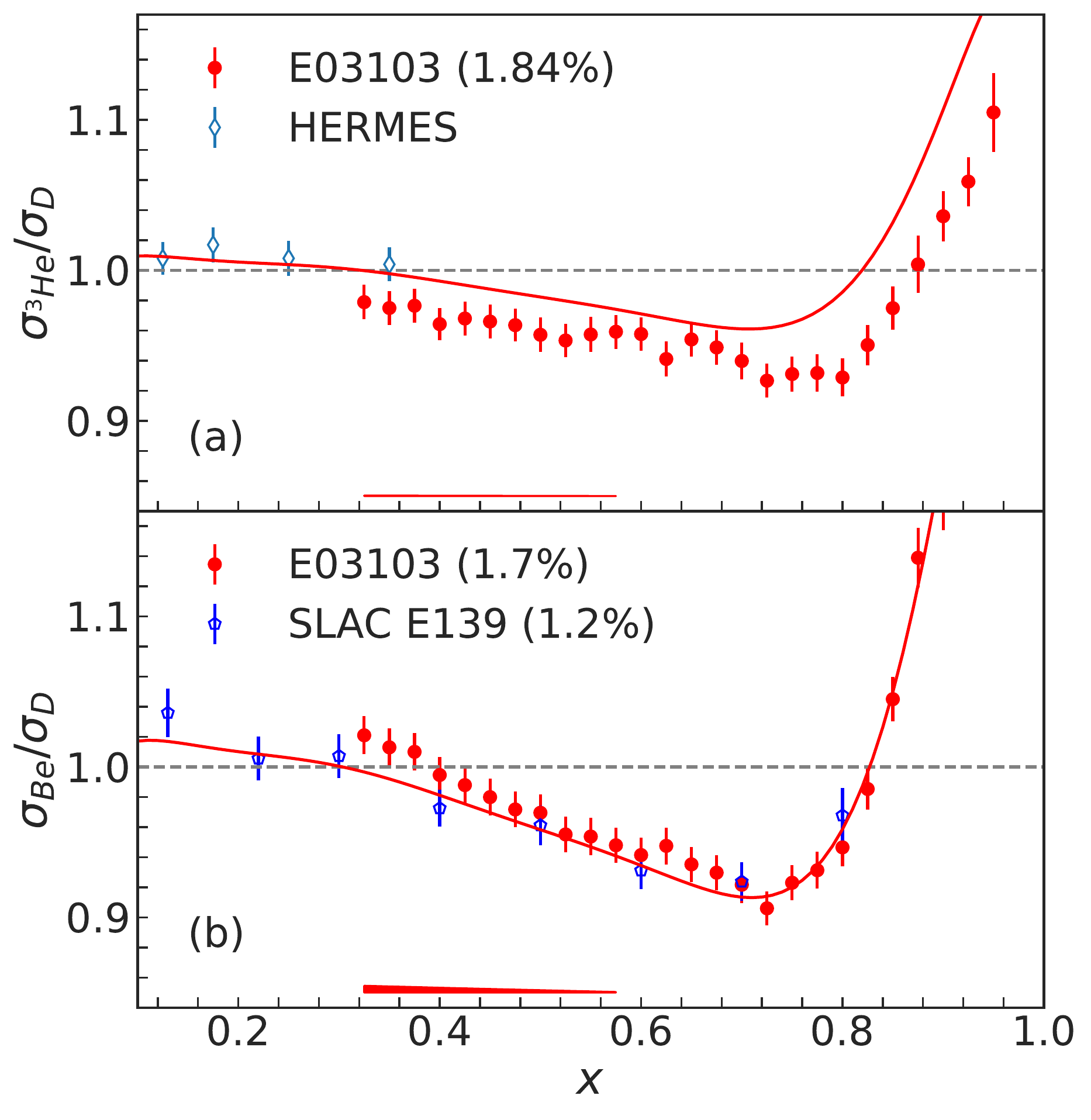}
\caption{(Color online) Isoscalar EMC ratios for $^3$He (a) and $^9$Be (b) for the 40 degree data. Uncertainties are as described in Fig.~\ref{emc_x_40deg_he4_fig}. 
Also shown are the
HERMES $^3$He data~\cite{hermes_ackerstaff,hermes_correction_airapetian} 
(updated to include our modified isoscalar correction). The solid curve shows an 
$A$ dependent parameterization~\cite{slace139} for the EMC effect.
\label{emc_x_40deg_he3_fig}}
\end{center}
\end{figure}

Figure~\ref{emc_x_40deg_he3_fig} shows the cross section ratios for $^3$He
and $^9$Be.  Both of these nuclei are light enough that the Coulomb corrections
are small, but require a proton (neutron) excess correction to obtain the
isoscalar EMC ratios (see section~\ref{iso.ssec}). The magnitude of this
correction is significant for $^3$He, ranging from about $5\%$ to $15\%$ for
our kinematics.  For $^9$Be, the correction is of the opposite sign and
roughly a factor of three smaller.  The $^3$He EMC ratios exhibit the general
shape observed for the cross section ratios for heavy nuclei.

\begin{figure}[htbp]
\begin{center}
\includegraphics[width=0.46\textwidth, trim={2mm 0mm 15mm 15mm}, clip]{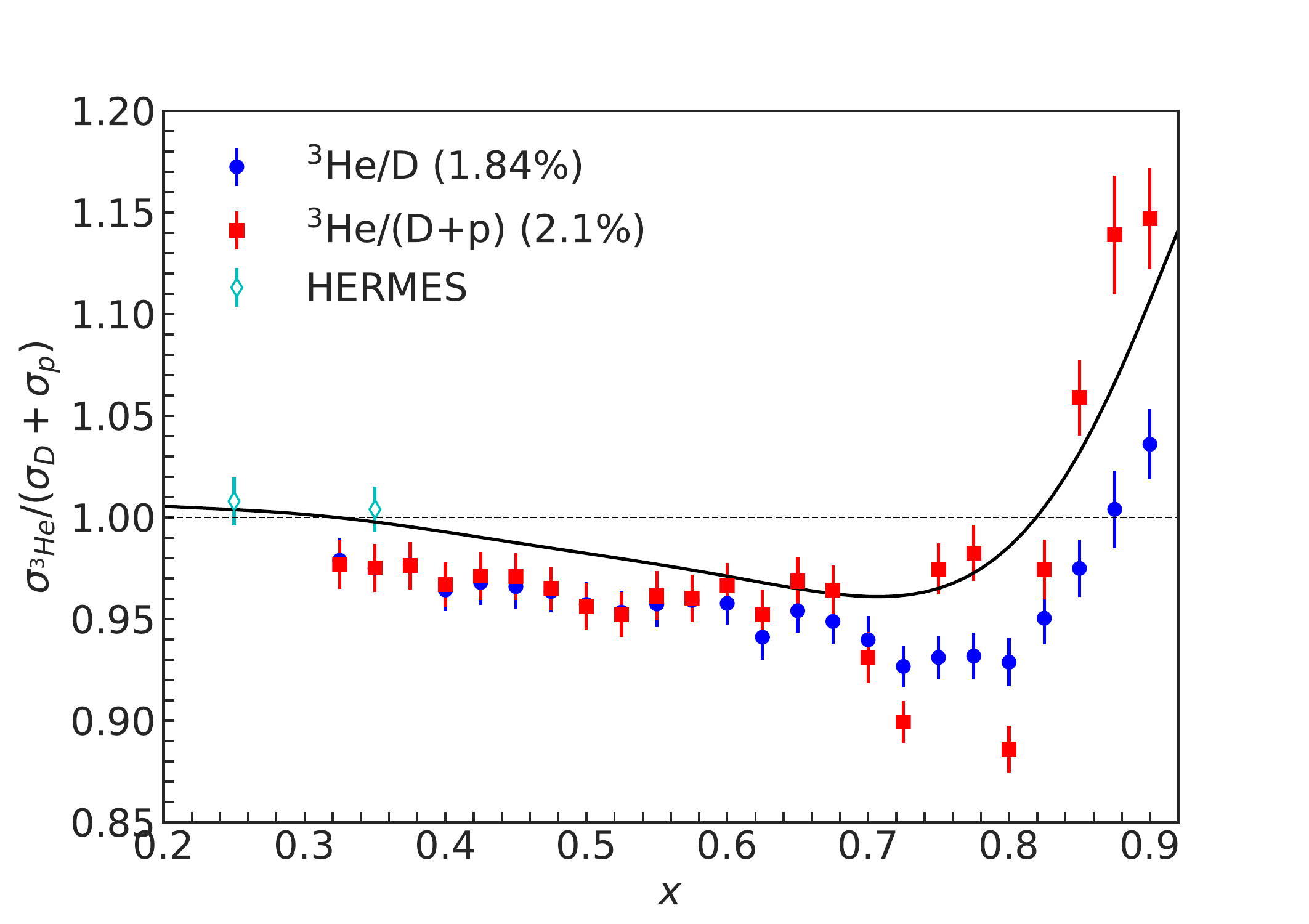}
\caption{(Color online) Comparison of the isoscalar-corrected $^3$He/D ratio (blue circles) to $^3$He/(D+p) (red squares).  The agreement is very good below $x=0.65$ (which corresponds to $W \approx 1.9$~GeV).  At larger $x$, the resonance structure in the free proton is evident.}
\label{fig:3he_over_dplusp}
\end{center}
\end{figure}

One can avoid the uncertainty associated with the isoscalar correction, and
thus better evaluate models of the EMC effect, by taking the ratio of $^3$He
to ($^2$H+$^1$H) which allows comparisons to calculations that are independent
of the neutron structure function. These ratios are extracted for our 40
degree setting and shown in Figure~\ref{fig:3he_over_dplusp} (red squares),
along with the isoscalar-corrected $^3$He/$^2$H ratios (blue circles).
The isoscalar-corrected $^3$He/$^2$H ratio and the $^3$He/($^2$H+$^1$H)
results are in good agreement below $x \approx 0.65$, but the resonance
structure at large $x$ in the proton is not washed out, and so the extended
scaling observed in nuclei~\cite{Arrington:2003nt} is not as effective,
limiting the useful range for this ratio to $x \ltorder 0.65$.

\begin{figure}[htbp]
\begin{center}
\includegraphics[width=0.46\textwidth,  trim=0mm 0mm 0mm 0mm, clip]{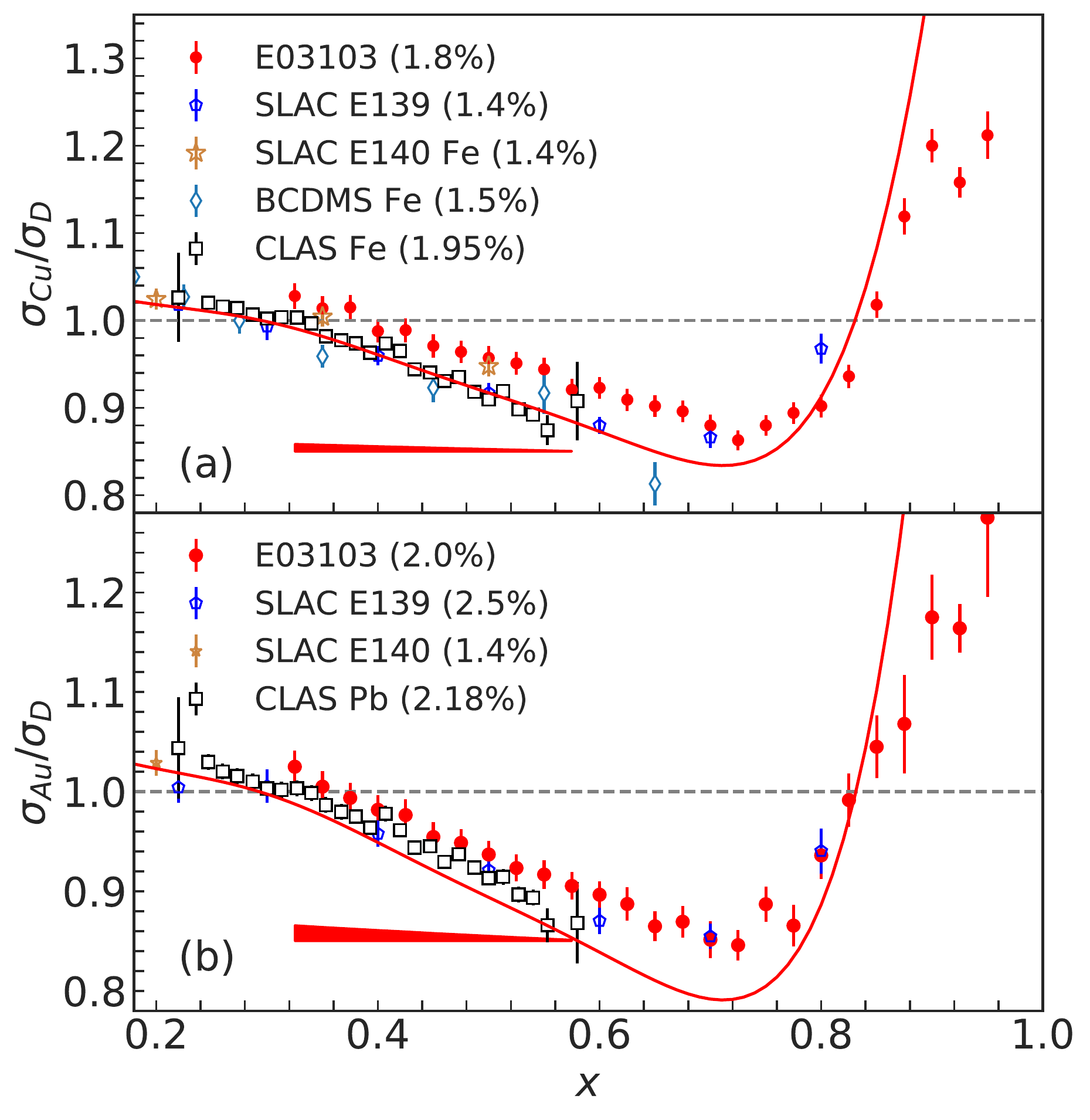}
\caption{(Color online) EMC ratios for Fe and Cu (a) and for Au and Pb (b)
as a function of $x$ for the 40 degree data. Uncertainties are as described in 
Fig.~\ref{emc_x_40deg_he4_fig}. The SLAC E139 and E140 data include updated 
Coulomb and isoscalar corrections, while the CLAS data has been updated with 
isoscalar corrections only since Coulomb corrections had already been applied. 
BCDMS~\cite{Benvenuti:1987az} Fe results are shown as published.
\label{emc_x_40deg_cu_fig}}
\end{center}
\end{figure}

Next, we examine the ratios for heavy nuclei in Fig.~\ref{emc_x_40deg_cu_fig}.
Several corrections to the data on heavy nuclei are larger or more uncertain
than for light nuclei. At low $x$, the radiative corrections and charge
symmetric background (see section~\ref{csbg.sssec}) are quite large. At high
$x$, Coulomb distortion becomes large for high-$Z$ targets; the correction for
Au ranges from 3\% at low $x$ to 12\% at high $x$ values for the 40$^\circ$
data.

\begin{figure}[htbp]
\begin{center}
\includegraphics[width=85mm, height=60mm ,trim=5mm 0mm 10mm 15mm, clip]{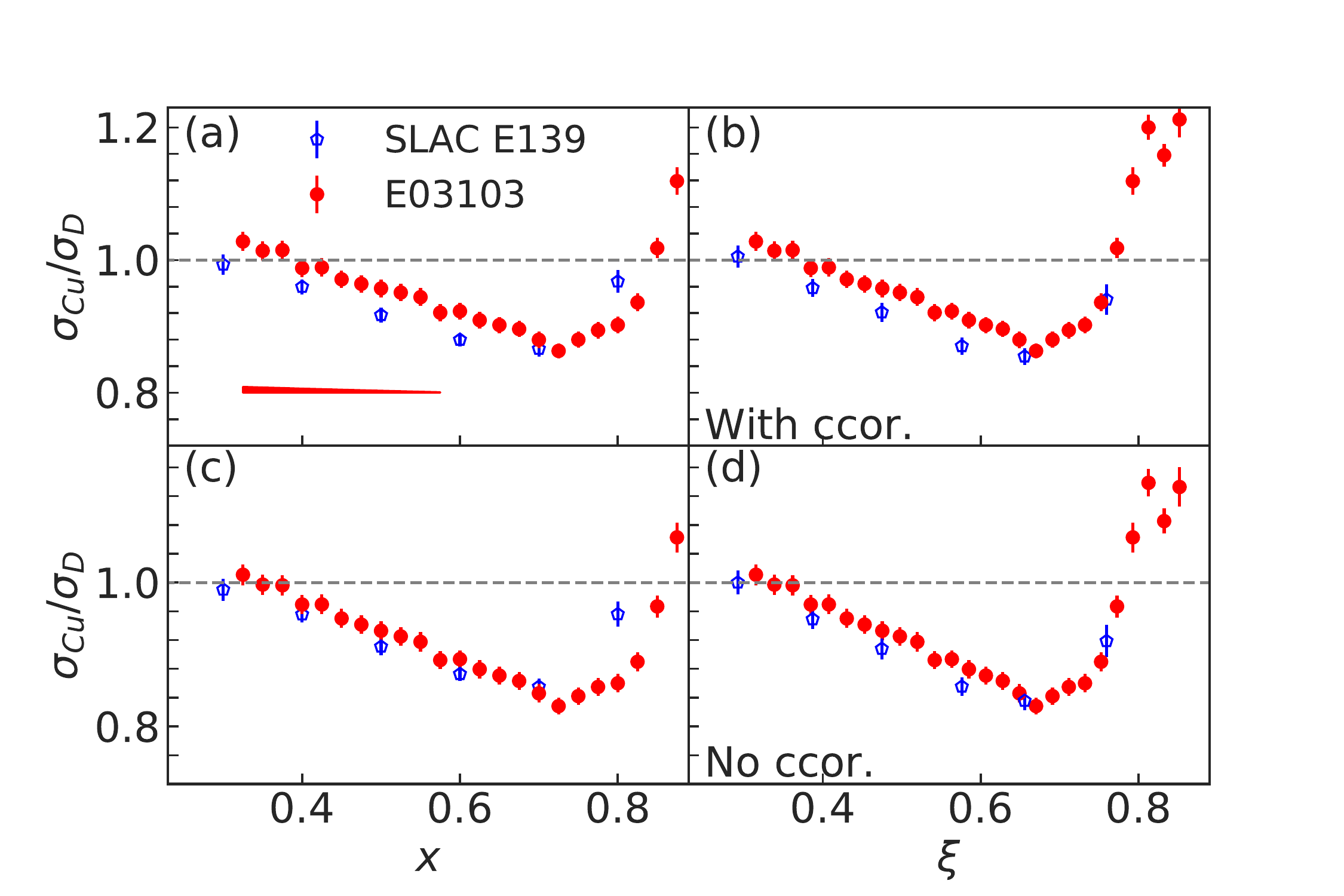}
\includegraphics[width=85mm , height=60mm , trim=5mm 0mm 10mm 15mm, clip]{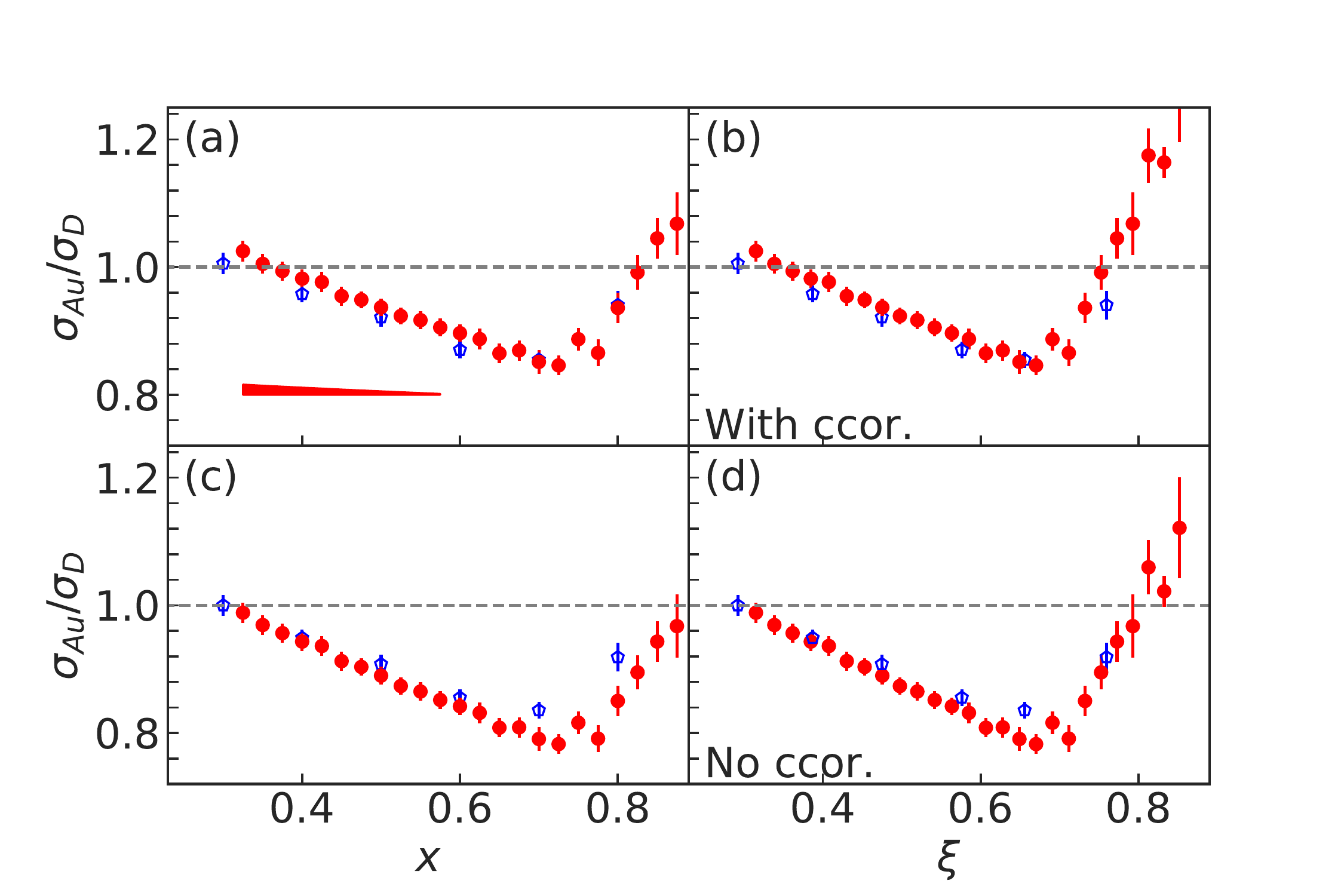}
\caption{(Color online) EMC ratios for our Cu and Au data compared
to the SLAC Fe and Au data, respectively, shown using four different sets of
corrections.  The panels on the left (right) side show the ratio vs $x$
($\xi$), while the panels on the top (bottom) show the ratios with (without)
Coulomb corrections applied. For each target, panel (b) shows the comparison
where one expects the best agreement between different measurements,
assuming that the Coulomb and so-called target mass corrections account for any 
$\theta$ and $Q^2$ dependence in the cross section ratios.  For all nuclei, 
high-$x$ SLAC and JLab results are in good agreement, after taking into account 
the scale uncertainties in the measurements.
\label{emc_slac_2by2zone_cu_fig}}
\end{center}
\end{figure}

Taking normalization uncertainties into account, our large-$x$ results are in
generally good agreement with the SLAC data, although the SLAC ratios at
$x=0.8$ are always slightly higher than our results.  This is possibly because the $x=0.8$ SLAC points were taken at higher $Q^2$ values ($Q^2=10$~GeV$^2$) than the E03103 data ($Q^2 \approx 6$~GeV$^2$), leading to a noticeable difference between the target mass corrections needed for the two data sets.
Figure~\ref{emc_slac_2by2zone_cu_fig} shows the points
plotted as a function of $x$ (left panels) and $\xi$ (right panels), where
plotting the ratio vs. $\xi$ provides the dominant part of the target mass
correction. The target mass correction shifts all points lower values of
$\xi$ with the largest shifts occurring at large $x$. When plotted as
a function of $\xi$, the EMC ratios are consistent within the scale uncertainties.

At small $x$ values, we find systematic disagreements with the SLAC
measurements.  While the light isoscalar nuclei are in relatively good
agreement with the E139 results, the $^3$He ratios are systematically lower
than HERMES for $x \leq 0.4$ (although the region of overlap is small), 
and the very heavy nuclei are systematically higher. Given the normalization 
uncertainties, it is difficult to conclude that there is a true inconsistency 
between the data sets, but we examine the pattern of disagreement to evaluate 
possible explanations for the small differences.

First, note that these nuclei have large isoscalar corrections, which
are of the opposite sign for $^3$He and the heavy nuclei.  However, the
low-$x$ region has the least uncertainty in the ratio of
$F_2^n/F_2^p$~\cite{accardi11, arrington12b}, and the correction becomes
smaller at low-$x$ values, where the $F_2^n/F_2^p$ becomes closer to unity. In
addition, the SLAC data as presented here include the updated isoscalar
correction that we apply to our data, and thus such a discrepancy would have
to be associated with the $Q^2$ dependence of the isoscalar correction.  It
therefore seems unlikely that it could be responsible for the difference between
data sets at small $x$.

The heavy nuclei also have significant corrections due to Coulomb distortion,
radiative corrections, and charge-symmetric backgrounds.  The charge-symmetric 
background is directly measured for all nuclei so it is unlikely this is the 
source of the discrepancy.  It is interesting to note that while effects due to 
Coulomb distortion tend to be smaller at low $x$, the agreement between the 
E03103 and SLAC results for heavy targets is apparently better with no Coulomb 
corrections applied to either data set.

\begin{figure}[htbp]
\begin{center}
\includegraphics[width=.44\textwidth,trim={6mm 4mm 15mm 15mm}, clip]{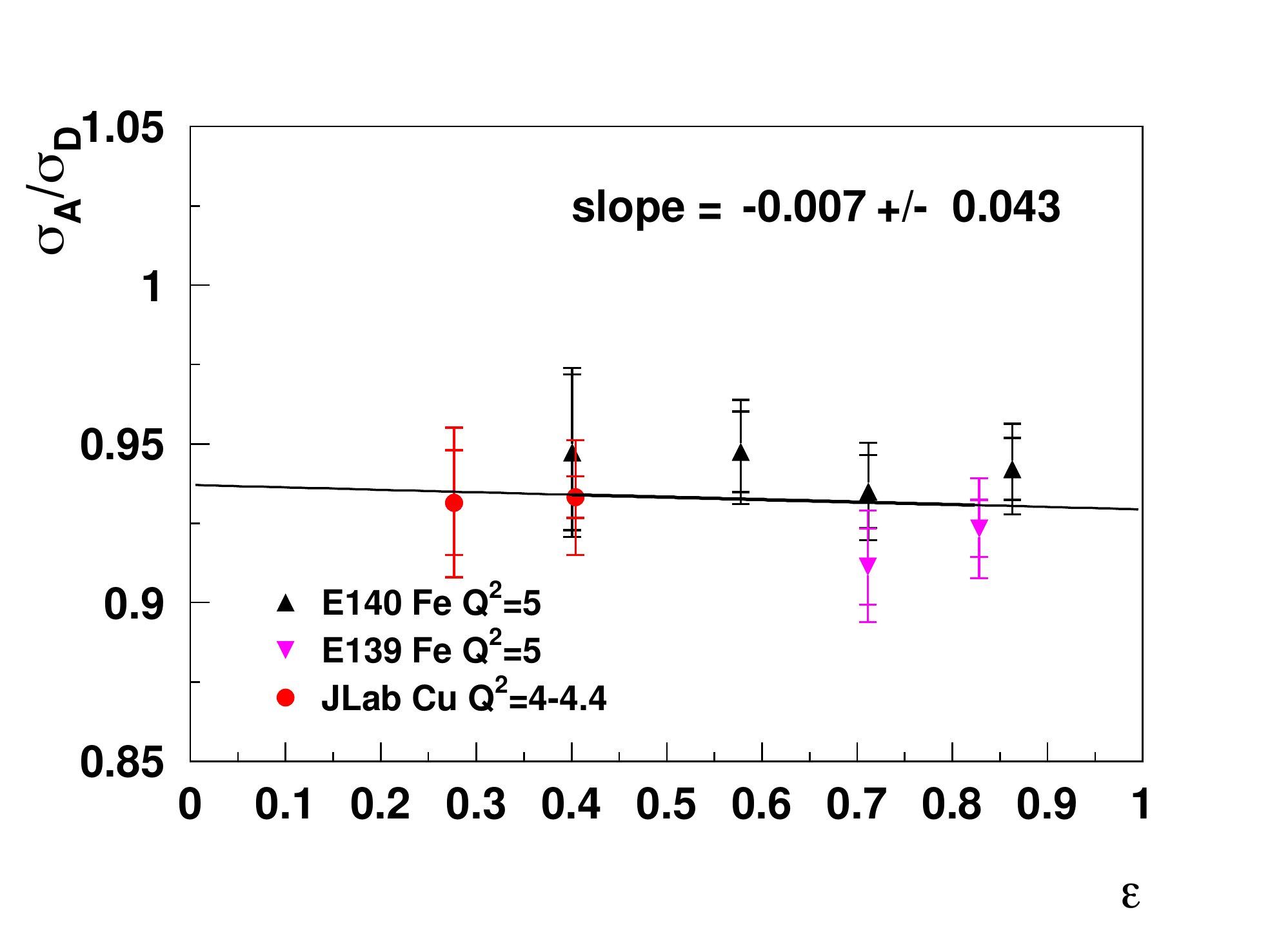}
\includegraphics[width=.44\textwidth,trim={6mm 4mm 15mm 15mm}, clip]{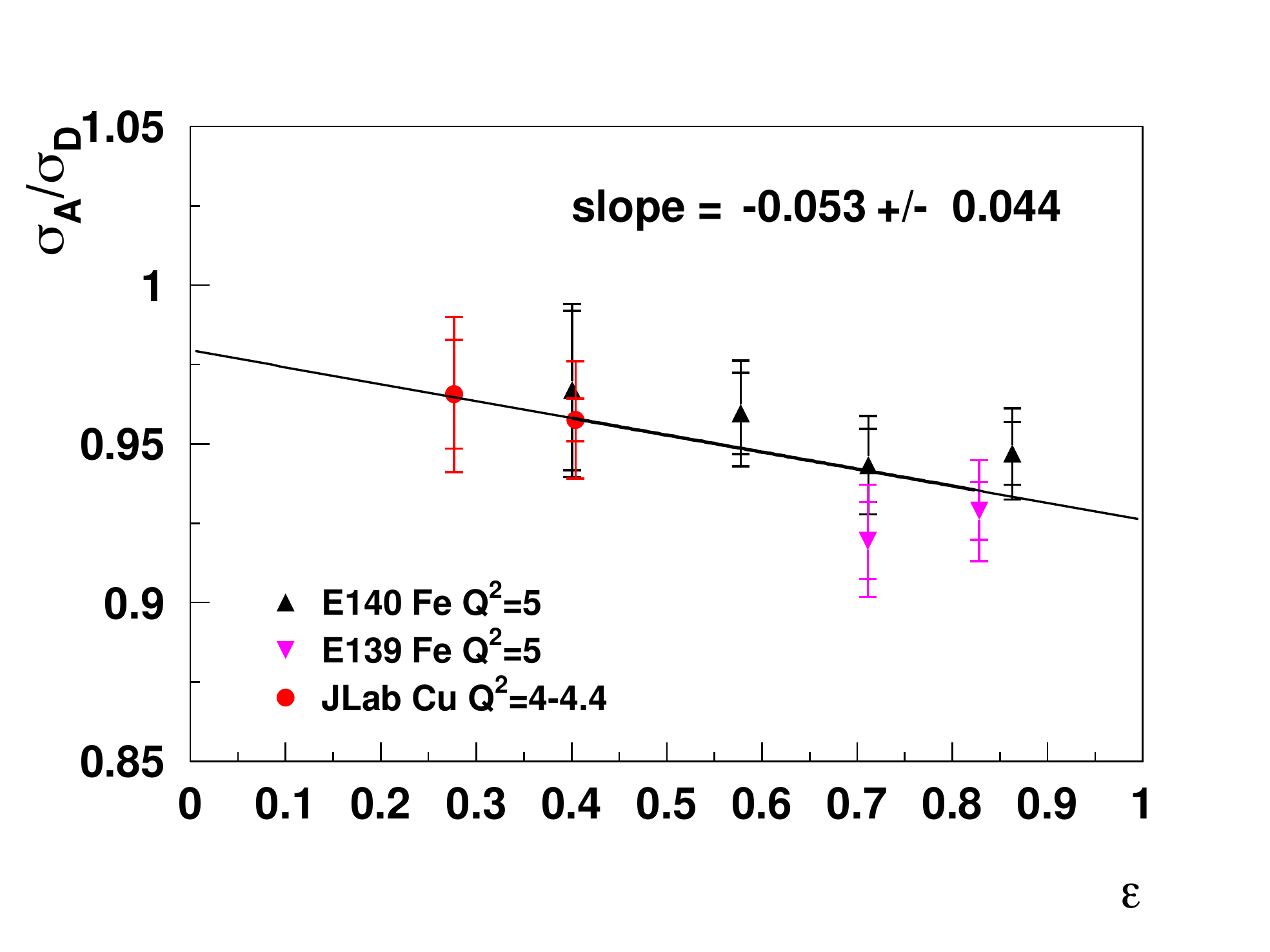}
\caption{(Color online) Extracted cross section ratios using the updated data 
from ~\cite{slace140,slace139} and E03103 experiment as a function of $\epsilon$
for the Fe/Cu targets for $x=0.5$ and $Q^2$ values as mentioned in the legend.  
The top (bottom) plot shows the target ratio without (with) Coulomb corrections 
applied. Inner error bars denote statistical and point-to-point uncertainties 
combined in quadrature while out error bars include contributions from
normalization uncertainties. The uncertainty on the slope is calculated from 
point-to-point errors as well as the experiment-dependent normalization uncertainties.
\label{epsdep_cu_fig}}
\end{center}
\end{figure}

Since the Coulomb correction factors (see section~\ref{cc.ssec}) are substantial 
for the heavy nuclei, it motivated us to further investigate the details of this 
correction; in particular the impact of its strong angular dependence. This
angular dependence could potentially affect the apparent $\epsilon$ dependence
of the cross section ratios. As mentioned in the introduction, the 
identification of the cross section ratio with the $F_2$ ratio, and thus the EMC 
effect, is valid only if $\epsilon=1$ or $R_{A_1} =R_{A_2}$ (identical ratio of 
longitudinal to transverse virtual-photon absorption cross section for the two 
nuclei). This idea was tested by SLAC E140~\cite{slace140}, which set limits on 
any possible nuclear dependence for $R$. They assumed the Coulomb distortion 
effects were small and did not include these corrections in their analysis. 
However, a re-examination of the SLAC 140~\cite{slace140}, SLAC 
E139~\cite{slace139} (including updated Coulomb and isoscalar corrections) and 
preliminary results for the Cu target from E03103 data suggested a non-zero 
nuclear dependence in $R_{A}-R_{D}$~\cite{Solvignon:2009it}.

Here we present an updated version of the analysis initially performed 
in~\cite{Solvignon:2009it}. Figure~\ref{epsdep_cu_fig} shows the $\epsilon$ 
dependence of the extracted cross section ratios for the Cu (Fe target for the 
SLAC experiments) target extracted for $x=0.5,$ $Q^2\sim 5$~GeV$^2$ point. In 
this analysis, the data at low $\epsilon$ values from the E03103 experiment are 
combined with the measurements from SLAC~\cite{slace140,slace139} to study the 
$\epsilon$ dependence of the cross section ratios. The slope derived using a 
linear fit after accounting for the appropriate normalization uncertainties 
between  different experimental data sets is found to be consistent with zero 
(see top plot in figure~\ref{epsdep_cu_fig}). However, after the application of 
Coulomb corrections there is a change in the slope (from -0.007$\pm$0.043
to -0.053$\pm$0.044). This analysis hints at the interesting possibility that 
there may be a non-trivial $\epsilon$ dependence for the cross section ratios, 
implying a detectable nuclear dependence of $R =\sigma_L/\sigma_T$ at large $x$. 

There have been other indications of possible $A$ dependence to 
$R$~\cite{Tao:1995uh, Tvaskis:2006tv, Mamyan:2012th, Alsalmi:2019sie}.  These 
previous results are consistent with a decrease in $R$ for nuclei with more 
neutrons, which could explain the observation of an increase in $\sigma_A/\sigma_D$ 
for $^3$He and a decrease for heavier nuclei with a significant neutron excess. 
However, we cannot exclude the possibility that these features are the result of 
errors in our knowledge of the thickness of these targets which give shifts in the
ratios which happen to vary with the $N/Z$ ratio of the nucleus. More definitive 
information with respect to a possible $A$ dependence of $R$ will be forthcoming
in the final analysis of Hall C experiments E02109~\cite{e02109_proposal} and
E04001~\cite{e08104_proposal}, which took data primarily (although not exclusively)
in the resonance region, and the future E12-14-002~\cite{e1214002_proposal}, 
which will emphasize measurements in the DIS region.

\subsection{$A$ dependence of the EMC effect}\label{adepresult.ssec}

\begin{figure}[htb]
\begin{center}
\includegraphics[width=0.48\textwidth,trim={5mm 5mm 15mm 15mm}, clip]{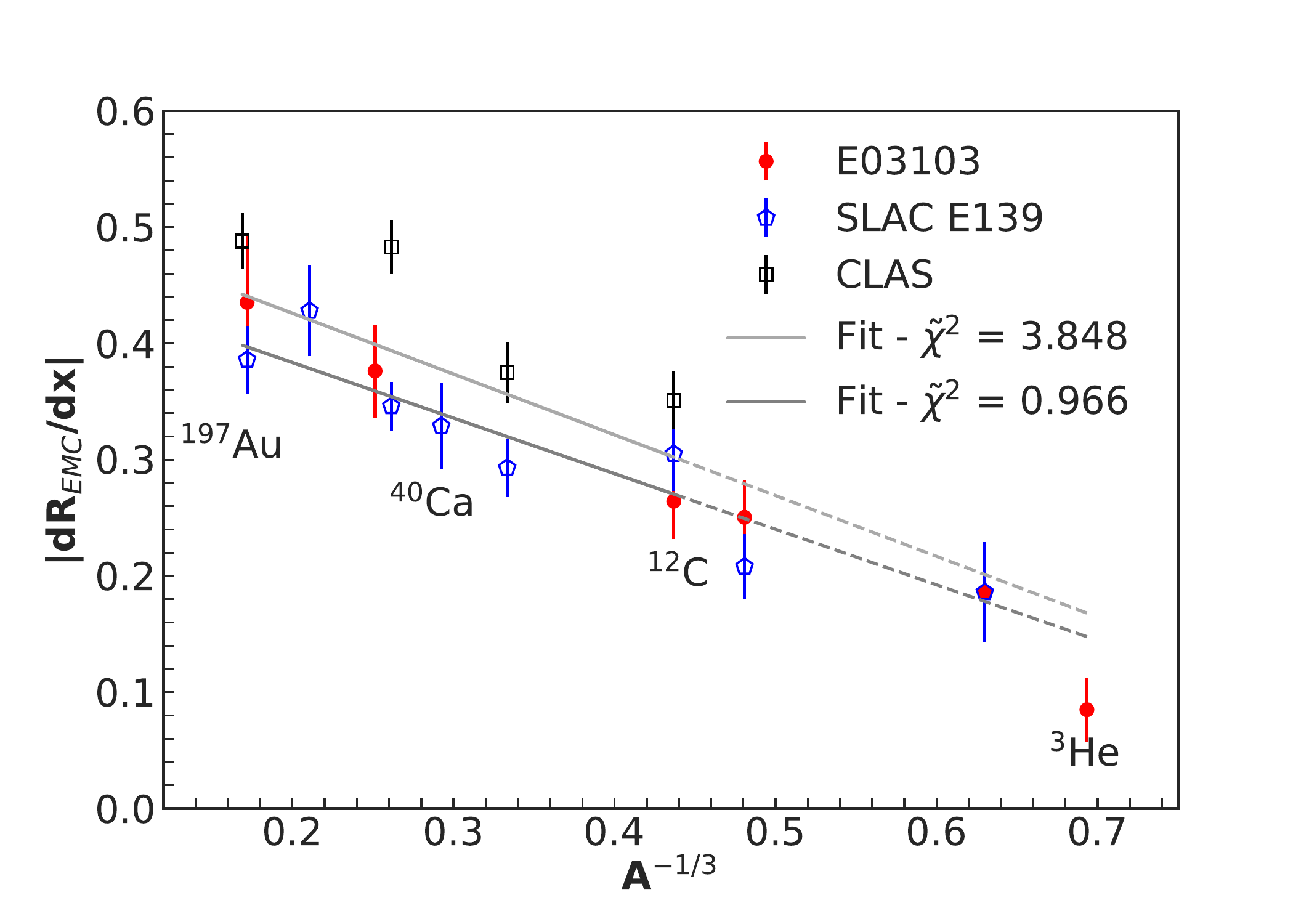}
\caption{(Color online) EMC slope vs. $A$ for JLab E03103 (this work), SLAC E139~\cite{slace139}, and CLAS~\cite{Schmookler19}. The linear fit excludes $A<12$ nuclei, with the upper fit (and reduced $\chi^2$ value including all data sets, and the lower excluding the CLAS data.}
\label{adep_slac_jlab}
\end{center}
\end{figure}

The overall size of the EMC effect is parameterized in terms of the $x$ dependence (slope) of the
EMC ratios, $R_\text{EMC}(x)$. Table~\ref{slope_emc_table} shows the EMC slopes,
$|dR_\text{EMC}/dx|$ for $0.3<x<0.7$, extracted from data from SLAC, CLAS and this
experiment. This table is an updated version of
table 1 provided in~\cite{arrington12c} which includes some of the updated
results from E03103 as well as the recent CLAS data~\cite{Schmookler19}.
The slopes are shown vs. $A$ in Fig.~\ref{adep_slac_jlab}. The CLAS slopes
are systematically higher than those from the other experiments. This, combined
with the fact that CLAS does not provide results on nuclei lighter than carbon,
means that a combination of the slopes for all nuclei (or $A\ge4$) will yield a
larger $A$ dependence than any of the individual data sets.  Each experiment uses a
single deuteron data set for all A/D ratios, so the deuteron uncertainties should 
be treated as a common normalization uncertainty for all ratios from a given 
experiment in a complete analysis of the $A$ dependence.

\begin{table}[htb]
\begin{center}
\caption{EMC slopes extracted from SLAC~\cite{slace139,arrington12c}, CLAS~\cite{Schmookler19}, and this experiment. Slopes are extracted using consistent isoscalar corrections for all three experiments, and with Coulomb corrections applied to all three data sets.}
\begin{tabular}{|l|l|l|l|}
\hline
A          & JLab E03103      & SLAC E139         & CLAS              \\ \hline
$^{3}$He   & 0.085 $\pm$ 0.027 & -                 & -                 \\
$^{4}$He   & 0.186 $\pm$ 0.030 & 0.186 $\pm$ 0.043 & -                 \\
$^{9}$Be   & 0.250 $\pm$ 0.032 & 0.208 $\pm$ 0.028 & -                 \\
$^{12}$C   & 0.264 $\pm$ 0.033 & 0.305 $\pm$ 0.032 & 0.351 $\pm$ 0.025 \\
$^{27}$Al  & -                 & 0.293 $\pm$ 0.025 & 0.375 $\pm$ 0.026 \\
$^{40}$Ca  & -                 & 0.329 $\pm$ 0.037 & -                 \\
$^{56}$Fe  & -                 & 0.346 $\pm$ 0.021 & 0.483 $\pm$ 0.023 \\
$^{63}$Cu  & 0.376 $\pm$ 0.040 & -                 & -                 \\
$^{107}$Ag & -                 & -                 & -                 \\
$^{197}$Au & 0.435 $\pm$ 0.059 & 0.386 $\pm$ 0.029 & -                 \\
$^{208}$Pb & -                 & -                 & 0.488 $\pm$ 0.024 \\ \hline
\end{tabular}

\label{slope_emc_table}
\end{center}
\end{table}

It is not clear why the CLAS EMC ratios yield larger slopes. This data set is taken 
at lower $Q^2$ than the E03103 and SLAC data, but target mass corrections yield a
larger slope (when fitting $F_2^{(0)}(x)$ rather than $F_2(x)$), and this increase 
is largest for CLAS because it is at lower $Q^2$. So applying target mass corrections would only 
increase the discrepancy between CLAS and the higher-$Q^2$ data sets. 
Ref.~\cite{Schmookler19} extracts the EMC ratios with a 
$Q^2$ cut of $Q^2>1.5$~GeV$^2$, but also examines the impact of other cuts. In their
analysis requiring $Q^2>2.0$~GeV$^2$, the average slope is decreased by 0.02 with 
little impact on the uncertainties, while $Q^2>2.5$~GeV$^2$ decreases the average 
slope by 0.035 but with much larger uncertainties. This suggests that inclusion of 
the lower $Q^2$ data may be increasing the slope, but it is difficult to quantify 
exactly how this impacts the comparison to the SLAC and JLab E03103 measurements.

Radiative corrections may also play a role in the difference in the CLAS EMC slopes. While CLAS, E03103, and SLAC all treat radiative effects based on the Mo and Tsai formalism~\cite{motsai_rc}, the detailed implementation and the cross section models used differ.  The radiative corrections program used by E03103 is based on that used for the earlier SLAC analysis, while CLAS uses the program described in~\cite{sargsian_inclusive}.  In particular, it is possible that differing approximations in the two approaches may result in systematic differences in the cross section and EMC slopes which can have a significant impact at smaller $x$ values.

The measurements on light nuclei, in particular for $^9$Be, show a clear deviation 
from scaling with density~\cite{seely09}, while the lightest nuclei show deviations 
from a smooth scaling with $A$.  It has been suggested that the local density
or the overlap of the struck nucleon with nearby neighbors may drive the
scaling of the EMC effect~\cite{seely09, arrington12c,arrington19}, or that 
off-shell effects in the highly virtual nucleons may in fact be
responsible~\cite{weinstein11,Schmookler19}.  In connection with these ideas, it has
been suggested that there may be both an $A$ dependence and an isospin dependence,
with additional modification in neutron-rich nuclei~\cite{arrington12c,
hen13, sargsian12}. So far, examinations of the $A$ dependence of the EMC ratios
under different assumptions about the isospin dependence are inconclusive, with the 
data being consistent with a significant flavor dependence based on the isospin 
structure of SRCs~\cite{Schmookler19}, but somewhat better described under the 
assumption of isospin independence~\cite{arrington12c, arrington19}. The additional 
data on heavy nuclei presented here and the small changes in the results for light 
nuclei do not significantly impact the 
conclusions of such comparisons, as a larger range of $N/Z$ is needed to increase 
the sensitivity~\cite{e12008_proposal}.

\section{conclusions}\label{concl.sec}

Deep inelastic scattering from $^{1,2}$H, $^{3,4}$He, Be, C, Cu, and Au 
targets was measured by the E03103 experiment at Jefferson Lab. The
ratios of inclusive nuclear cross sections with respect to the deuterium cross
section have been determined for $x>0.3$ for $Q^2$ values between 3 and 8
GeV$^2$. We include new data on heavy nuclei, not included in the original
results~\cite{seely09}, and provide a combined analysis of our results with previous
SLAC measurements~\cite{slace139} and recent CLAS data~\cite{Schmookler19},
applying consistent isoscalar and Coulomb corrections to the different data sets. 

E03103 addressed several of the limitations of previous measurements. We have
provided benchmark data for calculations of the EMC effect in light nuclei.
Predicted deviations from the $x$ dependence observed in heavy 
nuclei~\cite{smirnov99,burov99}
were not observed in $^3$He and $^4$He, but clear deviations from the simple
assumption of mass or density scaling of the EMC effect are observed.  At large $x$,
where binding and Fermi motion effects dominate, our new data for light and
heavy nuclei can serve as a base-line for traditional nuclear physics
calculations, including several few-body nuclei where structure related
uncertainties are minimal.

The data presented in this work will bridge the gap between
measurement of the EMC effect in light nuclei and medium heavy nuclei,
thus providing a comprehensive basis to test state of the art models that
attempt to explain the observed nuclear dependence.  For the moment, few
models provide an explicit prediction for the $A$ dependence, thus limiting the
ability to directly constrain these models without further effort on the
theory side.

While these data provide important new information about the EMC effect, there
are still limitations on how well these results could be used to constrain
explanations of the EMC effect. Some of these limitations will be addressed by
12 GeV experiments at Jefferson Lab~\cite{e12008_proposal, MARATHON}. This will
provide further information on the detailed behavior of the observed nuclear
nuclear dependence with an expanded set of light nuclei, including nuclei with
significant cluster structure and medium-to-heavy nuclei covering a range
of $N/Z$ to increase sensitivity to flavor-dependent effects.

\begin{acknowledgments}

This work was supported in part by the National Science Foundation, the U.S. Department of Energy, Office of Science, Office of Nuclear Physics, under contracts DE-AC02-05CH11231, DE-AC02-06CH11357, DE-AC02-05CH11231, DE-FE02-96ER40950 and DE-SC0013615, the South African NRF, and DOE contract DE-AC05-06OR23177, under which Jefferson Science Associates, LLC operates Jefferson Lab.

\end{acknowledgments}

\bibliography{main}

\end{document}